\documentclass[12pt]{article}
\usepackage{graphicx}
\usepackage{multicol} 
\usepackage{color}
\usepackage{amssymb}

\definecolor{rosso}{cmyk}{0,1,1,0.4}
\definecolor{rossos}{cmyk}{0,1,1,0.55}
\definecolor{rossoc}{cmyk}{0,0.5,1,0.2}
\definecolor{blu}{cmyk}{1,1,0,0.3}
\definecolor{blus}{cmyk}{1,1,0,0.6}
\definecolor{blucc}{cmyk}{1,0.4,0.2,0}
\definecolor{viola}{cmyk}{0,1,0,0.6}
\definecolor{viola2}{cmyk}{0,1,0.2,0.6}
\definecolor{verde}{cmyk}{0.92,0,0.59,0.25}
\definecolor{verdec}{cmyk}{0.92,0,0.59,0.15}
\definecolor{verdes}{cmyk}{0.92,0,0.59,0.4}
\font\tenrsfs=rsfs10 at 12pt
\font\sevenrsfs=rsfs7
\font\fiversfs=rsfs5
\newfam\rsfsfam
\textfont\rsfsfam=\tenrsfs
\scriptfont\rsfsfam=\sevenrsfs
\scriptscriptfont\rsfsfam=\fiversfs
\def\mathscr#1{{\fam\rsfsfam\relax#1}}

\newcommand{\mb}[1]{\mbox{\normalsize\boldmath $#1$}}
\newcommand{\rhob}{{\mb{\rho}}}

\newcommand{\fig}[1]{~\ref{fig:#1}}
\newcommand{\eq}[1]{~{\rm (\ref{eq:#1})}}

\newcommand{\GeV}{\,{\rm GeV}}
\newcommand{\TeV}{\,{\rm TeV}}
\def\circa#1{\,\raise.3ex\hbox{$#1$\kern-.75em\lower1ex\hbox{$\sim$}}\,}

\newcommand{\book}[5]{{\rm #1}, {\em #2} (#5) {\rm #4, #3}}

\newcommand{\NP}{Nucl. Phys.}
\newcommand{\PRL}{Phys. Rev. Lett.}
\newcommand{\PL}{Phys. Lett.}
\newcommand{\PR}{Phys. Rev.}
\newcommand{\beq}{\begin{equation}}
\newcommand{\eeq}{\end{equation}}
\newcommand{\bea}{\begin{eqnarray}}
\newcommand{\eea}{\end{eqnarray}}
\newcommand{\diag}{\hbox{diag}\,}

\newcommand{\nubarnu}{\raisebox{1ex}{\hbox{\tiny(}}\overline\nu\raisebox{1ex}{\hbox{\tiny)}}\hspace{-0.5ex}}
\newcommand{\ellbarell}{\raisebox{1.5ex}{\hbox{\tiny(}}\overline\ell\raisebox{1.5ex}{\hbox{\tiny)}}\hspace{-0.5ex}}
\def\circa#1{\,\raise.3ex\hbox{$#1$\kern-.75em\lower1ex\hbox{$\sim$}}\,}
\makeatletter

\def\art{\@ifnextchar[{\eart}{\oart}}
\def\eart[#1]#2#3#4#5#6{{\rm #2}, {\em #3 \rm #4} {\rm (#6) #5} [{#1}]}
\def\hepart[#1]#2{{\rm #2, #1}}
\newcommand{\oart}[5]{{\rm #1}, {\em #2 \rm #3} {\rm (#5) #4}}

\newcounter{alphaequation}[equation]
\def\thealphaequation{\theequation\hbox to
0.6em{\hfil\alph{alphaequation}\hfil}}
\def\eqnsystem#1{
\def\@eqnnum{{\rm (\thealphaequation)}}
\def\@@eqncr{\let\@tempa\relax \ifcase\@eqcnt \def\@tempa{& & &} \or
  \def\@tempa{& &}\or \def\@tempa{&}\fi\@tempa
  \if@eqnsw\@eqnnum\refstepcounter{alphaequation}\fi
\global\@eqnswtrue\global\@eqcnt=0\cr}
\refstepcounter{equation} \let\@currentlabel\theequation \def\@tempb{#1}
\ifx\@tempb\empty\else\label{#1}\fi
\refstepcounter{alphaequation}
\let\@currentlabel\thealphaequation
\global\@eqnswtrue\global\@eqcnt=0 \tabskip\@centering\let\\=\@eqncr
$$\halign to \displaywidth\bgroup \@eqnsel\hskip\@centering
$\displaystyle\tabskip\z@{##}$&\global\@eqcnt\@ne
\hskip2\arraycolsep\hfil${##}$\hfil& \global\@eqcnt\tw@\hskip2\arraycolsep
$\displaystyle\tabskip\z@{##}$\hfil
\tabskip\@centering&\llap{##}\tabskip\z@\cr}
\def\endeqnsystem{\@@eqncr\egroup$$\global\@ignoretrue} \makeatother

\newcommand{\eV}{\,\hbox{\rm eV}}

\newcommand{\km}{\,\hbox{\rm km}}

\newcommand{\sW}{s_{\rm W}}

\begin{document}

\thispagestyle{empty}

\begin{flushright}
{DFTT12/2005\\
IFUP--TH/2005-13\\
hep-ph/0506298\\

}
\end{flushright}
\vspace{1cm}

\begin{center}
{\LARGE \bf \color{rossos}
Spectra of neutrinos from\\
 dark matter annihilations}\\[1cm]

{
{\large\bf Marco Cirelli}$^a$,
{\large\bf Nicolao Fornengo}$^b$,
{\large\bf Teresa Montaruli}$^c$,\\
{\large\bf Igor Sokalski}$^d$,
{\large\bf Alessandro Strumia}$^e$,
{\large\bf Francesco Vissani}$^f$
}  
\\[7mm]
{\it $^a$ Physics Department, Yale University, New Haven, CT 06520, USA}\\[3mm]
{\it $^b$ Dipartimento di Fisica Teorica, Universit\`a di Torino\\ and INFN, Sez.\ di Torino, via P. Giuria 1, I-10125 Torino, Italia}\\[3mm]
{\it $^c$ University of Wisconsin, Chamberlin Hall, Madison, WI 53706, USA.\\ On leave of absence from Universit\` a di Bari\\ and INFN, Sez.\ di Bari, via Amendola 173, I-70126 Bari, Italia}\\[3mm]
{\it $^d$ INFN, Sez.\ di Bari, via Amendola 173, I-70126 Bari, Italia}\\[3mm]
{\it $^e$ Dipartimento di Fisica dell'Universit{\`a} di Pisa and INFN, Italia}\\[3mm]
{\it $^f$ INFN, Laboratori Nazionali del Gran Sasso, Assergi (AQ), Italia}\\[3mm]
\vspace{1cm}
{\large\bf\color{blus} Abstract}

\end{center}
\begin{quote}
{\large\noindent\color{blus}
We study  the fluxes of neutrinos from annihilations of dark matter particles in the Sun and the Earth. We give the spectra of all neutrino flavors 
for the main known annihilation channels: $\nu\bar\nu$, $b\bar b$, $\tau\bar\tau$, $c\bar{c}$, light quarks, $ZZ$, $W^+W^-$.
We present the appropriate formalism for computing the combined effect of oscillations,  absorptions, $\nu_\tau$-regeneration.
Total rates are modified by an ${\cal O}(0.1\div10)$ factor,
comparable to astrophysical uncertainties, that instead
negligibly affect the spectra.
We then calculate different signal topologies in neutrino telescopes: through-going muons, contained muons, showers, and study their capabilities 
 to discriminate a dark matter signal from  backgrounds.
We finally discuss how measuring the neutrino spectra can allow to reconstruct the fundamental properties of the dark matter: its mass and its annihilation branching ratios. 
}

\end{quote}

\newpage


\setcounter{footnote}{0}

\section{Introduction}
The most appealing scenario to explain the observed Dark Matter (DM) abundance  $\Omega_{\rm DM}\sim0.3$ consists in postulating that DM arises as the thermal relic of a new
stable neutral particle with mass $m_{\rm DM}$.
Assuming it has weak couplings $g\sim 1$, the right $\Omega_{\rm DM}$ is obtained for
$m_{\rm DM} \sim (T M_{\rm Pl})^{1/2}\sim \TeV$, where $T\sim 3\,{\rm K}$ is
the present temperature of the universe, and $M_{\rm Pl}\sim 10^{19}\GeV$ is the Planck mass~\cite{review}.
One motivated DM candidate is the lightest neutralino in 
supersymmetric extensions of the Standard Model with conserved matter parity,
that for independent reasons is expected to have a mass around the electroweak scale~\cite{GRS}.
Many other DM candidates have been proposed: 
we will generically have in mind a DM particle heavier than few tens of GeV, 
keeping the concrete connection to the neutralino as a guideline.
This scenario seems testable by DM search and by collider experiments:
one would like to see a positive signal in both kind of experiments
and to check if the same particle is responsible for both signals.
As emphasized in~\cite{Kane} this is an important but difficult goal.


A huge effort is currently put in experiments that hope to discover DM either directly (through the interaction of DM particles with the detector) or indirectly (through the detection of secondary products of DM annihilations). 
Among the indirect methods, a promising signal consists in neutrinos with energy $E_\nu \circa{<} m_{\rm DM}$ produced by annihilations of DM particles accumulated in the core of the Earth and of the Sun~\cite{idea,previous}, detected by large neutrino detectors. We will refer to them as `DM$\nu$'. 
IMB~\cite{IMB}, Kamiokande~\cite{Kamiokande}, Baksan~\cite{Baksan}, {\sc Macro}~\cite{MACRO}, Super-Kamiokande~\cite{SK},  AMANDA~\cite{AMANDA} and BAIKAL~\cite{BAIKAL} already obtained constraints on DM$\nu$ fluxes,
while experiments that are under construction, like ANTARES~\cite{ANTARES} and ICECUBE~\cite{ICECUBE}, or that are planned, like NEMO~\cite{NEMO}, NESTOR~\cite{NESTOR} and a Mton-scale water \v{C}erenkov detector~\cite{Mton}, will offer improved sensitivity. 



\medskip

We compute the spectra of neutrinos of all flavors generated by DM annihilations in the Earth and in the Sun.

Today, before a discovery, this can be used to convert experimental data into more reliable constraints on model parameter space and helps in identifying more relevant features of the DM$\nu$ signal searched for. For instance, we include in the analysis all main annihilation channels, we address the effect of neutrino oscillations and interactions with matter and we point out more experimental observables that those usually considered.

After a discovery the situation will be analogous to the solar neutrino anomaly: a natural source of neutrinos carries information about fundamental parameters and we must find realistic observables that allow to extract it.
As in that case, also in the DM case the total $\nu$ rate is the crucial parameter for discovery but is
plagued by a sizable ${\cal O}(10)$ astrophysical uncertainty. How can we then reconstruct the properties of the DM?

Astrophysical uncertainties negligibly affect the ratios between different neutrino flavors
and the neutrino energy spectra (as well as the closely related angular distributions~\cite{angdistribution}).
They depend on the DM mass $m_{\rm DM}$ and on the branching ratios
of the channels into which DM particles may annihilate:
$\nu\bar\nu$, $b\bar{b}$, $t\bar{t}$, $\tau^+\tau^-$, $W^+ W^-$, $ZZ$...
In order to extract these fundamental parameters from future data one needs to precisely compute
DM$\nu$ spectra taking into account the astrophysical environment,
where several processes are important.




\medskip

In section~\ref{Production} we motivate our phenomenological procedure and
compute the fluxes of electron, muon and tau neutrinos at production point:
the different density of the Earth and solar core affects energy loss of particles
that decay producing neutrinos.

\medskip

\begin{figure}[t]
$$\includegraphics[width=8cm]{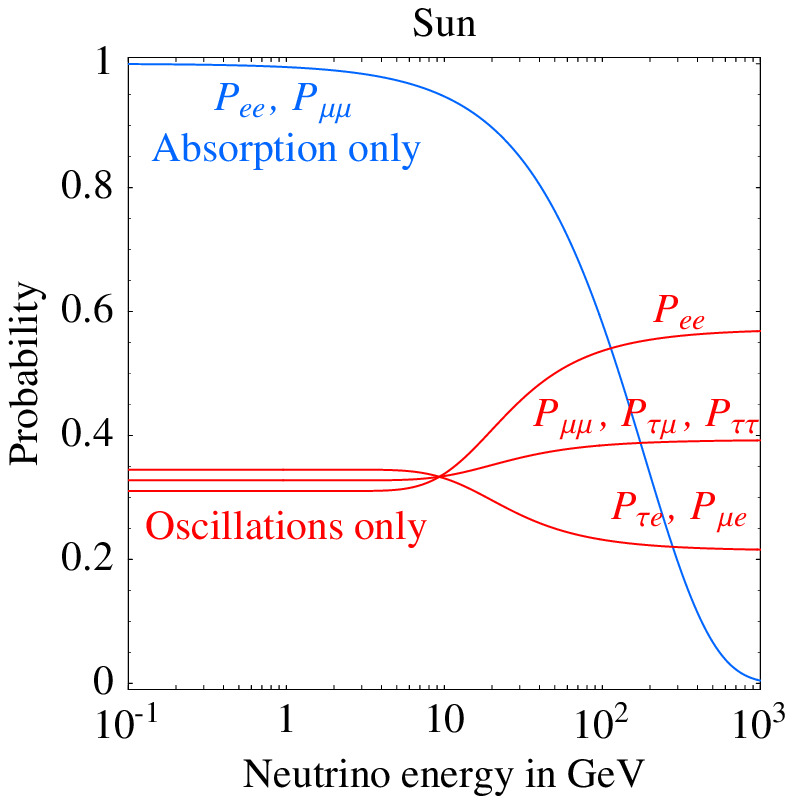}\qquad
\includegraphics[width=8cm]{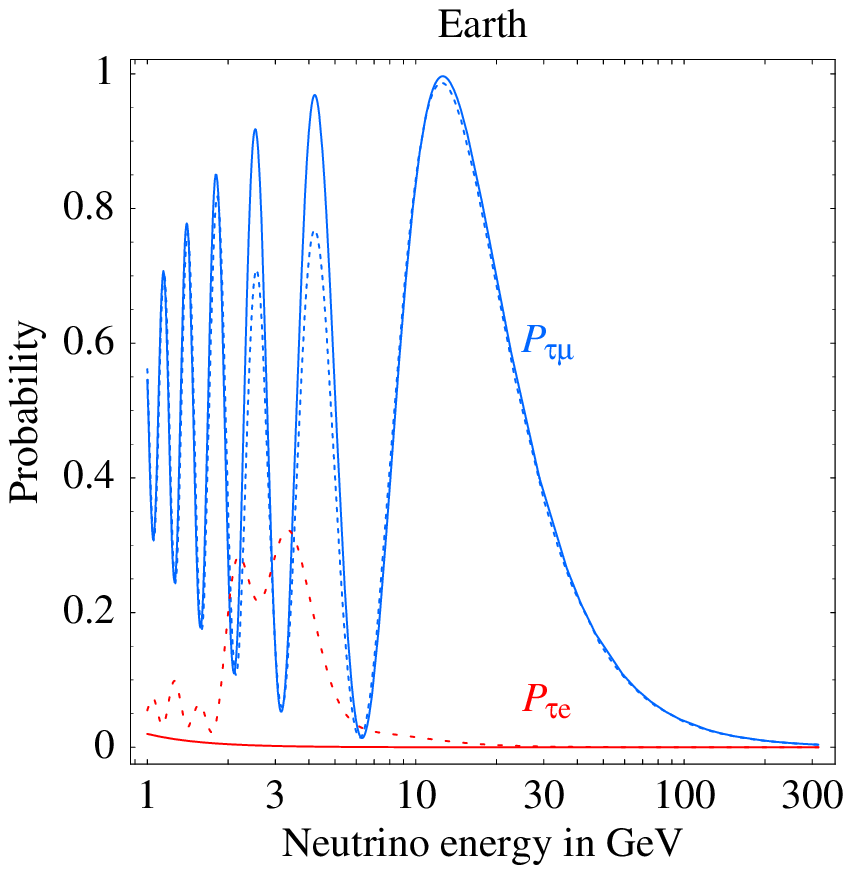}$$
\caption[X]{\label{fig:OscAbs}\em The left plot illustrates how oscillations and CC absorption
separately affect a flux of neutrinos produced in the center of the Sun.
The right plot shows the oscillation probabilities from the center of the Earth.
The continuous line applies to $\nu$ for $\theta_{13}=0$ and to $\bar\nu$ for
any allowed $\theta_{13}$, since matter effects suppress their mixing.
The dotted line applies to  $\nu$ for $\theta_{13}=0.1$ rad.
The average over the production point has been performed as appropriate for $m_{\rm DM}=100\GeV$. It is responsible for the damping effect visible at $E_{\nu}\circa{<}10\GeV$.
\label{fig:PEarth}}
\end{figure}

In section~\ref{formalism} we compute how propagation from the center of the Earth and of the Sun
affects the flavor and energy spectra.
At production, the neutrino flavor ratios from the DM annihilations are simply given by:
  $$
   \nu_e:\bar\nu_e:\nu_\mu:\bar\nu_\mu:
   \nu_\tau:\bar\nu_\tau=1:1:1:1:r:r \ .
  $$
In the Earth, the main effect is due to oscillations with `atmospheric' frequency: 
the neutrino oscillation length
$$\lambda_{\rm atm} = 4\pi E_\nu/|\Delta m^2_{\rm atm}| \approx 10^5\km (E_\nu/100\GeV)$$
is comparable to the Earth radius $R_\oplus = 6371\km$ if $E_\nu \lesssim 100\GeV$.
In the Sun, also the size of the production region of DM$\nu$ is of the same order.
Furthermore, in the Sun at $E_\nu \circa{>}10\GeV$ neutrino interactions start to be significant and solar oscillations cease to be adiabatically MSW-enhanced, as illustrated in fig.\fig{OscAbs}a.
Interactions manifest in several ways: absorption, re-injection of neutrinos of lower energy
(as produced by NC scatterings and $\nu_\tau$ CC scatterings), breaking of coherence among different flavors. 
These effects operate at the same time and with comparable importance: while previous works addressed the issues separately~\cite{nuDMosc,nuDMinterac}, the density-matrix formalism of  section~\ref{formalism} allows to take into account their combined action.

\medskip

In section~\ref{Earth} (\ref{Sun}) we give the resulting energy spectra of DM neutrinos of all flavors from the Earth (Sun).
We consider the standard through-going muon signal and point out
that other classes of events can be studied in realistic detectors and
have interesting features from the point of view of discriminating a DM$\nu$ signal
from the atmospheric background and of reconstructing DM properties.
This latter point is discussed in section~\ref{banana}.

\section{Neutrino production}\label{Production}
A flux of neutrinos is produced inside the Earth or the Sun as a
consequence of annihilation of dark matter particles which have been
gravitationally captured inside these celestial bodies
\cite{Gould:1987,capture1,edsjo}. The differential neutrino flux is:
\begin{equation}
\frac{dN_\nu}{dE_\nu} = \frac{\Gamma_{\rm ann}}{4\pi d^2} \, \sum_f {\rm BR}_f \frac{dN_f}{dE}
\label{eq:nuflux}
\end{equation}
where $f$ runs over the different final states of the DM annihilations
with branching ratios ${\rm BR}_f$, 
$d$ is the distance of the neutrino source from the
detector (either the Sun--Earth distance $r_{\rm SE}$ or the Earth radius $R_\oplus$) and
where the annihilation rate $\Gamma_{\rm ann}$ depends on the rate $\Gamma_{\rm capt}$ of
captured particles by the well known relation:
\begin{equation}
\Gamma_{\rm ann}=  \frac{\Gamma_{\rm capt}}{2} \tanh^2(t_0/\tau_A)
\label{eq:gamma}
\end{equation}
where $t_0=4.5$ Gyr is the age of the Earth and of the Sun and $\tau_A$
denotes a time-scale for the competing processes of capture and
annihilation, and it is proportional to the DM annihilation cross
sections (for explicit formul\ae{} see~\cite{Gould:1987,edsjo,helio}). For the present discussion we
just remind that the capture rate $\Gamma_{\rm capt}$ depends linearly on the
DM/nucleus scattering cross section and on the local dark matter
density $\rho_{{\rm DM}}$:
\begin{equation}
\Gamma_{\rm capt} \propto \sigma_{\rm scattering}\, \rho_{\rm DM}
\label{eq:capture}
\end{equation}
Eq. (\ref{eq:gamma}) shows that the two competing processes of capture
and annihilation may eventually reach an equilibrium situation when
the time scale $\tau_A$ is much smaller than the age of the body. While
this is usually the case for the Sun, it does not always occur for the
Earth, since in this case the gravitational potential, which is
responsible for the capture, is much smaller. 
Equilibrium is fulfilled only for large elastic
scattering cross sections.

\subsection{Observables with and without astrophysical uncertainties}
{}From the previous equations we see that the neutrino signal shares
both astrophysical and particle physics uncertainties. However, the
shape of neutrino spectra are virtually free from the astrophysical
ones, even in presence of oscillations -- as we shall discuss below -- and therefore they can be potentially used to study the
fundamental DM parameters, like its mass and annihilation
channels. This topic will be addressed in section~\ref{banana}.

\smallskip

A quantity which suffers from sizable astrophysical
uncertainties is the total DM$\nu$ flux, mainly due to the poor knowledge of the local DM density
$\rho_{\rm DM}$. The experimental indetermination on this parameter is
still large. Detailed analyses, performed assuming different DM
density profiles, find densities that vary by about one order of
magnitude~\cite{DMastro}. This translates into the same order of
magnitude uncertainty on the DM$\nu$ rate, due to the direct
proportionality between the neutrino signal and the local dark matter
density through the capture rate. The uncertainty on $\rho_{\rm DM}$
can also play a role in the setup of capture/annihilation equilibrium
in the Earth, giving an additional reduction effect.

An additional astrophysical uncertainty comes from the local DM
velocity distribution function. Since capture is driven by the
relation between the DM velocity and the escape velocity of the
capturing body ($11\km/{\rm sec}$ and $620\km/{\rm sec}$ at the surface of the
Earth and the Sun, respectively), the high--velocity tail of the DM
velocity distribution function may play a role. In the case of the
Earth, the actual motion of DM particle in the solar system is another
relevant ingredient which can alter significantly the predicted
capture rate and therefore the predicted DM$\nu$
rate. Recently this issue has been re--evaluated in~\cite{edsjo},
where it has been shown that in the Earth the capture rate of DM
particles heavier than a few hundreds of GeV may be considerably reduced.

\medskip

On the contrary, neutrino spectra can be considered as virtually free
from astrophysical uncertainties. The shape of the spectra depends on
the type of particle produced in the annihilation process and on its
subsequent energy--loss processes (remember that annihilation occurs in
a medium, not in vacuum) before decaying into neutrinos. 
As a consequence of the thermalization of the captured DM, the density
distribution of DM particles within the Sun or the Earth is predicted
to be~\cite{n(r)}:
\beq\label{eq:RDM}  n(r) = n_0 \exp(-r^2/R_{\rm DM}^2)
\qquad R_{\rm DM} =\frac{R}{\sqrt{\beta m_{\rm DM}}} 
\eeq 
where $r$ is the radial coordinate, $\beta = 2\pi G_{\rm N} \rho_0
R^2/3T_0$, $\rho_0$ and $T_0$ are the central density and temperature
of the body (Sun or Earth) and $R$ is its radius. These astrophysical
parameters are relatively well known, much better than the above mentioned galactic ones. Numerically:
\beq \beta = \left\{\begin{array}{ll} 1.76/\GeV &
\hbox{for the Earth,}\\ 98.3/\GeV & \hbox{for the Sun.}
\end{array}\right.
\eeq
This means that the size of the production region of DM neutrinos is
$\sim 500\km \sqrt{100\GeV/m_{\rm DM}}$ in the Earth and $\sim 0.01
\,R_\odot \sqrt{100\GeV/m_{\rm DM}}$ in the Sun.

The finite size can affect the spectra in two ways: 1) Different
DM$\nu$ originate in regions with different densities, so that hadrons
may loose different amounts of energy before decaying into
neutrinos. This, however, is not an important effect because the size
of the production region is small enough that the matter density can
be safely considered as constant where neutrinos are produced; 2)
Neutrino propagation: while in the Earth the production region has a size
much smaller than the atmospheric oscillation length, in the Sun the size is instead comparable. The resulting coherence between different flavors gets however washed--out by the much
longer eventual propagation up to the Earth.

In conclusion, the production regions are small enough that performing
the spatial average according to eq.\eq{RDM} gives a final total
spectrum negligibly different than the one obtained by just assuming
that all DM$\nu$ are produced at the center of the Earth or of the
Sun. We will  prove this statement in section~\ref{sec:average}, after discussing our treatment of
neutrino propagation.

\paragraph{DM$\nu$ spectra and fundamental parameters}
Since DM particles inside the Earth or the Sun are highly
non--relativistic,
their annihilations occur almost at rest and the main
phenomenological parameters that determine DM$\nu$ spectra are the DM
mass $m_{\rm DM}$ and the BR of the basic channels into which DM
particles may annihilate, as shown in eq.~(\ref{eq:nuflux}):
$q\bar{q}$, $\ell\bar{\ell}$, $\nu\bar\nu$, $W^+ W^-$, $ZZ$ and higgs
particles or mixed higgs/gauge boson final states~\cite{previous,ritz}. 
Besides the direct $\nu\bar\nu$ annihilation
channel, neutrinos originate from the decays of the particles produced
in the annihilation. In the case of quarks, hadronization will produce
hadrons whose subsequent decay may produce neutrinos. Also charged
leptons, apart from electrons, will produce neutrinos. In the
case of gauge bosons or higgs particles, their decay will produce again
leptons or quarks, which then follow the same evolutions just
described.

The basic ``building blocks'' we need in order to  calculate
DM$\nu$ fluxes are therefore the spectra produced by the
hadronization of quarks and by the decay of charged leptons in the Sun
and Earth cores. Among leptons, only the $\tau$ is relevant, since muons are
stopped inside the Earth and the Sun before they can decay \cite{ritz},
and therefore produce neutrinos of energy below experimental
thresholds for the signal topologies we will discuss later on (up-going
muons, contained muons and showers in large area neutrino
telescopes). For the present discussion we consider neutrino energies
above 0.5 GeV.
In all the other situations, which involve gauge and
higgs bosons, we can make use of the basic spectra discussed above and
calculate the neutrino spectra by just composing properly boosted
spectra originated from quarks or $\tau$, following the decay chain of
the relevant annihilation final state particle. The method is briefly
sketched in Appendix \ref{app:boost} for completeness.

In this paper we are interested in the discussion of the effect
induced by oscillations on the neutrino signal.  We therefore need to
calculate the spectra for all three neutrino flavors. We model the
hadronization and decay processes by means of a  PYTHIA Monte Carlo
simulation~\cite{pythia}, suitably modified in order to take into
account the relevant energy losses.  The neutrino spectra which we
obtain are presented in numerical form, but we also provide an
interpolating function for all the quark flavors and the $\tau$
lepton.

We will not consider effects on the neutrino fluxes arising from the
spin of the DM particle. In general, the DM spin may control the
polarization of primary particles produced in the annihilation. For
instance, if the DM is a Majorana fermion (such as the neutralino) it
can only decay into $\tau_L\bar\tau_R + {\rm h.c.}$ (with amplitude
proportional to the $\tau$ mass) while a scalar can decay into
$\tau_L\bar\tau_L$ and $\tau_R\bar\tau_R$ with different branching
ratios. Only if the branching ratios are the same the DM$\nu$ spectrum is
equal to the Majorana case. We will assume that this is the case,
studying a single $\tau\bar\tau$ channel rather than two slightly
different $\tau_L\bar\tau_L$ and $\tau_R\bar\tau_R$ channels.
Furthermore, when discussing direct annihilation into neutrinos
(possible for a scalar DM) we will assume that the flux is equally
divided among the three flavors.

\bigskip

We now discuss the calculation of the spectra for the relevant final
states and their distinctive features. In this paper we will not focus
on a specific DM candidate, rather we will attempt a more general
phenomenological analysis. Our results can therefore be used for any DM
candidate, by using the basic spectra of primary annihilation
particles given here. The full spectrum for a specific candidate in a specific model is
then easily constructed by summing up these building blocks
implemented by the information on the annihilation branching ratios ${\rm BR}_f$ in that model.

\subsection{Annihilation into light fermions}

The direct ${\rm DM~DM}\to \nu\bar\nu$ channel (if allowed with a
reasonable branching ratio) usually gives the dominant contribution to
DM$\nu$ signals: its spectrum is a line at $E_\nu = m_{\rm DM}$ so it gives the neutrinos with highest multiplicity and energy. If the DM is a Majorana fermion the ${\rm DM~DM}\to f\bar{f}$ annihilation
amplitude is proportional to $m_f$, so that the $\nu\bar\nu$ channel
is irrelevant and the most important fermions are the heaviest ones:
$b\bar{b}$, $\tau\bar\tau$ $c\bar c$ and, if kinematically accessible,
$t\bar{t}$ ({\em i.e.}\ if $m_{\rm DM}> m_t$). Even in the context of
SUSY models the relative weight of their branching ratios should be
considered as a free parameter: significant deviations from the
qualitative expectation $\sigma({\rm DM~DM}\to b\bar{b})/\sigma({\rm
DM~DM}\to \tau\bar\tau) = 3m_b^2/m_\tau^2$, which exactly holds
in the case of a dominant higgs exchange, can arise if staus are much
lighter than sbottoms.

Once a quark is produced, it will hadronize and produce a large number
of mesons and baryons, which will then decay and eventually produce
neutrinos.  
We calculate the $\nu_e$, $\nu_\mu$ and $\nu_\tau$ fluxes originated by
quark hadronization and lepton decay in the medium by following and
properly adapting the method of~\cite{ritz}. We improve on previous analyses \cite{previous} by calculating full
spectra for all neutrino flavors (usually only $\nu_\mu$ were
considered since the main signal is upgoing muons and oscillations
have been neglected, except in a few seminal cases~\cite{nuDMosc}). 
We also provide here neutrino fluxes coming from
light quarks: their contribution is usually neglected
since they mostly hadronize into pions, which are stopped in the medium and
do not produce neutrinos in an interesting energy range. 
We show below
that for relatively large DM masses neutrinos from light quarks should be taken into account
in a precise computation.
Their main contribution occurs through the excitation of $c$ quarks in the hadronization process and
subsequent decay of $c$ mesons.

For completeness, we include also the case of DM annihilation into
gluons, which may be relevant for some DM candidate. For instance, in
the case of neutralinos, gluons can be produced at one loop level: even
though this channel is usually subdominant, it can provide some
contribution in specific portions of the SUSY parameter space,
especially for light neutralinos.

\paragraph{Annihilation inside the Earth}

As previously discussed, the annihilation process occurs primarily in
the center of the Earth, where the density is $\rho = 13$ g cm$^{-3}$. 
Therefore the particles produced in DM annihilations may undergo energy
loss before decay. 


In the case of charged leptons, the energy loss process is
calculated by means of the Bethe--Bloch equation. The typical stopping
time is of the order of $\tau_{\rm stop} = 2\cdot 10^{-10}$ sec. This
has to be compared to the boosted lifetime $\gamma \tau_{\rm dec}$,
where $\tau_{\rm dec} = 2.2\cdot 10^{-6}$ sec for the muon and $\tau_{\rm
dec} = 3\cdot 10^{-13}$ sec for the tau. We see that for leptons with
energies up to 1 TeV, muons are always stopped before their decays,
while taus may decay as if they were in vacuum~\cite{ritz}. 
In order to take into account the small deviations from the limit described above we 
adapted the PYTHIA code to allow free lepton decay if
$\gamma \tau_{\rm dec} < \tau_{\rm stop}$, 
otherwise the lepton is
stopped and then it decays.


As for the hadrons, the situation is different if the jets are
produced by light or heavy quarks. The  interaction time in a material with density $\rho$ is~\cite{ritz} 
\begin{equation}
\tau_{\rm int} = [n\ \sigma_{\rm int} v]^{-1} = 5\cdot 10^{-35} [\rho\ \sigma_{\rm int} \beta]^{-1} {\rm sec}
\label{eq:inttime}
\end{equation}
where $\sigma_{\rm int}$ denotes the typical interaction cross section for a hadron.

For hadrons made of light quarks $\sigma_{\rm int}\sim20$ mbarn~\cite{cross}, 
which implies $\tau_{\rm int} \simeq 2\cdot 10^{-10}$ sec. 
 Since the typical lifetime of $\pi$ and $K$ mesons is of the order
of $10^{-8}$ s, light hadrons are usually stopped before decay,
unless they are very relativistic. We implemented this process in
the PYTHIA code by letting the hadron decay freely when
$\gamma \tau_{\rm dec} < \tau_{\rm int}$, otherwise it is
stopped. With this modification of the code we take into account the
actual lifetime of any hadron and the actual energy it has in the
fragmentation process. When a very energetic light hadron is produced,
we therefore do not neglect its decay. This situation however is not
very frequent and it can occur only for very energetic injected jets.

In the case of heavy hadrons one has
$\sigma_{\rm int}\sim14$ mbarn for a $c$ or $b$ meson and 
$\sigma_{\rm int}\sim24$ mbarn for a $c$ or $b$ hadron~\cite{cross},
giving $\tau_{\rm int} \sim (2\div3) \cdot 10^{-10}$ sec. 
The typical lifetime for these
hadrons is $\tau_{\rm dec} \sim  10^{-12}$ sec,
or less. We therefore may assume that they decay before loosing a
significant part of their energy. We again implemented a modification
of the PYTHIA code which is similar to the case of leptons and which takes
into account the relevant time scales.

In addition to energy losses, we should also take into account that
interaction of hadrons with the medium could lead to the
production of additional hadrons. For instance, a heavy-hadron
collision with the medium may produce additional light hadrons.
However these additional light hadrons of lower energies are
easily stopped, as discussed before, and therefore give a negligible
contribution to the neutrino flux in our relevant energy range from
this process. We therefore ignore here this possibility, a consistent assumption under our approximations.

\begin{figure}[p]
$$\hspace{-8mm}\includegraphics[width=17.5cm]{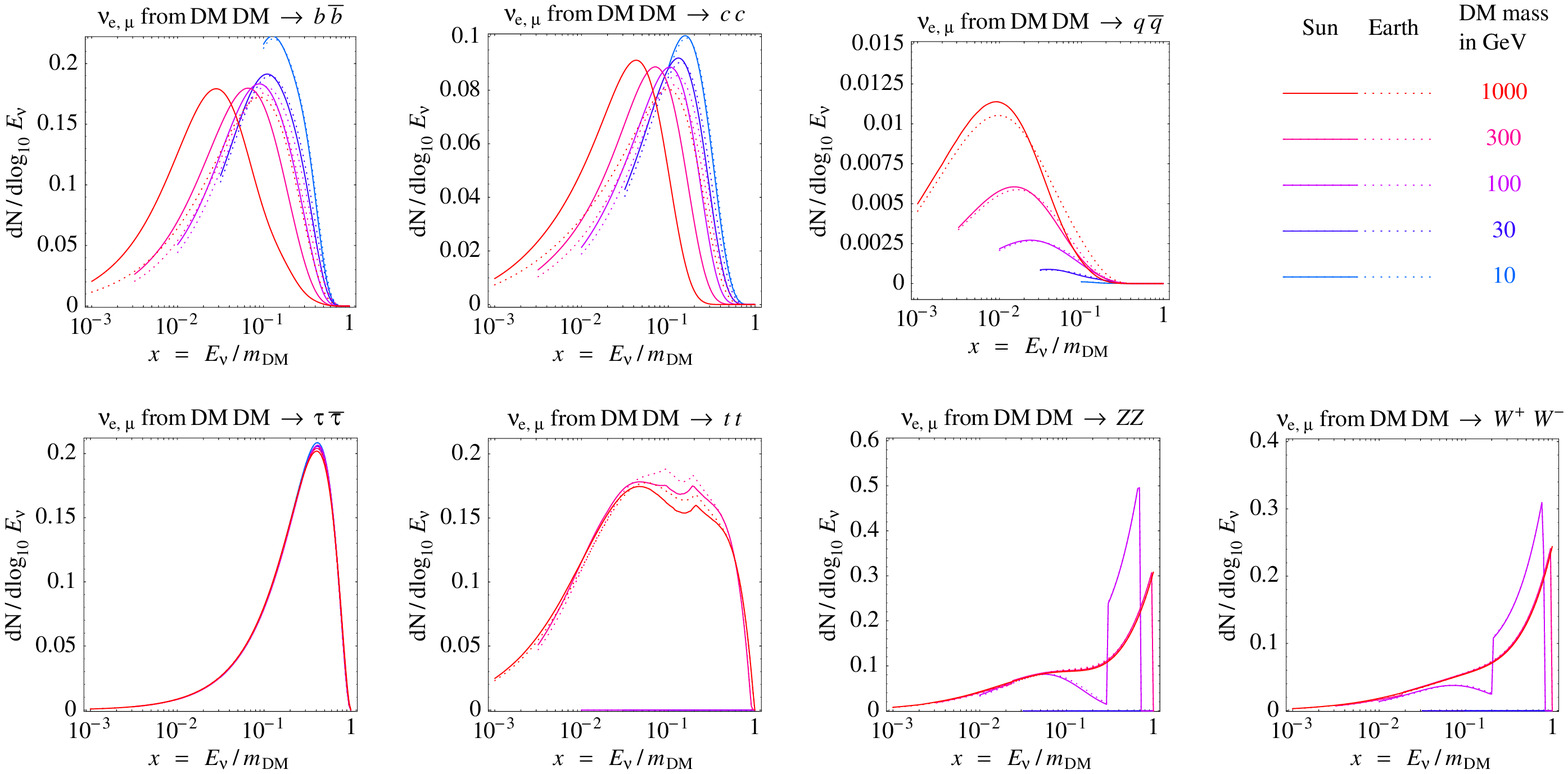}$$
$$\hspace{-8mm}\includegraphics[width=17.5cm]{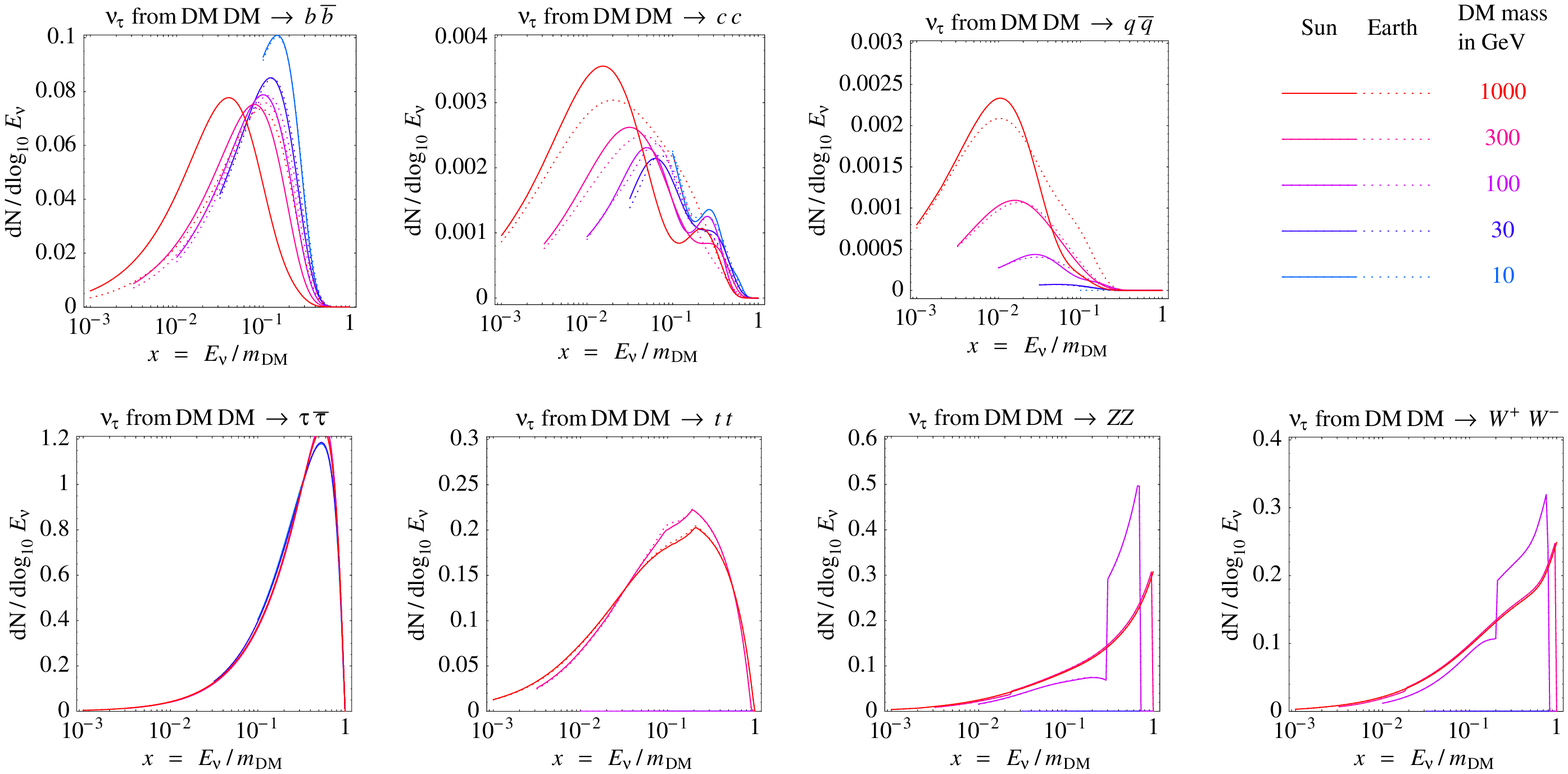}$$
\caption[X]{\label{fig:Prim}\em Neutrino spectra at production. Upper half: the fluxes of electron and muon neutrinos, for the seven main annihilation channels and for different masses of the parent DM particle (different colors). The solid lines apply to the case of the Sun, the dotted of the Earth. In all cases, the spectra of antineutrinos are the same as those of neutrinos. Lower half: the same for $\nubarnu_\tau$.}
\end{figure}

\bigskip

Our results on the neutrino spectra from annihilations in the Earth are shown in fig.~\ref{fig:Prim} as dotted lines. 
Each spectrum refers to the
flux of {\em neutrinos} for a given $q\bar q$ or $\tau\bar\tau$ pair
and for different values of $m_{\rm DM}$, equal to the energy of the primary jet or $\tau$.
Antineutrinos are not summed up and their fluxes are the same as those of neutrinos, since
the initial state is neutral respect to all quantum numbers. The plots
are shown as a function of  $x=E/m_{\rm DM}$,
which is defined in the interval $[0,1]$. 
The curves start from the $x$ corresponding to the minimal neutrino energy that we consider,
$E_\nu=0.5\GeV$.

The spectra at production of $\nu_e$ and
$\nu_\mu$ are equal, since light hadrons and muons do not contribute to
the DM$\nu$ fluxes,  and since $\tau,c,b$ produce an equal amount of $\nu_e$ and $\nu_\mu$.
This equality would not hold for neutrinos produced from $\pi$ or $\mu$.

We also see that light quarks contribute to the neutrino fluxes, even
though light hadrons are stopped. This is due to the fact that a $u$,
$d$ or $s$ quark has a non vanishing probability of splitting into a
$c$ quark in the fragmentation process and this process is favored
for larger energies (for details, see~\cite{pythia} and
references therein). We therefore have $c$ hadrons in the outgoing
jets also when we inject a light quark. The decay of these hadrons
produces neutrino fluxes in the interesting energy range. We see that
at low neutrino energies around 1 GeV and for $m_{\rm DM}$ larger than
about 500 GeV the contribution coming from light quarks can even be
the dominant one.  This effect was neglected in previous analyses.

\begin{table}[tp]
$$
\tiny\hspace{-5mm}
\vspace{-5mm}
\begin{array}{|c|cccccccccc|}
\hline
m_{\rm DM} & a_0 & a_1 & a_2 & a_3 & a_4 & a_5 & b & c_0 & c_1 & c_2 \\
\hline\hline
\multicolumn{11}{|c|}{\hbox{DM annihilations into $b\bar{b}$}}\\ 
\hline
10 & {34.5}/{55.7} & {4.15}/{4.34} & {7.47}/{8.03}
 & {6.83}/{7.52} & {3.16}/{3.55} & {0.594}/{0.677} & {7.98}/{8.75}
 & {0}/{0} & {0}/{0} & {0}/{0}\\
30 & {10.1}/{16.6} & {3.27}/{3.43} & {4.99}/{5.25}
 & {3.69}/{3.95} & {1.34}/{1.46} & {0.192}/{0.214} & {7.13}/{8.18}
 & {0}/{0} & {0}/{0} & {0}/{0}\\
50 & {7.17}/{9.60} & {3.01}/{3.00} & {4.40}/{4.27}
 & {3.04}/{2.90} & {1.01}/{0.962} & {0.132}/{0.126} & {6.96}/{8.00}
 & {0}/{0} & {0}/{0} & {0}/{0}\\
70 & {4.46}/{7.42} & {2.76}/{2.85} & {4.07}/{3.98}
 & {2.70}/{2.58} & {0.848}/{0.803} & {0.104}/{0.097} & {6.49}/{7.99}
 & {0}/{0} & {0}/{0} & {0}/{0}\\
100 & {6.07}/{8.28} & {2.87}/{2.77} & {4.12}/{3.68}
 & {2.74}/{2.26} & {0.878}/{0.657} & {0.110}/{0.072} & {7.05}/{8.67}
 & {0}/{0} & {0}/{0} & {0}/{0}\\
200 & {1.82}/{3.28} & {2.46}/{2.90} & {4.26}/{4.56}
 & {2.75}/{2.87} & {0.830}/{0.838} & {0.098}/{0.094} & {5.74}/{7.90}
 & {0}/{0} & {0}/{0} & {0}/{0}\\
300 & {1.61}/{2.36} & {2.37}/{2.79} & {4.17}/{4.45}
 & {2.61}/{2.45} & {0.770}/{0.613} & {0.089}/{0.058} & {5.73}/{9.12}
 & {0}/{0} & {0}/{0} & {0}/{0}\\
500 & {0.918}/{10.9} & {1.73}/{2.87} & {4.08}/{3.74}
 & {2.54}/{2.22} & {0.770}/{0.635} & {0.092}/{0.071} & {5.62}/{12.3}
 & {0}/{1.20} & {0}/{1.72} & {0}/{8.06}\\
700 & {0.893}/{4.21} & {1.68}/{3.64} & {4.12}/{5.23}
 & {2.62}/{2.93} & {0.827}/{0.772} & {0.101}/{0.079} & {5.78}/{8.99}
 & {0}/{0} & {0}/{0} & {0}/{0}\\
1000 & {0.779}/{31.3} & {1.39}/{2.76} & {4.13}/{2.99}
 & {2.77}/{1.47} & {0.934}/{0.345} & {0.117}/{0.032} & {5.99}/{18.7}
 & {0}/{1.20} & {0}/{1.72} & {0}/{8.06}\\
\hline\hline
\multicolumn{11}{|c|}{\hbox{DM annihilations into $\tau\bar{\tau}$}}\\ \hline 
10 & {1.09}/{0.903} & {1.33}/{0.654} & {0.795}/{-0.946}
 & {-0.404}/{-2.42} & {-0.728}/{-1.80} & {-0.253}/{-0.462} & {2.11}/{2.02}
 & {0}/{0} & {0}/{0} & {0}/{0}\\
30 & {1.07}/{0.839} & {1.48}/{0.763} & {1.54}/{-0.058}
 & {0.926}/{-0.687} & {0.312}/{-0.460} & {0.046}/{-0.096} & {2.07}/{1.95}
 & {0}/{0} & {0}/{0} & {0}/{0}\\
50 & {1.17}/{0.994} & {1.72}/{1.30} & {2.06}/{1.13}
 & {1.41}/{0.499} & {0.519}/{0.105} & {0.078}/{0.0070} & {2.12}/{2.03}
 & {0}/{0} & {0}/{0} & {0}/{0}\\
70 & {1.10}/{0.962} & {1.56}/{1.20} & {1.68}/{0.942}
 & {1.01}/{0.353} & {0.323}/{0.060} & {0.043}/{0.0036} & {2.09}/{2.01}
 & {0}/{0} & {0}/{0} & {0}/{0}\\
100 & {1.03}/{1.25} & {1.40}/{1.81} & {1.35}/{2.12}
 & {0.715}/{1.35} & {0.203}/{0.446} & {0.024}/{0.059} & {2.06}/{2.16}
 & {0}/{0} & {0}/{0} & {0}/{0}\\
200 & {0.895}/{1.25} & {1.08}/{1.80} & {0.781}/{2.08}
 & {0.257}/{1.30} & {0.029}/{0.412} & {-0.0010}/{0.052} & {1.97}/{2.18}
 & {0}/{0} & {0}/{0} & {0}/{0}\\
300 & {1.09}/{0.378} & {1.48}/{-1.73} & {1.45}/{-3.67}
 & {0.744}/{-2.90} & {0.195}/{-1.00} & {0.020}/{-0.128} & {2.11}/{1.53}
 & {0}/{0} & {0}/{0} & {0}/{0}\\
500 & {1.04}/{1.09} & {1.39}/{1.46} & {1.30}/{1.39}
 & {0.639}/{0.701} & {0.162}/{0.183} & {0.017}/{0.019} & {2.08}/{2.11}
 & {0}/{0} & {0}/{0} & {0}/{0}\\
700 & {0.958}/{1.13} & {1.23}/{1.53} & {1.05}/{1.55}
 & {0.469}/{0.826} & {0.107}/{0.225} & {0.0099}/{0.024} & {2.02}/{2.13}
 & {0}/{0} & {0}/{0} & {0}/{0}\\
1000 & {1.01}/{0.686} & {1.32}/{0.526} & {1.18}/{0.057}
 & {0.548}/{-0.146} & {0.129}/{-0.066} & {0.012}/{-0.0084} & {2.06}/{1.81}
 & {0}/{0} & {0}/{0} & {0}/{0}\\
\hline\hline
\multicolumn{11}{|c|}{\hbox{DM annihilations into $c\bar{c}$}}\\ \hline
10 & {0.703}/{0.654} & {-1.66}/{-1.46} & {-2.96}/{-1.06}
 & {-1.68}/{1.55} & {-0.408}/{1.79} & {-0.058}/{0.492} & {6.27}/{6.40}
 & {0}/{0} & {0}/{0} & {0}/{0}\\
30 & {0.233}/{0.674} & {-7.56}/{-3.07} & {-8.62}/{-5.38}
 & {-3.09}/{-3.29} & {0.353}/{-0.691} & {0.297}/{0.0012} & {6.35}/{7.86}
 & {0}/{0} & {0}/{0} & {0}/{0}\\
50 & {0.513}/{0.828} & {-2.06}/{-2.93} & {-2.90}/{-5.87}
 & {-1.62}/{-4.35} & {-0.323}/{-1.38} & {-0.0032}/{-0.154} & {6.81}/{8.84}
 & {0}/{0} & {0}/{0} & {0}/{0}\\
70 & {0.328}/{0.962} & {-2.66}/{-3.78} & {-0.903}/{-8.23}
 & {1.82}/{-6.68} & {1.66}/{-2.40} & {0.388}/{-0.324} & {6.55}/{9.91}
 & {0}/{0} & {0}/{0} & {0}/{0}\\
100 & {0.685}/{1.60} & {-1.92}/{-1.70} & {-4.19}/{-4.59}
 & {-3.65}/{-3.87} & {-1.42}/{-1.39} & {-0.208}/{-0.185} & {7.57}/{11.1}
 & {0}/{0.160} & {0}/{6.11} & {0}/{12.8}\\
200 & {0.532}/{1.41} & {-2.17}/{-2.30} & {-3.51}/{-4.73}
 & {-2.49}/{-3.42} & {-0.727}/{-1.06} & {-0.074}/{-0.120} & {7.58}/{12.9}
 & {0}/{0.158} & {0}/{6.11} & {0}/{12.8}\\
300 & {0.279}/{1.72} & {-3.52}/{-2.68} & {-2.93}/{-4.98}
 & {-1.03}/{-3.28} & {0.118}/{-0.919} & {0.072}/{-0.094} & {6.98}/{16.8}
 & {0}/{0.153} & {0}/{6.11} & {0}/{12.8}\\
500 & {0.363}/{1.35} & {-3.02}/{-3.24} & {-3.31}/{-4.89}
 & {-1.61}/{-2.74} & {-0.136}/{-0.613} & {0.029}/{-0.045} & {7.56}/{17.5}
 & {0}/{0.153} & {0}/{6.11} & {0}/{12.8}\\
700 & {0.476}/{2.90} & {-1.60}/{-3.10} & {-1.64}/{-5.64}
 & {-0.663}/{-3.50} & {0.088}/{-0.909} & {0.044}/{-0.085} & {7.77}/{23.4}
 & {0}/{0.152} & {0}/{6.11} & {0}/{12.8}\\
1000 & {0.420}/{2.19} & {-2.08}/{-3.80} & {-1.65}/{-5.49}
 & {-0.195}/{-2.84} & {0.408}/{-0.584} & {0.095}/{-0.040} & {8.01}/{26.4}
 & {0}/{0.151} & {0}/{6.11} & {0}/{12.8}\\
\hline\hline
\multicolumn{11}{|c|}{\hbox{DM annihilations into $q\bar{q}$}}\\ \hline
10 & {0.0024}/{0.919} & {-2.17}/{1.04} & {-4.97}/{1.73}
 & {-6.24}/{5.85} & {-4.68}/{6.21} & {-1.38}/{2.00} & {16.0}/{42.2}
 & {0}/{0.025} & {0}/{2.59} & {0}/{11.2}\\
30 & {0.038}/{0.871} & {-0.632}/{3.50} & {-5.46}/{4.96}
 & {-7.21}/{3.52} & {-3.64}/{1.26} & {-0.649}/{0.183} & {17.3}/{20.4}
 & {0}/{0} & {0}/{0} & {0}/{0}\\
50 & {0.020}/{0.405} & {2.78}/{3.44} & {5.17}/{4.79}
 & {5.22}/{3.18} & {3.26}/{1.03} & {0.771}/{0.132} & {12.0}/{20.4}
 & {0}/{0} & {0}/{0} & {0}/{0}\\
70 & {0.017}/{0.0057} & {2.28}/{-2.11} & {5.65}/{-0.717}
 & {5.80}/{-1.42} & {3.44}/{0.398} & {0.733}/{0.208} & {14.2}/{15.5}
 & {0}/{0} & {0}/{0} & {0}/{0}\\
100 & {0.019}/{0.017} & {2.25}/{-0.973} & {6.88}/{-2.07}
 & {8.24}/{-0.842} & {5.16}/{1.31} & {1.10}/{0.494} & {14.6}/{15.3}
 & {0}/{0} & {0}/{0} & {0}/{0}\\
200 & {0.017}/{0.012} & {2.28}/{-0.863} & {6.86}/{-1.64}
 & {7.97}/{-1.60} & {5.49}/{1.65} & {1.14}/{0.561} & {14.3}/{15.4}
 & {0}/{0} & {0}/{0} & {0}/{0}\\
300 & {0.015}/{0.041} & {2.05}/{-1.38} & {6.28}/{-2.16}
 & {7.67}/{0.967} & {6.01}/{2.60} & {1.23}/{0.635} & {14.3}/{17.3}
 & {0}/{0} & {0}/{0} & {0}/{0}\\
500 & {0.065}/{0.047} & {3.44}/{-1.73} & {7.82}/{-1.49}
 & {8.11}/{2.53} & {4.25}/{3.41} & {0.708}/{0.747} & {15.5}/{19.7}
 & {0}/{0} & {0}/{0} & {0}/{0}\\
700 & {0.093}/{0.043} & {3.57}/{-1.50} & {7.95}/{-1.57}
 & {8.14}/{2.47} & {4.04}/{3.63} & {0.638}/{0.754} & {15.7}/{19.7}
 & {0}/{0} & {0}/{0} & {0}/{0}\\
1000 & {0.061}/{0.050} & {2.90}/{-2.07} & {8.48}/{-2.74}
 & {10.2}/{1.18} & {5.57}/{2.90} & {0.873}/{0.579} & {16.3}/{23.5}
 & {0}/{0} & {0}/{0} & {0}/{0}\\
\hline\hline
\multicolumn{11}{|c|}{\hbox{DM annihilations into gluons}}\\ \hline
10 & {0.050}/{0.017} & {-0.286}/{0.645} & {-1.43}/{7.23}
 & {-2.12}/{1.20} & {-0.727}/{-4.95} & {-0.011}/{-2.21} & {9.31}/{9.15}
 & {0}/{0} & {0}/{0} & {0}/{0}\\
30 & {0.042}/{0.012} & {-2.49}/{0.394} & {-0.522}/{8.54}
 & {3.44}/{-2.01} & {5.60}/{-1.81} & {2.00}/{0.043} & {8.88}/{7.62}
 & {0}/{0} & {0}/{0} & {0}/{0}\\
50 & {0.011}/{0.802} & {-4.03}/{0.211} & {3.80}/{-2.76}
 & {-0.409}/{-3.88} & {4.15}/{-1.88} & {1.90}/{-0.310} & {7.10}/{14.5}
 & {0}/{0} & {0}/{0} & {0}/{0}\\
70 & {0.013}/{0.532} & {-0.695}/{0.030} & {3.66}/{-2.72}
 & {-3.72}/{-3.74} & {2.55}/{-1.73} & {1.56}/{-0.271} & {6.06}/{13.8}
 & {0}/{0} & {0}/{0} & {0}/{0}\\
100 & {0.353}/{1.01} & {-2.00}/{-0.413} & {-6.65}/{-3.37}
 & {-6.74}/{-3.66} & {-2.61}/{-1.47} & {-0.352}/{-0.205} & {13.1}/{17.4}
 & {0}/{0} & {0}/{0} & {0}/{0}\\
200 & {0.082}/{0.555} & {-4.58}/{-0.850} & {-6.82}/{-3.38}
 & {-5.00}/{-3.39} & {-0.326}/{-1.18} & {0.207}/{-0.139} & {11.1}/{17.4}
 & {0}/{0} & {0}/{0} & {0}/{0}\\
300 & {0.052}/{0.227} & {-4.17}/{-2.12} & {-6.56}/{-3.38}
 & {-5.60}/{-2.52} & {0.399}/{-0.196} & {0.433}/{0.073} & {8.88}/{17.9}
 & {0}/{0.090} & {0}/{2.57} & {0}/{8.13}\\
500 & {0.063}/{0.814} & {-5.59}/{-1.86} & {-6.83}/{-4.34}
 & {-3.22}/{-3.20} & {1.88}/{-0.802} & {0.619}/{-0.062} & {10.4}/{23.9}
 & {0}/{1.43} & {0}/{2.41} & {0}/{13.6}\\
700 & {0.069}/{0.453} & {-3.50}/{-2.42} & {0.232}/{-4.26}
 & {4.58}/{-2.75} & {5.20}/{-0.358} & {1.05}/{0.023} & {11.0}/{24.0}
 & {0}/{59.7} & {0}/{4.77} & {0}/{16.4}\\
1000 & {0.235}/{0.328} & {3.74}/{-2.99} & {8.11}/{-4.19}
 & {7.39}/{-2.08} & {3.63}/{0.246} & {0.561}/{0.123} & {9.14}/{24.0}
 & {0}/{481.} & {0}/{5.52} & {0}/{20.6}\\
 \hline
\end{array}$$
\caption{\em Fit parameters for the expression $g(x)$ in eq.~(\ref{eq:fit}). 
Give the $\nu_e=\nu_\mu\,\, = \bar\nu_e=\bar\nu_\mu$ spectra at production from annihilations in the Earth/Sun. 
DM masses are in {\rm GeV}. These data are available at~\cite{www}.\label{tab:params1}}
\end{table}%

\begin{table}[tp]
$$
\tiny\hspace{-12mm}
\vspace{-5mm}
\begin{array}{|c|cccccccccc|}
\hline
m_{\rm DM} & a_0 & a_1 & a_2 & a_3 & a_4 & a_5 & b & c_0 & c_1 & c_2 \\
\hline\hline
\multicolumn{11}{|c|}{\hbox{DM annihilations into $b\bar{b}$}}\\ \hline
10 & {5.30}/{12.9} & {-7.48}/{-4.22} & {-29.0}/{-20.8}
 & {-37.9}/{-30.1} & {-21.7}/{-18.8} & {-4.69}/{-4.36} & {12.3}/{13.6}
 & {0}/{0} & {0}/{0} & {0}/{0}\\
30 & {0.629}/{0.927} & {-14.8}/{-10.5} & {-35.4}/{-26.1}
 & {-32.9}/{-24.7} & {-13.8}/{-10.5} & {-2.20}/{-1.70} & {10.2}/{10.9}
 & {0}/{0} & {0}/{0} & {0}/{0}\\
50 & {0.387}/{0.766} & {-17.3}/{-11.5} & {-36.5}/{-26.7}
 & {-30.9}/{-23.8} & {-11.9}/{-9.59} & {-1.75}/{-1.45} & {9.86}/{11.2}
 & {0}/{0} & {0}/{0} & {0}/{0}\\
70 & {0.290}/{0.326} & {-18.9}/{-16.4} & {-37.0}/{-28.3}
 & {-29.7}/{-19.6} & {-10.9}/{-6.00} & {-1.51}/{-0.668} & {9.65}/{10.7}
 & {0}/{0} & {0}/{0} & {0}/{0}\\
100 & {0.184}/{0.284} & {-21.6}/{-16.8} & {-37.3}/{-28.1}
 & {-27.6}/{-19.5} & {-9.41}/{-6.13} & {-1.21}/{-0.715} & {9.08}/{10.8}
 & {0}/{0} & {0}/{0} & {0}/{0}\\
200 & {0.143}/{0.381} & {-23.9}/{-15.8} & {-37.4}/{-28.1}
 & {-25.8}/{-20.3} & {-8.10}/{-6.60} & {-0.959}/{-0.806} & {9.05}/{12.4}
 & {0}/{0} & {0}/{0} & {0}/{0}\\
300 & {0.116}/{0.103} & {-26.0}/{-26.0} & {-37.4}/{-25.7}
 & {-24.0}/{-11.6} & {-6.99}/{-2.21} & {-0.760}/{-0.129} & {9.01}/{12.3}
 & {0}/{0} & {0}/{0} & {0}/{0}\\
500 & {0.090}/{0.190} & {-28.6}/{-20.9} & {-36.8}/{-27.6}
 & {-21.7}/{-16.3} & {-5.71}/{-4.33} & {-0.551}/{-0.430} & {9.00}/{13.9}
 & {0}/{980.} & {0}/{5.78} & {0}/{15.3}\\
700 & {0.073}/{0.085} & {-30.7}/{-28.9} & {-36.1}/{-23.6}
 & {-20.0}/{-8.85} & {-4.84}/{-1.06} & {-0.419}/{0.030} & {8.84}/{14.1}
 & {0}/{       4
5.29 10} & {0}/{6.72} & {0}/{22.0}\\
1000 & {0.087}/{0.086} & {-29.3}/{-35.8} & {-36.9}/{-16.3}
 & {-21.2}/{-0.119} & {-5.32}/{1.85} & {-0.483}/{0.344} & {9.37}/{23.1}
 & {0}/{9.00} & {0}/{1.78} & {0}/{15.3}\\
\hline\hline
\multicolumn{11}{|c|}{\hbox{DM annihilations into $\tau\bar{\tau}$}}\\ \hline \

10 & {3.75}/{3.90} & {2.45}/{2.58} & {5.73}/{6.14}
 & {6.85}/{7.49} & {4.01}/{4.49} & {0.915}/{1.05} & {1.20}/{1.22}
 & {0}/{0} & {0}/{0} & {0}/{0}\\
30 & {2.34}/{2.57} & {0.855}/{1.20} & {1.79}/{2.52}
 & {1.82}/{2.60} & {0.867}/{1.26} & {0.156}/{0.233} & {0.996}/{1.04}
 & {0}/{0} & {0}/{0} & {0}/{0}\\
50 & {4.49}/{4.28} & {2.09}/{2.00} & {3.09}/{2.93}
 & {2.34}/{2.22} & {0.892}/{0.848} & {0.134}/{0.128} & {1.28}/{1.25}
 & {0}/{0} & {0}/{0} & {0}/{0}\\
70 & {4.13}/{4.05} & {1.87}/{1.81} & {2.59}/{2.45}
 & {1.84}/{1.70} & {0.654}/{0.593} & {0.092}/{0.082} & {1.24}/{1.23}
 & {0}/{0} & {0}/{0} & {0}/{0}\\
100 & {3.99}/{4.02} & {1.78}/{1.80} & {2.38}/{2.42}
 & {1.62}/{1.66} & {0.552}/{0.568} & {0.074}/{0.077} & {1.22}/{1.23}
 & {0}/{0} & {0}/{0} & {0}/{0}\\
200 & {3.44}/{3.43} & {1.40}/{1.39} & {1.62}/{1.62}
 & {0.943}/{0.943} & {0.272}/{0.274} & {0.031}/{0.031} & {1.15}/{1.15}
 & {0}/{0} & {0}/{0} & {0}/{0}\\
300 & {3.24}/{3.21} & {1.25}/{1.23} & {1.38}/{1.34}
 & {0.749}/{0.723} & {0.202}/{0.193} & {0.021}/{0.020} & {1.12}/{1.12}
 & {0}/{0} & {0}/{0} & {0}/{0}\\
500 & {3.20}/{3.25} & {1.22}/{1.25} & {1.34}/{1.38}
 & {0.722}/{0.751} & {0.193}/{0.202} & {0.020}/{0.021} & {1.12}/{1.12}
 & {0}/{0} & {0}/{0} & {0}/{0}\\
700 & {2.96}/{3.50} & {1.05}/{1.42} & {1.07}/{1.65}
 & {0.540}/{0.944} & {0.136}/{0.265} & {0.013}/{0.029} & {1.08}/{1.17}
 & {0}/{0} & {0}/{0} & {0}/{0}\\
1000 & {2.97}/{3.00} & {1.06}/{1.08} & {1.13}/{1.14}
 & {0.586}/{0.588} & {0.151}/{0.150} & {0.015}/{0.015} & {1.09}/{1.09}
 & {0}/{0} & {0}/{0} & {0}/{0}\\
\hline\hline
\multicolumn{11}{|c|}{\hbox{DM annihilations into $c\bar{c}$}}\\ \hline
10 & {-9.25}/{0.434} & {1.33}/{0.371} & {-3.55}/{-8.04}
 & {-8.46}/{-16.3} & {-6.08}/{-11.7} & {-1.48}/{-2.93} & {32.0}/{10.3}
 & {0.014}/{       3
3.76 10} & {0.348}/{12.7} & {4.75}/{11.4}\\
30 & {-0.681}/{0.064} & {1.79}/{-0.626} & {-0.550}/{-9.81}
 & {-2.61}/{-16.2} & {-1.66}/{-9.84} & {-0.328}/{-2.05} & {21.9}/{7.18}
 & {0.010}/{0} & {0.200}/{0} & {4.90}/{0}\\
50 & {-2.19}/{0.012} & {3.30}/{-0.739} & {4.15}/{-9.93}
 & {2.55}/{-16.5} & {0.767}/{-8.97} & {0.091}/{-1.59} & {22.2}/{4.11}
 & {0.011}/{0} & {0.245}/{0} & {5.14}/{0}\\
70 & {-0.087}/{0.652} & {-0.198}/{3.83} & {-5.46}/{5.72}
 & {-6.28}/{4.05} & {-2.72}/{1.39} & {-0.414}/{0.186} & {14.1}/{9.87}
 & {0.051}/{0} & {0.966}/{0} & {6.81}/{0}\\
100 & {-4.27}/{1.05} & {3.33}/{3.67} & {4.24}/{5.23}
 & {2.63}/{3.57} & {0.799}/{1.18} & {0.095}/{0.151} & {26.6}/{11.1}
 & {0.015}/{0} & {0.307}/{0} & {5.71}/{0}\\
200 & {0.015}/{0.031} & {1.59}/{0.326} & {-2.99}/{-4.78}
 & {-6.99}/{-7.54} & {-3.50}/{-3.60} & {-0.546}/{-0.566} & {7.48}/{5.94}
 & {0.0073}/{0} & {-0.067}/{0} & {5.20}/{0}\\
300 & {-0.027}/{0.0075} & {0.630}/{0.650} & {-2.12}/{-5.05}
 & {-2.18}/{-7.97} & {-1.10}/{-1.92} & {-0.194}/{-0.026} & {8.78}/{6.00}
 & {0.022}/{232.} & {0.512}/{4.74} & {6.42}/{18.9}\\
500 & {0.323}/{0.022} & {4.35}/{-1.40} & {7.00}/{-8.57}
 & {5.33}/{-10.1} & {1.98}/{-3.80} & {0.297}/{-0.483} & {10.5}/{9.25}
 & {0.00083}/{0.069} & {-1.10}/{2.02} & {3.13}/{6.79}\\
700 & {0.0066}/{0.013} & {0.048}/{-2.95} & {-3.54}/{-9.83}
 & {-3.66}/{-8.34} & {0.714}/{-1.50} & {0.352}/{0.017} & {7.27}/{8.55}
 & {0.012}/{0.561} & {0.169}/{2.66} & {6.12}/{10.4}\\
1000 & {0.0083}/{0.035} & {-2.01}/{-2.47} & {-1.13}/{-9.60}
 & {2.26}/{-8.66} & {3.59}/{-2.50} & {0.748}/{-0.240} & {13.5}/{8.54}
 & {0.010}/{0} & {0.083}/{0} & {6.05}/{0}\\
\hline\hline
\multicolumn{11}{|c|}{\hbox{DM annihilations into $q\bar{q}$}}\\ \hline
10 & {0.00084}/{0.012} & {1.76}/{2.20} & {0.493}/{2.52}
 & {-0.811}/{1.60} & {-0.585}/{0.527} & {-0.108}/{0.066} & {18.8}/{21.2}
 & {0}/{0} & {0}/{0} & {0}/{0}\\
30 & {0.0032}/{0.012} & {1.06}/{2.20} & {-0.294}/{2.52}
 & {-1.33}/{1.60} & {-0.723}/{0.527} & {-0.114}/{0.066} & {16.4}/{21.2}
 & {0}/{0} & {0}/{0} & {0}/{0}\\
50 & {0.0049}/{0.042} & {2.79}/{2.45} & {3.19}/{2.57}
 & {1.06}/{1.38} & {0.042}/{0.429} & {-0.017}/{0.066} & {14.8}/{21.2}
 & {0}/{0} & {0}/{0} & {0}/{0}\\
70 & {0.0053}/{0.071} & {2.49}/{2.46} & {3.28}/{2.56}
 & {1.57}/{1.38} & {0.469}/{0.431} & {0.081}/{0.064} & {14.8}/{21.2}
 & {0}/{0} & {0}/{0} & {0}/{0}\\
100 & {0.0094}/{0.217} & {1.42}/{2.72} & {-0.133}/{2.71}
 & {-1.71}/{1.12} & {-0.834}/{0.156} & {-0.111}/{-0.0047} & {14.9}/{21.3}
 & {0}/{0} & {0}/{0} & {0}/{0}\\
200 & {0.0066}/{0.036} & {1.33}/{2.66} & {-0.235}/{2.66}
 & {-1.74}/{1.14} & {-0.306}/{0.362} & {0.048}/{0.063} & {14.9}/{21.3}
 & {0}/{0} & {0}/{0} & {0}/{0}\\
300 & {0.044}/{0.088} & {0.368}/{-3.12} & {-1.68}/{-8.93}
 & {-2.04}/{-7.62} & {-0.696}/{-2.63} & {-0.076}/{-0.328} & {23.0}/{32.8}
 & {0}/{0} & {0}/{0} & {0}/{0}\\
500 & {0.012}/{0.026} & {-0.660}/{-8.28} & {-1.80}/{-8.91}
 & {-1.04}/{-3.36} & {0.566}/{-0.0037} & {0.201}/{0.128} & {23.0}/{66.9}
 & {0}/{1.50} & {0}/{1.39} & {0}/{34.0}\\
700 & {0.014}/{0.080} & {-0.566}/{-7.99} & {-1.84}/{-9.32}
 & {-1.09}/{-2.99} & {0.570}/{0.129} & {0.186}/{0.121} & {23.0}/{66.9}
 & {0}/{1.46} & {0}/{1.62} & {0}/{34.0}\\
1000 & {0.114}/{0.018} & {0.00048}/{-8.85} & {-1.94}/{-7.69}
 & {-1.58}/{-4.58} & {-0.342}/{-0.993} & {-0.018}/{-0.080} & {23.1}/{67.0}
 & {0}/{1.48} & {0}/{1.75} & {0}/{34.0}\\
\hline\hline
\multicolumn{11}{|c|}{\hbox{DM annihilations into gluons}}\\ \hline
10 & {0.054}/{0.057} & {-1.40}/{-1.05} & {-5.22}/{-5.92}
 & {-4.66}/{-7.03} & {-1.73}/{-3.33} & {-0.235}/{-0.498} & {19.0}/{18.7}
 & {0}/{4.65} & {0}/{6.26} & {0}/{12.8}\\
30 & {0.131}/{0.607} & {0.249}/{-0.739} & {-2.65}/{-5.88}
 & {-4.08}/{-7.25} & {-2.17}/{-3.56} & {-0.392}/{-0.630} & {15.4}/{21.8}
 & {0}/{0.719} & {0}/{3.64} & {0}/{12.7}\\
50 & {0.236}/{0.590} & {-0.0054}/{-1.37} & {-3.05}/{-6.41}
 & {-3.97}/{-6.82} & {-1.89}/{-2.97} & {-0.312}/{-0.467} & {17.1}/{23.5}
 & {0}/{0.394} & {0}/{2.27} & {0}/{12.7}\\
70 & {0.568}/{0.638} & {1.48}/{-1.47} & {0.407}/{-6.39}
 & {-0.625}/{-6.55} & {-0.440}/{-2.74} & {-0.081}/{-0.416} & {18.4}/{23.9}
 & {0}/{0.029} & {0}/{0.431} & {0}/{12.7}\\
100 & {0.696}/{0.568} & {1.74}/{-1.28} & {1.03}/{-5.89}
 & {0.0057}/{-6.04} & {-0.154}/{-2.49} & {-0.033}/{-0.367} & {18.7}/{22.7}
 & {0}/{0.245} & {0}/{2.25} & {0}/{12.7}\\
200 & {0.750}/{0.118} & {2.03}/{-3.84} & {1.75}/{-6.39}
 & {0.682}/{-3.73} & {0.139}/{-0.686} & {0.014}/{-0.0021} & {17.8}/{22.3}
 & {0}/{0.198} & {0}/{1.70} & {0}/{12.7}\\
300 & {0.094}/{0.420} & {-3.76}/{-2.79} & {-8.88}/{-6.76}
 & {-7.42}/{-5.10} & {-2.38}/{-1.55} & {-0.262}/{-0.168} & {16.1}/{27.5}
 & {0}/{0.110} & {0}/{1.28} & {0}/{12.7}\\
500 & {0.480}/{0.160} & {0.831}/{-4.43} & {-0.339}/{-6.73}
 & {-0.728}/{-3.55} & {-0.240}/{-0.547} & {-0.023}/{0.0066} & {18.9}/{27.5}
 & {0}/{0.201} & {0}/{1.52} & {0}/{12.7}\\
700 & {0.123}/{0.112} & {-4.50}/{-4.53} & {-9.65}/{-6.57}
 & {-7.07}/{-3.70} & {-1.90}/{-0.582} & {-0.172}/{0.0030} & {16.8}/{27.5}
 & {0}/{0.058} & {0}/{0.698} & {0}/{12.7}\\
1000 & {0.596}/{0.071} & {1.80}/{6.05} & {1.48}/{12.5}
 & {0.613}/{8.52} & {0.198}/{3.03} & {0.028}/{0.387} & {16.9}/{24.3}
 & {0}/{0.081} & {0}/{0.750} & {0}/{12.7}\\
 \hline
\end{array}$$
\caption{\em Fit parameters for the expression $g(x)$ in eq.~(\ref{eq:fit}). 
Give the $\nu_\tau= \bar\nu_\tau$ spectra at production from annihilations in the Earth/Sun. 
DM masses are in {\rm GeV}. These data are available at~\cite{www}.\label{tab:params2}}
\end{table}%

\medskip

We provide analytical fitted formul\ae{} for the spectra. We fitted
the MC results with the following expression, which proved to be suitable:
\beq 
g(x) =\frac{dN}{dx} = a_0 (1 + a_1 w + a_2 w^2 + a_3 w^3 + a_4 w^4 +
a_5 w^5)  (1-x)^b + c_0 x^{c_1}  (1-x)^{c_2} 
\label{eq:fit}
\eeq
where $w={\rm log}_{10} x$. The values of the parameters are shown in
table \ref{tab:params1} and table~\ref{tab:params2} for a sample of center-of-mass energies $m_{\rm DM}$ of the primary quark or $\tau$, and are also available at~\cite{www}. The fitted functions reproduce the MC result at a level better than a few percent in all the relevant energy range, from $0.5\GeV$ up to $m_{\rm DM}$.
The functions $g(x)$ should not be used outside this range.

\paragraph{Annihilation inside the Sun.}

The density of the core of the Sun is $\rho = 140$ g cm$^{-3}$,
about 10 times larger than in the Earth, so that
energy loss processes are more important than in the Earth case.

In the case of charged leptons, the stopping time is now 
 $\tau_{\rm stop} \sim 10^{-11}$ sec. Our modification of the PYTHIA code
takes into account this situation, as described previously.

The situation for the light-quark hadrons is similar to the case
of the Earth: they are mostly stopped and therefore they do not produce
neutrinos in the energy range of interest. In the case of
hadrons made by heavy quarks, the situation is now more subtle~\cite{ritz}. 
Their typical interaction time gets reduced by
an order of magnitude: $\tau_{\rm int} \sim (2\div3) \cdot 10^{-11}$ sec, and becomes
comparable to the typical heavy-hadron lifetime
$\tau_{\rm dec} \sim 10^{-12}$ sec (some hadrons decay faster).  
We must now be careful, since these hadrons may loose a fraction of their
energy before decaying. In order to take into account this effect, we
follow~\cite{ritz} where the average energy loss
of a heavy hadron in a dense medium was studied. 
For details about the analysis, we refer to~\cite{ritz}. 
Here we just recall the relevant results, implemented in our analysis.

A $c$ or $b$ hadron of initial energy $E_0$ after energy losses emerges
with an average energy:
\begin{equation}
\langle E \rangle = E_c\, \exp(x_0)\,E_1(x_0)
\label{eq:aveenergy}
\end{equation}
where $E_c=M_{\rm hadron}\,\tau_{\rm stop}/\tau_{\rm dec}$, $x_0=E_c/E_0$ and
the function $E_1$ is defined as:
\begin{equation}
E_1(x_0) = \int_{x_0}^\infty \frac{e^{-x}}{x}dx.
\end{equation}
The quantity $\tau_{\rm stop}$ is defined as $\tau_{\rm
stop}=\tau_{\rm int}/(1-Z)$ where $Z=x_i\,z_j$ and $\tau_{\rm int}$ is
the interaction time defined in eq.~(\ref{eq:inttime}). The quantity 
$x_i$ denotes the ratio between the quark and the hadron mass:
$x_i = m_i/M_{\rm hadron}$, for $i=c,b$ and $z_j=0.6$ for a $c$ hadron and
$z_j=0.8$ for a $b$ hadron.

We modified the PYTHIA code in order to take into account the energy
loss discussed above: when a $c$ or $b$ hadron is produced, we first
reduce its energy according to eq.~(\ref{eq:aveenergy}), and then it
is propagated and decayed by the PYTHIA routines.

\medskip

Our results for the neutrino spectra from annihilations in the Sun are shown in fig.~\ref{fig:Prim} as solid lines.
We see that the spectra are a little softer than in the Earth case,
due to hadron energy losses. The effect is more pronounced for larger
center-of-mass energies, since in this case hadrons loose a larger
fraction of their initial energy. Also in this case we fitted the
distributions with the same fitting formula of eq.~(\ref{eq:fit}), and
reported the parameters in table \ref{tab:params1} and
table \ref{tab:params2}.
They are again available at~\cite{www}.

\subsection{Annihilations into $W^+W^-$ and $ZZ$}
The lifetime of $W,Z$ gauge bosons is short enough that their energy losses can be neglected.
They therefore decay into quarks and leptons as in vacuum, but then
their decay products hadronize and decay, loosing energies as
discussed in the previous paragraphs. We can therefore calculate the
neutrino spectra by applying the results for quarks and leptons and by
using the formul\ae\ given in Appendix~\ref{app:boost}.
To the resulting spectra we than have to add the production of `prompt' neutrinos by the
decays $W \rightarrow \nu_\ell \bar{\ell}$ and $Z \rightarrow
\nu_\ell\bar{\nu_\ell}$, that give neutrino lines in the
reference frame of the decaying boson. When the boson is produced with
an energy $E_B$, the neutrino line is boosted to a flat spectrum:
\beq
\frac{dN}{dx} = \frac{{\rm BR}_i}{\beta}\qquad \hbox{in the range}\qquad
\frac{1-\beta}{2} \le x \le
\frac{1+\beta}{2}
 \label{eq:range} 
\eeq
where ${\rm BR}_i$ is the branching ratio for the prompt decay of the gauge boson and
$\beta$ is velocity of the gauge boson. As a check to our calculation, we produced a few sample cases of
neutrino spectra from $W$ and $Z$ with the PYTHIA code and compared
them to our analytical results. The agreement is well under the MC
statistical error.

\medskip

Our results are shown in fig.~\ref{fig:Prim} as dotted lines for the Earth and as solid lines for the Sun. 
DM annihilations into vector bosons produce a harder DM$\nu$
 spectrum as compared to  annihilations into $\tau\bar{\tau}$,
 $b\bar{b}$, $c\bar{c}$, $q\bar{q}$,
 thanks to prompt neutrino production.
 This is a dominant feature in the spectrum as long as
 $E_B=m_{\rm DM}$
 is not too much larger than $M_W,M_Z$.

\subsection{Annihilation into $t\bar{t}$}

The lifetime of the top quark is extremely short too, which allows us
to consider it decaying before any energy loss is operative. Also in
this case we build the spectra for the $t\bar{t}$ case as described in
Appendix \ref{app:boost}, by using the decay chain: $t \rightarrow
b+W$ followed by $W$ decay, as discussed in the previous paragraph.

Notice that we consider  a pure SM decay for the top
quark. In two--higgs doublet models like {\em e.g} in supersymmetric
extensions of the SM, there may be additional final states for the top
decay, due to the presence of a charged higgs: $t\rightarrow b + H^+$,
followed by $b$ hadronization and $H^+$ decay. 
Similarly, we do not consider DM  annihilations into new particles, like e.g.\ $H^+H^-$.


\subsection{Annihilation into higgs bosons or higgs$+$gauge bosons}

DM$\nu$ can also be generated by channels involving higgs particles in
the annihilation final state. We can safely ignore energy losses also
for the higgses and directly apply the method of Appendix~\ref{app:boost}. 
We do not explicitely provide results for this case, because even within the
SM the Higgs decays remain significantly uncertain until the Higgs mass is unknown.
Furthermore, Higgs decays can be affected by new physics: e.g.\ in SUSY models
the tree-level Higgs/fermions couplings differ from their SM values.

Higgs decays do not produce prompt neutrinos (because of the extremely
small neutrino masses) so that only soft neutrinos are generated,
even softer that in the $q\bar{q}$ case. Whenever a higgs is produced
in conjunction with a gauge boson   
prompt neutrinos from the gauge boson will be present.

\section{Neutrino propagation: oscillations, scatterings,...}\label{formalism}

We need to follow the contemporary effect on the neutrino fluxes from DM annihilations (presented in the previous section) of coherent flavor oscillations and of interactions with matter.

The appropriate formalism for this, that marries in a quantum-mechanically consistent way these two aspects, consists in studying the spatial evolution of the $3\times 3$ matrix of densities of neutrinos, $\rhob(E_\nu)$, and of anti-neutrinos, $\bar\rhob(E_\nu)$.
We will indicate matrices in bold-face and use the flavor basis.
The diagonal entries of the density matrix represent the population of the corresponding flavors, whereas the off-diagonal entries quantify the quantum superposition of flavors.
Matrix densities are necessary because
scatterings damp such coherencies, so that neutrinos are not in a pure state.
The formalism is readapted from~\cite{formalism}, where it was developed for
studying neutrinos in the early universe.

The evolution equation, to be evolved from the production point to the detector, has the form
\beq \label{eq:drho}\frac{d\rhob}{dr} =
- i [\mb{H},\ \rhob] +
\left.\frac{d\rhob}{dr}\right|_{\rm CC}+
\left.\frac{d\rhob}{dr}\right|_{\rm NC}+
\left.\frac{d\rhob}{dr}\right|_{\rm in}
\eeq
with an analogous equation for $\bar\rhob$.
The first term describes oscillations in vacuum or in matter.
The second and the third term describe the absorption and re-emission 
due to CC and NC scatterings, in particular including the effect of $\nu_\tau$ regeneration.
The last term represent the neutrino injection due to the annihilation of DM particles.
The average over the size of the production region can be approximately performed as described below.
Note that there is no neutrino-neutrino effect (i.e.\ the evolution equation is linear in $\rhob$) because neutrino fluxes are weak enough that they negligibly modify the surrounding environment. In particular Pauli blocking (namely: the suppression of neutrino production that occurs due to fermion statistics if the environment is already neutrino-dense), important in the early universe and in supernov\ae, can here be neglected.

We will discuss each term in detail in the following sections.

\bigskip

In the case of neutrinos from the center of the Earth, the formalism  simplifies: indeed, neutrino interactions with Earth matter only become relevant above $10\TeV$.
Since typical DM particles have the correct abundance for $m_{\rm DM}\circa{<}\TeV$, we can ignore such interactions in the Earth and only oscillations need to be followed. 
Moreover, taking into account that the initial spectra $\Phi^0$ do not distinguish $e$ from $\mu$,
$\Phi_e^0=\Phi^0_\mu \equiv \Phi^0_{e,\mu}$ (as discussed in sec.~\ref{Production}), and that
the oscillation probabilities obey $P_{\tau e}+P_{\tau\mu}+P_{\tau \tau}=1$,
the oscillated fluxes are given by
\beq
\label{eq:PhiEarth} 
\Phi_\ell(E_\nu) =  \Phi^0_{e,\mu}(E_\nu) + P_{\tau \ell}(E_\nu)[\Phi^0_\tau(E_\nu) - \Phi^0_{e,\mu}(E_\nu)]\qquad 
\ell=\{e,\mu,\tau\}
\eeq
An analogous result holds for anti-neutrinos.
$P_{\tau \ell}$, the conversion probability of a $\nu_\tau$ into a neutrino of flavor $\ell$, is easily computed with the standard oscillation formalism described below and is plotted in fig.\fig{PEarth}b.

\subsection{Oscillations}
\label{oscillations}

Oscillations are computed including the vacuum mixing and the MSW matter effect~\cite{MSW}.
The effective Hamiltonian reads
\beq\label{eq:H}
 \mb{H} =\frac{\mb{m}^\dagger \mb{m}}{2E_\nu } +
\sqrt{2} G_{\rm F}\bigg[N_e\ \diag\mb{(}1,0,0\mb{)} -\frac{N_n}{2}\ \diag\mb{(}1,1,1\mb{)}\bigg]\ ,
\eeq
where $\mb{m}$ is the $3 \times 3$ neutrino mass matrix.
One has $\mb{m}^\dagger \mb{m} = \mb{V}\cdot{\rm diag}(m_1^2,m_2^2,m_3^2)\cdot \mb{V}^\dagger$
where $m_{1,2,3}>0$ are the neutrino masses and $\mb{V}$ is the neutrino mixing matrix.
We define the solar mixing angle as $\tan\theta_{\rm sun}=|V_{e1}/V_{e2}|$,
the atmospheric mixing angle as $\tan\theta_{\rm atm}=|V_{\mu 3}/V_{\tau 3}|$ and
$\sin\theta_{13}=|V_{e3}|$.
$N_e(r)$ and $N_n(r)$ are the number density of electrons and neutrons in the matter, as predicted by solar and Earth models~\cite{BP00,PREM}.
The above Hamiltonian applies to neutrinos; for anti-neutrinos
one has to replace $\mb{m}$ with its complex conjugate
and flip the sign of the MSW term.
The difference between the matter potential for $\nu_\mu$ and $\nu_\tau$~\cite{Vmutau}, that arises only at one loop order, becomes relevant only at $E_\nu > \hbox{few}\TeV$ so we can neglect it.
Finally, notice that matter effects suppress oscillations of $\bar\nu_e$, since they 
encounter no MSW level crossings.

In the following we assume the present best fit values for the mixing parameters (from~\cite{NuFit})
$$\tan^2\theta_{\rm sun}=0.45,\qquad 
\theta_{\rm atm}=45^\circ,$$
$$\Delta m^2_{\rm sun}=8.0~10^{-5}\eV^2,\qquad
|\Delta m^2_{\rm atm}|=2.5~10^{-3}\eV^2.$$
We assume $\theta_{13}=0$ and we will later comment on how
a non-zero $\theta_{13}$ would marginally modify our results.


\bigskip

\subsection{Average over the production region}
\label{sec:average}
Neutrinos are produced in the core of the body (Earth or Sun) over a region of size $R_{\rm DM}$, as discussed in section~\ref{Production}, so in principle their propagation baseline is different depending on where they originate.
However, since $R_{\rm DM}$ turns out to be smaller than the size of the object, 
to a good approximation one can take into account oscillation effects assuming that all
neutrinos are produced at the center of the production region, with the following effective density matrix\footnote{We sketch the proof. To leading order in $R_{\rm DM}$
the distribution of neutrinos as a function of their path-length $L+\delta$ is
$n(\delta) \propto \exp(-2\delta^2/R_{\rm DM}^2)$, 
where $L$ is the distance of the detection point from the center of the body and $\delta$ spans the production region. The factor of 2 accounts for the two DM particles in the annihilation initial state. 
Oscillations can be decomposed as $U(L+\delta) = U(L)\cdot U(\delta)$, with $U$ the time evolution operator.
Averaging over $\delta$ gives eq.\eq{rhoeff}.
}
\beq \label{eq:rhoeff}
 (\rho_0^{\rm eff})_{ij} =\sum_{m,n,i',j'} (U_{im}U_{i'm} ^*) (\rho_0)_{i'j'} e^{-R_{\rm DM}^2(H_m-H_n)^2/8}
(U_{jn}^* U_{j'n})\eeq
where $\rho_0$ is the diagonal matrix of the total initial fluxes $\rho_0(E_\nu) = \diag\mb{(}\Phi_e^0,\Phi_\mu^0,\Phi^0_\tau\mb{)}$.
$\mb{U}$ and $H$ are the energy-dependent neutrino mixing matrix and hamiltonian eigenvalue
at the center of the body, to be computed diagonalizing the Hamiltonian in matter, eq.\eq{H}.

For a better intuitive understanding of the physical meaning of eq.\eq{rhoeff} one can neglect the small oscillation effects driven by $\theta_{13}$ and $\Delta m^2_{\rm sun}$ (since the oscillation length of the latter is much larger than the production region) and keep only the oscillations driven by $\Delta m^2_{\rm atm}$, thus reducing the oscillation to a ``$\nu_\mu\leftrightarrow\nu_\tau$'' case. Such oscillations are not affected by matter effects so that 
$\mb{U}=\mb{V}$  and the eigenvalues $H_{2,3}$ reduce to $m_{2,3}^2/E_\nu$.
Then the effect of the exponential factor in eq.\eq{rhoeff} is to damp the coherence between the $\nu_2$ and $\nu_3$ mass eigenstates, i.e. the off-diagonal elements of the density matrix (which express the superposition of different states) are suppressed. In other words, in the limit of a complete damping the effective density matrix is diagonal and composed of an average of the initial $\nu_\mu$ and $\nu_\tau$ fluxes. This is exactly the case for neutrinos with very small $E_\nu$. At larger energies, the damping effect is milder and indeed one can follow the fast oscillations. In fig.\fig{PEarth}b the result is well visible.
In short: the spatial average over the slightly different baselines produces some partial flavor equilibration and some loss of coherence.

\begin{figure}[t]
$$\includegraphics[width=9cm]{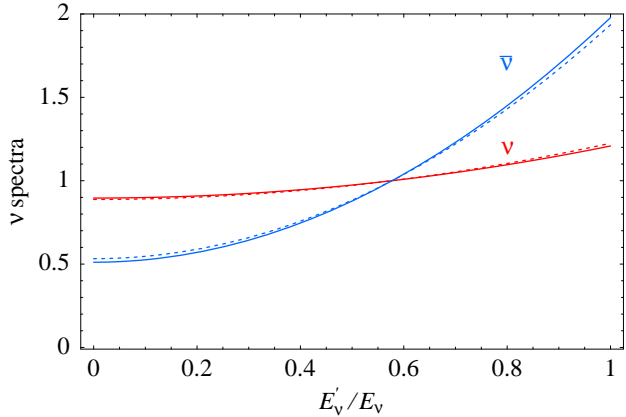}$$
\caption[X]{\label{fig:NC}\em Energy distributions of $\nu$ (red) and $\bar\nu$ (blue) produced 
with energy $E'_\nu$ by one NC DIS interaction of a $\nubarnu$ with energy $E_\nu$.
The energy is plotted in units of $E_\nu$.
Continuous line: in normal matter, where $N_p \approx N_n$. 
Dotted line: around the center of the Sun, where $N_p\approx 2 N_n$.}
\end{figure}

\begin{figure}[t]
$$\includegraphics[width=\textwidth]{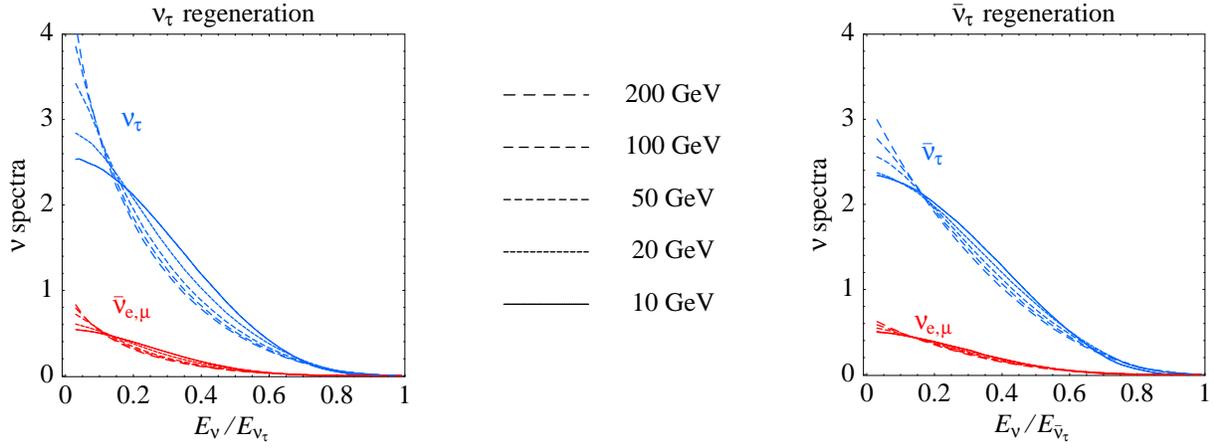}$$
\caption[X]{\label{fig:TauReg}\em Energy distributions of neutrinos
regenerated by CC scatterings of a $\nubarnu_\tau$ of given energy $E_{\nu_\tau}$,
produced by one $\nubarnu_\tau$/nucleon scattering.
The blue upper curves are $f_{\tau\to\tau}(E_{\nu_\tau},E'_\nu)$, 
and the red lower curves are $f_{\tau\to e,\mu}(E_{\nu_\tau},E'_\nu)$,
plotted for several values of the incident $\nu_\tau$ energy $E_{\nu_\tau}$.
}
\end{figure}

\subsection{NC scatterings}
NC scatterings $\nubarnu N \leftrightarrow \nubarnu N$  effectively remove a neutrino from the flux and re-inject it with a lower energy. 
So they contribute to the evolution equation as:
\beq\label{eq:NC}\left.\frac{d\rhob}{dr}\right|_{\rm NC} = - \int_0^{E_\nu} dE'_\nu 
\frac{d\Gamma_{\rm NC}}{dE'_\nu} (E_\nu,E'_\nu) \rhob(E_\nu)+
\int_{E_\nu}^\infty dE'_\nu 
\frac{d\Gamma_{\rm NC}}{dE_\nu} (E'_\nu,E_\nu) \rhob(E'_\nu)\eeq
where
\beq\label{eq:Gamma}
\Gamma_{\rm NC}(E_\nu,E'_\nu) = N_p(r)\ \diag\sigma(\nu_\ell p\to \nu_\ell' X)
+N_n(r)\ \diag\sigma(\nu_\ell n\to \nu_\ell' X)\eeq
 The first term describes the absorption:
 the integral over $E'_\nu$ just gives the total NC cross section.
  The second term describes the reinjection of lower energy neutrinos:
  their spectrum is plotted in fig.\fig{NC}.
  We see that it negligibly depends on the chemical composition $N_p/N_n$,
where $N_p$ and $N_n$ are the number densities of protons and neutrons.
We use the $N_e(r)=N_p(r)$ and $N_n(r)$ profiles predicted by solar and Earth models~\cite{BP00,PREM}.
In the Sun $N_p/N_n$ varies from the BBN value, $N_p/N_n\sim 7$ present in the outer region 
$r/R_\odot \circa{>}0.3$,
down to $N_p/N_n\sim2$ in the central region composed of burnt $^4$He.
The Earth is mostly composed by heavy nuclei, so that $N_p$ and $N_n$ are roughly equal.

\subsection{CC absorptions and $\nu_\tau$ regeneration}\label{CCsec}

The effect of CC interactions to the evolution of the neutrino fluxes can be intuitively pictured as follows. The deep inelastic CC process on a nucleon ($\nubarnu N \to \ellbarell X$) effectively removes a neutrino from the flux and produces an almost collinear charged lepton. The $\tau^\pm$ produced by $\nubarnu_\tau$ decays promptly, before loosing a significant part of its energy in interactions with the surrounding matter, and therefore re-injects secondary fluxes of energetic neutrinos~\cite{taureg idea, taureg comput}: 
\beq
\begin{array}{llll}
\nu_\tau ~\rightarrow~ \tau^- & \to X\, \nu_\tau \qquad  & \bar\nu_\tau ~\rightarrow~ \tau^+ & \to X\, \bar\nu_\tau\\
& \to e^-\, \bar\nu_e\, \nu_\tau\, \qquad &  & \to e^+\, \nu_e\, \bar\nu_\tau\\
& \to \mu^-\, \bar\nu_\mu\, \nu_\tau\, \qquad &  & \to \mu^+\, \nu_\mu\, \bar\nu_\tau
\end{array}
\eeq
with branching ratios ${\rm BR}_{X}=64.8\%$, ${\rm BR}_{e}=17.84\%$ and 
${\rm BR}_{\mu}=17.36\%$ for hadronic, electron and muonic decay 
modes respectively. In this way, besides $\nubarnu_\tau$ that always re-appears from 
$\tau^\pm$ decays, in $\approx 35\%$ of cases also $\nubarnu_{e}$ or 
$\nubarnu_{\mu}$ are produced, enlarging the total neutrino 
flux that reaches a detector. 
Note that $e$, $\mu$ and hadrons produced by CC scattering and by $\tau$ 
decays loose essentially all their energy in the matter and
are absorbed or decay into neutrinos with negligibly small
energy for our purposes.

\medskip

The CC contribution to the evolution equation of the density matrices is therefore
\begin{eqnsystem}{sys:CC}
\left.\frac{d\rhob}{dr}\right|_{\rm CC} &=& - \frac{\{\mb{\Gamma}_{\rm CC},\rhob\}}{2}+
\int \frac{dE^{\rm in}_\nu}{E^{\rm in}_\nu}  
 \bigg[ \mb{\Pi}_\tau \rho_{\tau\tau}(E^{\rm in}_\nu) \Gamma_{\rm CC}^\tau(E^{\rm in}_\nu) 
 f_{\tau\to\tau}({E_\nu^{\rm in}},{E_\nu})\\
 & &\qquad\nonumber
 + \mb{\Pi}_{e,\mu} \bar\rho_{\tau\tau} (E^{\rm in}_\nu) \bar\Gamma_{\rm CC}^{\tau} (E^{\rm in}_\nu)
 f_{\bar\tau\to e,\mu}( E^{\rm in}_\nu, E_\nu)
\bigg],  \\
\left.\frac{d\bar\rhob}{dr}\right|_{\rm CC} &=& - \frac{\{\bar{\mb{\Gamma}}_{\rm CC},\bar\rhob\}}{2}+
\int \frac{dE^{\rm in}_\nu}{E^{\rm in}_\nu}  
 \bigg[\mb{\Pi}_\tau\bar\rho_{\tau\tau}(E^{\rm in}_\nu) \bar\Gamma_{\rm CC}^\tau(E^{\rm in}_\nu)
  f_{\bar\tau\to\bar\tau}(E_\nu^{\rm in},{E_\nu})\\
  &&\qquad+\mb{\Pi}_{e,\mu} \rho_{\tau\tau} (E^{\rm in}_\nu) \Gamma_{\rm CC}^{\tau}(E^{\rm in}_\nu)
 f_{\tau\to \bar{e},\bar\mu}({E_\nu^{\rm in}},{E_\nu})\nonumber
\bigg].
\end{eqnsystem}
The first terms describe the absorption; their anticommutator arises because loss terms 
correspond to an anti-hermitian effective Hamiltonian  such that 
the usual commutator gets replaced by an anticommutator 
(see the full formalism in~\cite{formalism}).
The second terms describe the `$\nu_\tau$ regeneration'. 
We explicitely wrote the equations for neutrinos and for anti-neutrinos because they are coupled
by the second terms.

In the formul\ae\ above, $\mb{\Pi}_\ell$ is the projector on the flavor $\nu_\ell$: e.g.\ $\mb{\Pi}_\tau = \diag\mb{(}0,0,1\mb{)}$.
The $\mb{\Gamma}_{\rm CC}$, $\bar{\mb{\Gamma}}_{\rm CC}$ matrices express the rates of absorption due to the CC scatterings and are given by
\begin{eqnsystem}{sys:Gamma}
\mb{\Gamma}_{\rm CC}(E_\nu) = \diag\mb{(}\Gamma_{\rm CC}^e,
\Gamma_{\rm CC}^\mu,\Gamma_{\rm CC}^\tau\mb{)},\quad
\Gamma_{\rm CC}^\ell=
N_p(r)\  \sigma(\nu_\ell p\to \ell X)
+N_n(r)\  \sigma(\nu_\ell n\to \ell X),\\
\bar{\mb{\Gamma}}_{\rm CC}(E_\nu) = \diag\mb{(}\bar\Gamma_{\rm CC}^e,
\bar\Gamma_{\rm CC}^\mu,\bar\Gamma_{\rm CC}^\tau\mb{)},\quad
\bar\Gamma_{\rm CC}^\ell=
N_p(r)\  \sigma(\bar\nu_\ell p\to \bar\ell X)
+N_n(r)\  \sigma(\bar\nu_\ell n\to \bar\ell X).
\end{eqnsystem}
Deep inelastic scatterings of $\nubarnu$ on nucleons are the dominant process at the energies involved ($E_\nu \gg \GeV$), so the cross sections $\sigma(\nubarnu N \to \ellbarell X)$ (reported in Appendix~\ref{Cross}) are the only ones relevant. Scatterings on electrons have a cross section which is $\sim m_e/m_N$ smaller and would become relevant only at energies $E_\nu \circa{>}\TeV$~\cite{sigmaonelectrons}.

Notice that the matrix $\mb{\Gamma}_{\rm CC}$ is not proportional to the unit matrix because
at the relevant neutrino energies the $\nubarnu_\tau N$ cross sections~\cite{sigmatau} are suppressed with respect to the corresponding $\nubarnu_{e,\mu} N$ cross sections by the kinematical effect of the $\tau$ mass.
 E.g.\ at $E_\nu =100\GeV$ $m_\tau$ gives a $30\%$ suppression.
In particular, this implies that the coherence among $\nubarnu_\tau$ and $\nubarnu_{e,\mu}$ is broken by the CC interactions and the formalism is taking this into account.
A non trivial consequence (interactions increase the oscillation length) is discussed in appendix~\ref{B}.


The functions $f(E_\nu, E'_\nu)$ are the energy distributions of secondary neutrinos
produced by a CC scattering of an initial neutrino with energy $E_{\nu_\tau}$. They have been precisely computed numerically as described in~\cite{taureg comput}.
In the computation of $\tau$ decay spectra we have taken into account the sizable widths of final state hadrons, which produce a significant smearing with respect to fig.~10 of~\cite{Crotty} where such widths are neglected. 
Fig.\fig{TauReg} shows our result for the neutrino spectra from $\nu_\tau$ regeneration.
The integrals of the $\nu_\tau$ curves equal to one, because $\nu_\tau$ are completely regenerated,
with lower energy (the curves are peaked at small $E_\nu/E_{\nu_\tau}$).
The integrals of the $\bar\nu_{e,\mu}$ curves have a value smaller than one,
equal to the branching ratio of leptonic $\tau$ decays.
The curves depend, but quite mildly, on the incident neutrino energy
$E_{\nu_\tau}$, mainly due to the finite value of $m_\tau$:
neutrinos with lower energy loose a smaller fraction of their energy,
because the energy stored in the $\tau$ mass becomes more important at lower energy.
We assumed that $\tau^-$ and $\tau^+$ have exact left and right helicity respectively;
this approximation fails at energies comparable to $m_\tau$
 (say $E_\nu,E_\tau\circa{<}20\GeV$~\cite{sigmatau}), where
absorption and regeneration due to
CC scatterings become anyhow negligible.

The $f$ functions do not significantly depend on the chemical composition:
in the plot we assumed $N_p/N_n=2$ as appropriate in the center of the Sun.
Writing $x = E'_\nu/E_{\nu_\tau}$, these functions are normalized to the branching ratios of $\tau^{\pm}$ decays given above:
$$ \int_0^1 dx~f_{\tau\to \tau}(E_\nu,x E_\nu) = 1,\qquad
 \int_0^1 dx~f_{\tau\to e,\mu}(E_\nu,xE_\nu)\approx 0.175.$$

\medskip

Given the ingredients above, it should now be apparent how the second terms in eq.s~(\ref{sys:CC}) incorporate the CC processes.
In words, focussing for definiteness on the case of neutrinos (antineutrinos follow straightforwardly): a $\nu_\tau$ of energy $E_\nu^{\rm in}$, described by the density $\rho_{\tau\tau}$, interacts with a rate $\Gamma_{\rm CC}^\tau$ and produces secondary $\nu_\tau$, $\bar\nu_e$ and $\bar\nu_\mu$ with energy $E_\nu$, that contribute to the corresponding diagonal entries of the density matrices $\mb{\rho}$ and $\mb{\bar\rho}$. Integrating over $E_\nu^{\rm in}$ gives the total regeneration contribution.

\medskip

\begin{figure}[p]
$$\hspace{-8mm}\includegraphics[width=18cm]{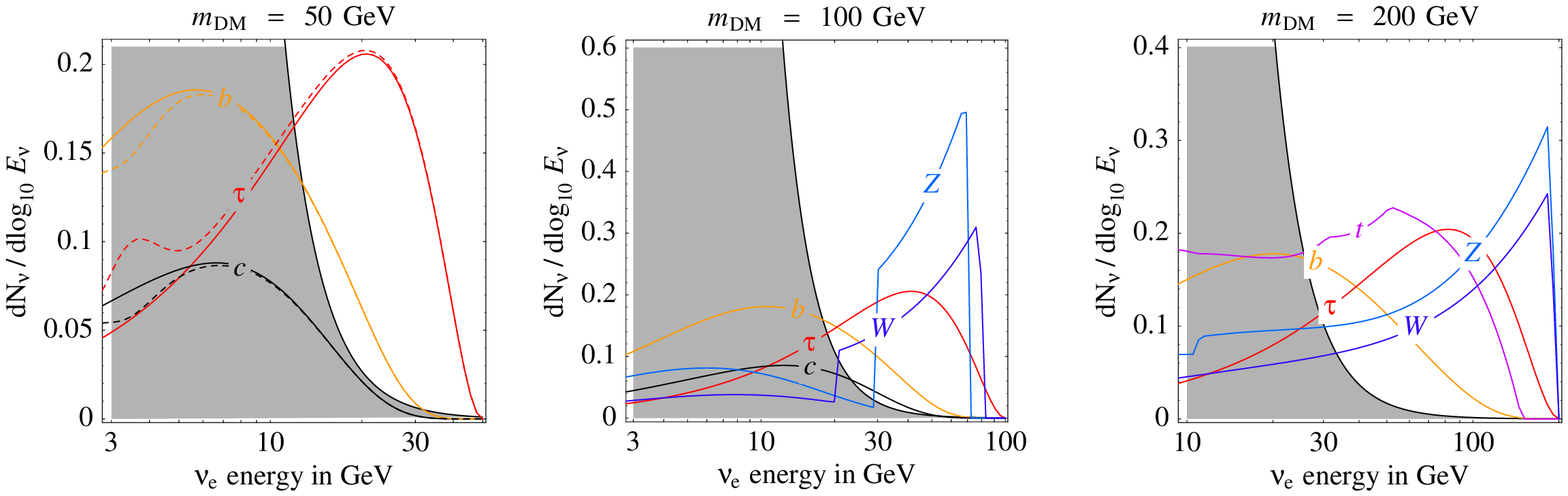}$$
$$\hspace{-8mm}\includegraphics[width=18cm]{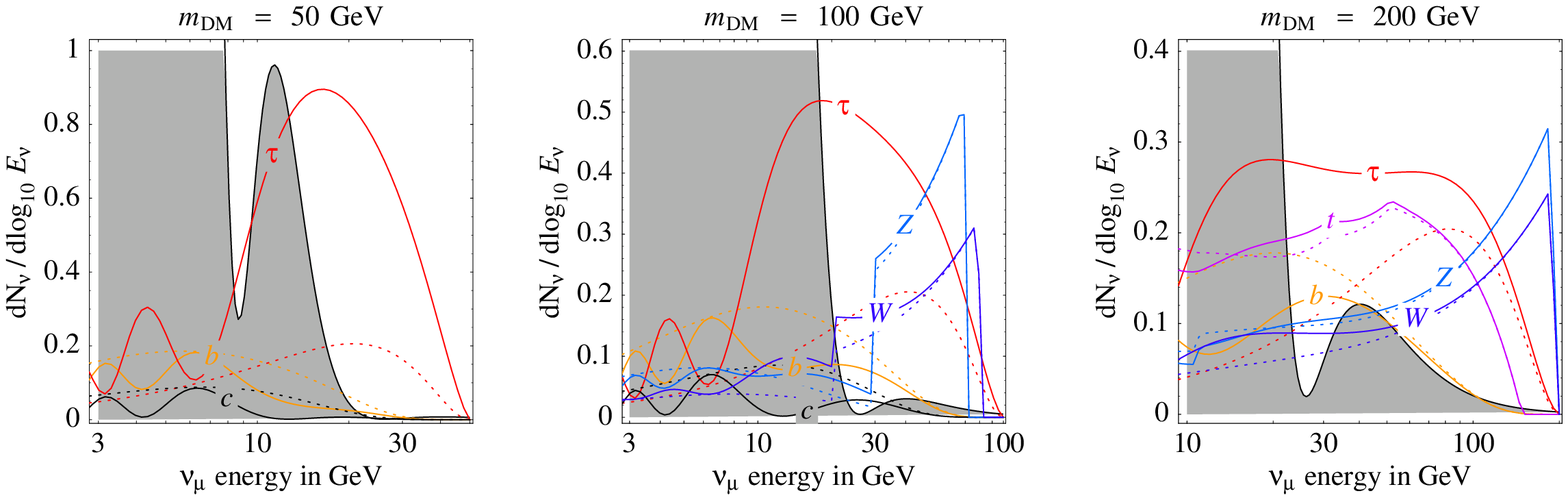}$$
$$\hspace{-8mm}\includegraphics[width=18cm]{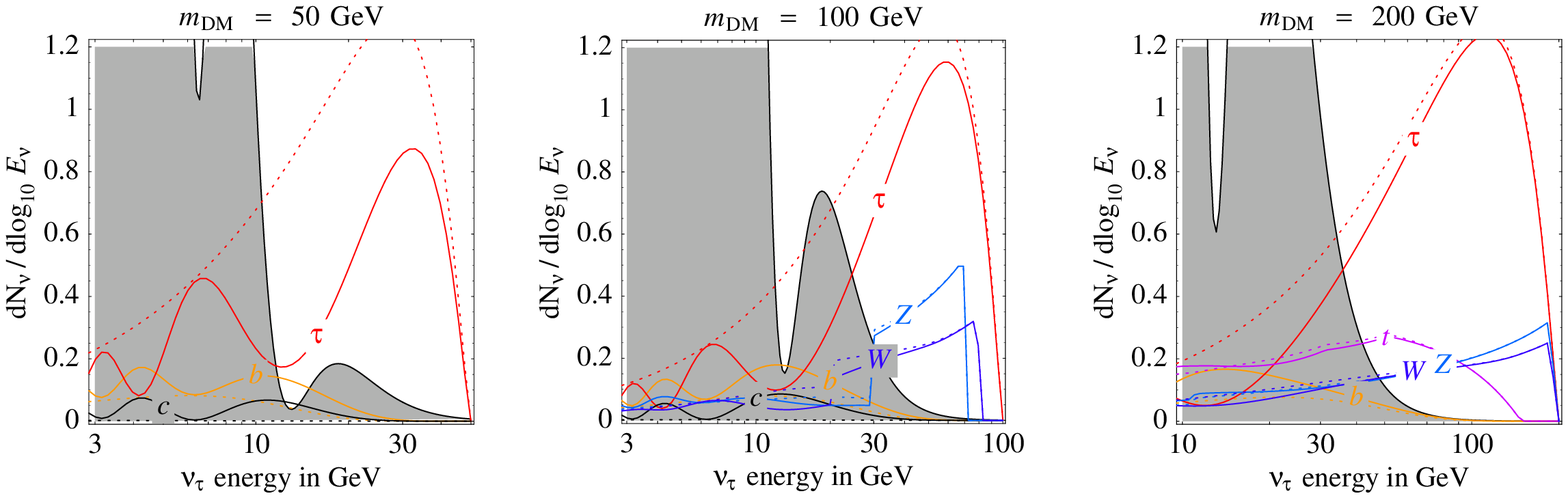}$$
\caption{\em Neutrino spectra generated by one DM annihilation
around the center of Earth.
The plots show the spectra of the three neutrino flavors (the three rows) and assume different DM masses (the three columns). Each plot shows the open annihilation channels ${\rm DM~DM}\to$  $b\bar{b}$, $\tau^+\tau^-$, $c \bar c$, $t\bar{t}$, $W^+ W^-$, $ZZ$.
The ${\rm DM~DM}\to\nu\bar\nu$ channel (not shown) would produce a line at $E_\nu = m_{\rm DM}$.
The dotted lines show the spectra without oscillations while solid lines are the final results after oscillations. The dashed lines in the upper-left panel have been computed with $\theta_{13}=0.1$ rad for illustration (see text); all the other results assume $\theta_{13}=0$.
Neutrino and anti-neutrino spectra are roughly equal:
we here show $(2\Phi_\nu + \Phi_{\bar\nu})/3$, in view of 
$\sigma(\nu N) \sim 2 \sigma(\bar\nu N)$.
The shaded region is the atmospheric background, normalized relative to DM$\nu$
as assumed in eq.\eq{EarthNorm}.
\label{fig:EarthNu}}
\end{figure}

\section{Neutrinos from DM annihilations in the Earth}\label{Earth}
In this section we show the results concerning the signal from DM annihilations around the center of the Earth: the energy spectra at detector of neutrinos and antineutrinos of all flavors  
and the energy spectra of the main classes of events that they produce.
\medskip

Fig.\fig{EarthNu} displays the neutrino spectra $dN/dE_\nu$, from the main annihilation channels {DM DM} $\to$  $b\bar{b}$, $\tau^+\tau^-$, $c \bar c$, $t\bar{t}$, $W^+ W^-$, $ZZ$, normalized to a single DM annihilation.\footnote{These fluxes are available at~\cite{www}.}
A linear combination of these basic spectra, weighted according to the BRs predicted by the specific DM model of choice and rescaled by the appropriate geometric factors, will give the actual neutrino signal at a detector: 
\beq \frac{dN}{dt\,dS~dE_\nu}= \frac{\Gamma_{\rm ann}}{4\pi R^2_\oplus}\sum_i {\rm BR}_i \frac{dN_i}{dE_\nu}
=\frac{0.2}{{\rm sec}\cdot {\rm m}^2}\
\frac{\Gamma_{\rm ann}}{10^{14}/{\rm sec}}\
\sum_i {\rm BR}_i \frac{dN_i}{dE_\nu}\, ,
\eeq
where the sum runs over the annihilation channels with branching ratios ${\rm BR}_i$, $R_\oplus$ is the Earth radius, and $\Gamma_{\rm ann}$ is the total number of DM annihilations per unit time.
As already discussed, this latter quantity is strongly dependent on the particle physics model under consideration and also on astrophysics, and can carry a large uncertainty. 
When we need to assume a value for it, e.g.\ to compare with the background or with the existing limits, we choose
\beq\label{eq:EarthNorm}
\left.\Gamma_{\rm ann}\right|_{\rm Earth}=\frac{10^{14}}{\rm sec}\left(\frac{100\GeV}{m_{\rm DM}}\right)^2.\eeq
In the neutralino case, samplings of the MSSM parameter space find a wide range of
 $10^{4\div 15}$ annihilations per second, that decreases for increasing $m_{\rm DM}$.
 So our assumption is realistically optimistic.
 
\medskip

We show plots for three different values of the DM mass,
which give qualitatively different results and
(in the case of the signal from Earth) well represent the general situation:
\begin{enumerate}
\item $m_{\rm DM}=50\GeV< M_{W,Z}$
so that only annihilations into leptons and quarks (other than the top) are allowed.  
Varying $m_{\rm DM}$ in this range the unoscillated fluxes rescale trivially; 
oscillated fluxes also rescale but of course keeping their first dip and peak at fixed energy,
as described below.

\item $M_{W,Z}<m_{\rm DM}=100\GeV\circa{<}m_t$ so that annihilations into
vector bosons are kinematically allowed, 
with kinetic energy comparable to their mass.
As explained in section~\ref{Production} this gives a characteristic threshold feature:
direct decays of $W,Z$ give neutrinos in the energy range of eq.\eq{range}
(producing the peaks in fig.s\fig{EarthNu}),
and neutrinos with lower energies are produced by secondary decay chains
(producing the tails).\label{123}

\item  $m_{\rm DM}=200\GeV > m_t$ so that also annihilations into top quarks
are allowed. Since the subsequent decay $t\to b \ell\nu$ is a 3-body process, it does not give
threshold features.  
$Z,W$ bosons are so energetic that their threshold features are minor.
No new notable features appear going to higher $m_{\rm DM}$.
If the DM is a neutralino only annihilations into $W^+W^-,ZZ,t\bar{t}$
(and possibly higgses and SUSY particles) are relevant.

\end{enumerate}

\medskip

\paragraph{The atmospheric neutrino background.} In all our figures,
the shaded region is the background of atmospheric neutrinos,
computed as predicted by FLUKA~\cite{FLUKA} (at the SuperKamiokande site)
and taking into account atmospheric oscillations.
The unknown DM$\nu$ signal is compared with the known magnitude of the background assuming
the annihilation rate in eq.\eq{EarthNorm}.

Since the signal comes from the center of the Earth, the background of atmospheric neutrinos can be suppressed exploiting directionality: in the figures
we applied an energy-dependent cut on the zenith-angle, keeping only neutrinos (and, later, events) with incoming direction that deviates from
the vertical direction by less than
\beq
\label{eq:cut} 
| \vartheta |  < \sqrt{\frac{m_N}{E}} = 5.7^\circ \sqrt{\frac{100\GeV}{E}}
\eeq
where $E$ is the energy of the detected particle and $m_N\approx\GeV$ is the nucleon mass.
Such a choice can be understood as follows.
First, the finite size of the DM annihilation region implies that the signal comes from a characteristic angular opening $\delta\vartheta \sim {R_{\rm DM}}/{R_\oplus} \sim \sqrt{m_N/E_\nu}$, where the last relation makes use of eq.\eq{RDM}  and of the fact that $E_\nu \sim m_{\rm DM}/{\rm few}$.
Furthermore, 
the kinematical angle $\delta\vartheta \approx  0.30\ (m_N/E_\nu)^{0.48}$~\cite{ANTARES}
between the incident neutrino and the produced lepton must be taken into account
and gives a comparable effect.
Finally, to these angles the effect of the angular resolution of
detectors should be added.
For \v{C}erenkov neutrino telescopes under construction
such as ANTARES, ICECUBE and the future km$^3$ detector in the
Mediterranean this resolution is $\delta\theta \circa{<} 1^\circ$.
For AMANDA and Super-Kamiokande the mean angular resolution is $\delta\theta \sim 2^\circ$ or larger, hence the angular cut may be larger than what we apply.

In summary, a more realistic dedicated analysis of the angular (and energy) spectrum will be certainly 
necessary to disentangle in the best possible way the signal from the atmospheric background, but  our approximation in eq.\eq{cut} is a reasonable cut applicable to many experiments.
We stress that the atmospheric background in the small cone around the vertical can be accurately and reliably estimated by interpolation of the measured rates in the adjacent angular bins
where no DM$\nu$ signal is present.

\medskip

\paragraph{The effect of oscillations.} 
In fig.\fig{EarthNu} the dotted lines show the spectra without (i.e.\ before) oscillations: these spectra have been already described in section~\ref{Production}. 
The final spectra (solid lines) are in many cases significantly different.
This is also illustrated  in fig.\fig{OscNoOsc}a for a few selected cases.

\begin{table}[t]
$$\begin{array}{|c|cccccccc|}\hline
\hbox{DM mass} & \multicolumn{8}{|c|}{\hbox{DM annihilation channels in the Earth/Sun}}\\
m_{\rm DM} & \nu\bar\nu & b\bar b & \tau\bar \tau &c\bar{c} & q\bar{q} & t\bar t & ZZ & W^+W^-\\
\hline
50\GeV & 1 / 0.75 & 0.50 / 0.67 & 3.9 / 3.2 & 0.32 / 0.59 & 0.48 / 0.66 & - / - & - / - & - / -\\ 
100\GeV & 1 / 0.55 & 0.70 / 0.63 & 2.0 / 2.7 & 0.49 / 0.55 & 0.45 / 0.63 & - / - & 1.0 / 0.75 & 1.1 / 0.75\\ 
200\GeV & 1 / 0.30 & 0.86 / 0.55 & 1.3 / 1.9 & 0.75 / 0.50 & 0.55 / 0.58 & 1.0 / 0.64 & 1.0 / 0.45 & 1.0 / 0.47\\ 
400\GeV & 1 / 0.1 & 0.95 / 0.44 & 1.1 / 0.91 & 0.91 / 0.42 & 0.77 / 0.51 & 1.0 / 0.31 & 1.0 / 0.19 & 1.0 / 0.20\\
1000\GeV & 1 / 0.02 & 0.99 / 0.32 & 1.0 / 0.28 & 0.98 / 0.34 & 0.93 / 0.42 & 1.0 / 0.11 & 1.0 / 0.05 & 1.0 / 0.06\\
\hline
\end{array}$$
\caption[X]{\em Ratios of through-going muon rates `with' over `without' the effects of the neutrino propagation, for DM annihilations around the center of the Earth/Sun. 
E.g.\ the bottom-right entry means that, for $m_{\rm DM}=1000\GeV$,
the rate is unaffected if ${\rm DM}\,{\rm DM}\to W^+W^-$ annihilations occur in the Earth,
and the rate gets reduced to $0.06$ of its value if annihilations occur in the Sun.
\label{tab:RatioFluxes}}
\end{table}

Oscillations driven by $\Delta m^2_{\rm atm}$ and $\theta_{\rm atm}$ are the main effect at work in fig.\fig{EarthNu}.
They convert $\nubarnu_\tau\leftrightarrow\nubarnu_\mu$ at $E_\nu\circa{<} 100 \GeV$
(at larger energies the oscillation length is larger than the Earth radius)
and thus are of the most importance when the initial $\nubarnu_\mu$ fluxes are significantly
different from the $\nubarnu_\tau$ fluxes.
This happens e.g.\ in the case of the {DM DM}$\to \tau\bar\tau$ annihilation channel:
a $\tau$ decay produces one $\nu_\tau$, 
and just about $0.2\ \nu_\mu$ with little energy; 
oscillations subsequently convert $\nu_\tau \to \nu_\mu$ and significantly enhance the rate of $\mu$ events.
For instance, neglecting oscillations the $\chi\chi \to b \bar b$ annihilation channel for neutralinos $\chi$ is regarded as a more significant source of $\nu_\mu$ than the $\chi\chi \to \tau\bar\tau$ channel, because of the relative branching ratio of $3(m_b/m_\tau)^2$  (the precise value depending on stau and sbottom masses). Oscillations partly compensate this factor
as quantitatively shown in table~\ref{tab:RatioFluxes}, 
that summarizes the relative enhancements or reductions due to oscillations, on the rate of through-going muon events (see below) for different annihilation channels and for different DM masses $m_{\rm DM}$.

\begin{table}[t]
\vspace{0.7cm}
$$\begin{array}{|c|cccccccc|}\hline
\hbox{DM mass} & \multicolumn{8}{|c|}{\hbox{DM annihilation channels in the Earth/Sun}}\\
m_{\rm DM} & \nu\bar\nu & b\bar b & \tau\bar \tau &c\bar{c} & q\bar{q} & t\bar t & ZZ & W^+W^-\\
\hline
50\GeV & 100 / 90 & 11 / 11 & 32 / 33 & 12 / 11 & 5.7 / 4.8 & - / - & - / - & - / -\\ 
100\GeV & 100 / 80 & 11 / 9.1 & 25 / 31 & 11 / 8.8 & 3.2 / 3.8 & - / - & 33 / 29 & 34 / 31\\ 
200\GeV & 100 / 62 & 12 / 7.3 & 22 / 26 & 12 / 6.9 & 2.8 / 2.8 & 15 / 12 & 33 / 22 & 34 / 24\\ 
400\GeV & 100 / 35 & 11 / 5.5 & 22 / 19 & 12 / 5.0 & 2.2 / 2.1 & 15 / 8.9 & 33 / 15 & 35 / 17\\ 
1000\GeV & 100 / 9.5 & 10 / 2.8 & 24 / 9.6 & 9.9 / 2.9 & 3.9 / 1.3 & 15 / 5.0 & 33 / 7.0 & 36 / 8.0\\
\hline
\end{array}$$
\caption[X]{\em Average percentage energies in units of $m_{\rm DM}$ of $\nubarnu_\mu$ 
produced by DM annihilations around the center of the Earth/Sun,
computed for various annihilation channels and for various DM masses. 
E.g.\ the bottom-right entry means that ${\rm DM}\,{\rm DM}\to W^+W^-$ annihilations with $m_{\rm DM}=1000\GeV$ produce $\nubarnu_\mu$ with average energy equal to
$36\%\cdot m_{\rm DM}=360\GeV$ 
if occurring in the Earth and to $8\%\cdot m_{\rm DM}$ if  in the Sun.
In the Earth the dependence on $m_{\rm DM}$ 
is due to oscillations (more important at lower $m_{\rm DM}$) and to
energy losses of primary particles (more important at higher $m_{\rm DM}$).
In the Sun oscillations give sizable effects for any $m_{\rm DM}$,
and absorption is significant for $m_{\rm DM}>100\GeV$.
\label{tab:Emean}}
\end{table}

Oscillations also distort the energy spectrum of neutrinos, in the ways precisely shown in fig.\fig{EarthNu}. 
Table~\ref{tab:Emean} reports the mean $\nubarnu_\mu$ energy after oscillations, for different DM masses and different annihilation channels. Notice that, when kinematically open, the $ZZ$, $W^+W^-$ channel remains harder than $\tau \bar\tau$, 
that is usually quoted as the source of a hard neutrino spectrum.

It is also worth noticing that the DM neutrino signal comes from a distance $L\simeq R_\oplus$, while
the background of up-going atmospheric neutrinos from $L\simeq 2 R_\oplus$. 
Indeed oscillations produce dips in the background atmospheric  $\nu_\mu$'s at energies
$E_\nu \approx \Delta m^2_{\rm atm} R_\oplus/2\pi (n-1/2)$
where $n=1,2,3,\ldots$ ($E_\nu \approx 26\GeV$ for $n=1$)
and in the background of atmospheric  $\nu_\tau$'s at
$E_\nu \approx \Delta m^2_{\rm atm} R_\oplus/2\pi n\approx 13\GeV/n$. 
Since uncertainties on the determination of $|\Delta m^2_{\rm atm}|$ from atmospheric experiments are still significant, all above energies could have to be rescaled by up to $\pm30\%$, so that our results would be somewhat affected.

Oscillations driven by $\Delta m^2_{\rm sun}$ and $\theta_{\rm sun}$ 
have a little effect (at variance of what will happen for DM annihilations in the Sun).
Finally, let us comment on the small effect of a non vanishing $\theta_{13}$ on the fluxes.
If $\theta_{13}=0$, $\nubarnu_e$ are decoupled from the oscillations driven by $\Delta m^2_{\rm atm}$ and $\theta_{\rm atm}$, so their spectra are not affected: the solid lines 
(oscillated results) are actually superimposed to the dotted ones (no oscillations) in fig.\fig{EarthNu}. 
Choosing instead $\theta_{13}=0.1$ rad, we plot for illustration in the upper-left panel of fig.\fig{EarthNu} the resulting spectra (dashed lines). It is evident that the modifications are small and mainly concentrated at low energies, where the atmospheric background is dominant. 
This behavior is readily understandable in terms of eq.\eq{PhiEarth} and fig.\fig{PEarth}, that plots the conversion probabilities as function of the energy.
With the same tools, one sees that for the other flavors or for more energetic neutrinos, the effects of $\theta_{13}$ are even smaller or completely negligible, so we go back to the assumption $\theta_{13}=0$ in the other panels of fig.\fig{EarthNu} and in all other results from now on.

\bigskip

Let us summarize  the impact of oscillations, making reference to
fig.s~\ref{fig:EarthNu} and\fig{OscNoOsc}a:  the $\nu_e$ flux is unchanged;
the $\nu_\mu$ flux is significantly increased only for the $\tau$
annihilation channel; the $\nu_\tau$ flux
increases in the case of annihilation into $b$ or $c$. We next compute 
the energy spectra of the main topologies of events that 
contribute to the measured rates in detectors.

\subsection{Through-going muons}
Through-going $\mu^\pm$ are the events dominantly generated by up-going $\nubarnu_\mu$
scattering with the water or (more importantly) with the
rock  below the detector and that run across the detector.
Their rate and spectrum negligibly depends on the composition of the material: 
for definiteness we consider the rock case.
We compute their spectra by considering all the muons produced by neutrinos in the rock underneath the detector base, following the energy loss process in the rock itself~\cite{muonEloss} and collecting all $\mu^\pm$ that reach (with a degraded energy) the detector base.
We ignore through-going muons produced
by $\nubarnu_\tau$ scatterings with the matter below the detector
that produce $\tau^\pm$ that decay into $\mu^\pm$, as these give only a small
contribution ($\circa{<}10\%$) to the total rate.

\begin{figure}[p]\vspace{-1cm}
\centerline{Through-going $\mu^\pm$ from the Earth}\vspace{-10mm}
$$\includegraphics[width=17cm]{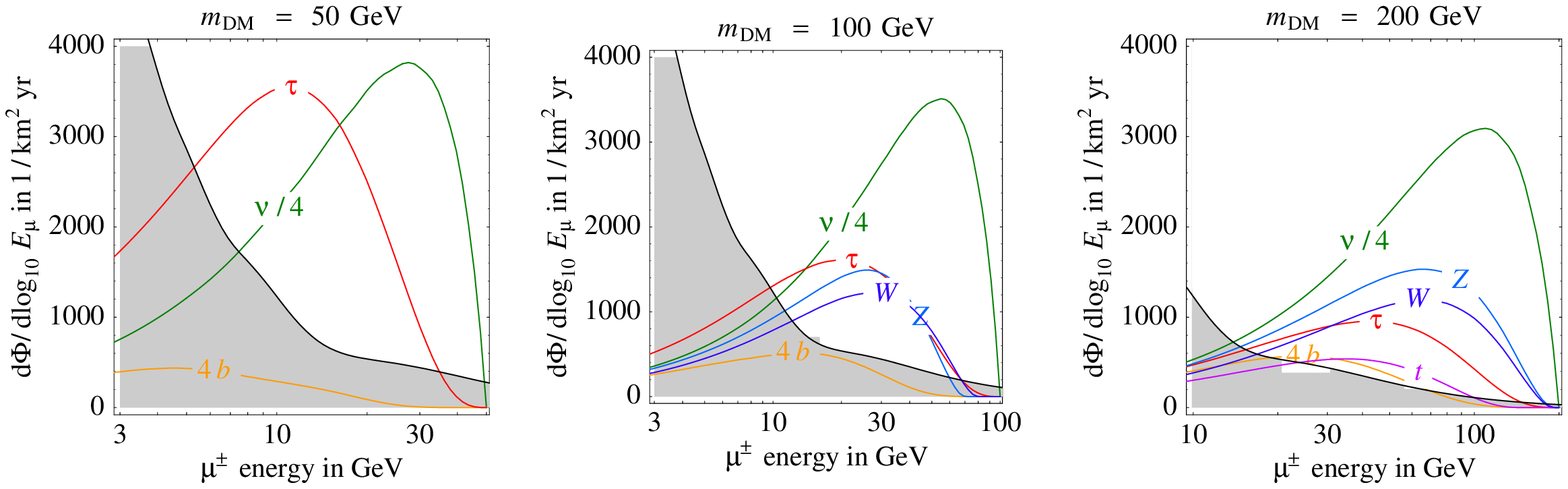}$$\vspace{-8mm}
\caption{\em Spectra of through-going $\mu^+$ and $\mu^-$ (summed) generated by DM annihilations
around the center of Earth.
The plots assume the DM annihilation rate of eq.\eq{EarthNorm},
different DM masses and show the main
annihilation channels.
For better illustration, some channels have been rescaled by the indicated factor.
The shaded region is the atmospheric background.
\label{fig:EarthMuTG}}
\vspace{5mm}
\centerline{Fully contained $\mu^\pm$ from the Earth}\vspace{-10mm}
$$\includegraphics[width=17cm]{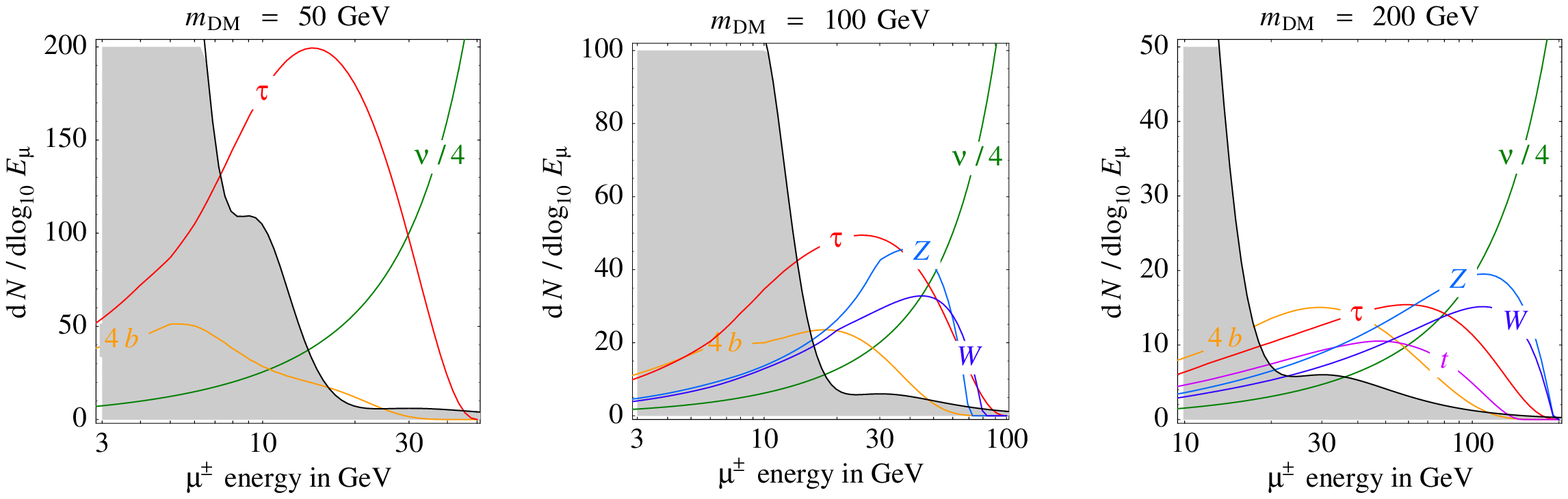}$$\vspace{-10mm}
\caption{\em Spectra of fully contained $\mu^+$ and $\mu^-$ (summed)
 generated by DM annihilations
around the center of Earth.
We assume the DM annihilation rate of eq.\eq{EarthNorm}  and a detector with {\rm Mton$\cdot$year} exposure.
The $\nu\bar\nu$ channel gives a $\mu^\pm$ 
spectrum peaked at $E_\mu \sim m_{\rm DM}$,
due to the monochromatic parent spectrum.
\label{fig:EarthMuC}}
\vspace{5mm}
\centerline{Showers from the Earth}\vspace{-10mm}
$$\includegraphics[width=17cm]{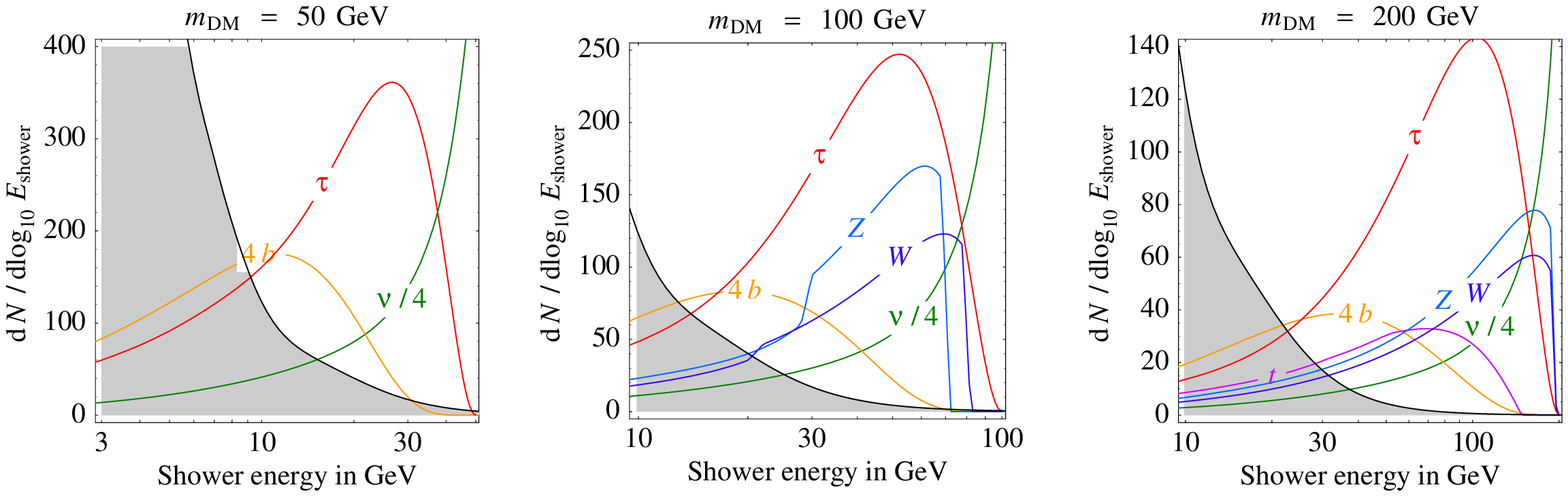}$$\vspace{-10mm}
\caption{\em (Idealized) energy spectra  of showers generated by DM annihilations
around the center of Earth, in a detector with {\rm Mton$\cdot$year} exposure and unit detector efficiency, assuming the DM annihilation rate of eq.\eq{EarthNorm}.
\label{fig:EarthShower}}
\end{figure}

Fig.\fig{EarthMuTG} shows their expected spectrum in  $\km^{-2}{\rm yr}^{-1}$.
Note that, since both the scattering cross section and muon path-length are
roughly proportional to the neutrino energy,
assuming the annihilation rate of eq.\eq{EarthNorm}
we get a total flux that roughly does not depend on $m_{\rm DM}$.
Also, note that due to the strong increase of the $\mu^\pm$ flux with the energy of the neutrino,
annihilation channels that produce very energetic neutrinos (such as {DM DM}$\to \nu\bar\nu$)
give a much larger flux than channels that produce soft neutrinos (such as {DM DM}$\to b\bar{b}$).
Therefore in fig.\fig{EarthMuTG} we had to rescale these fluxes by appropriate factors.

In Table~\ref{tab:RatioFluxes} we present the ratio of the rates of through-going muons with and without oscillations, showing how oscillations affect such experimental observable.

Present data constrain the total $\mu^\pm$ flux 
to be below $10^{2\div 5}/{\rm km}^2\cdot {\rm yr}$~\cite{AMANDA}, the constraint being stronger when $\mu^\pm$ are more energetic. 
The fluxes in fig.\fig{EarthMuTG} obey such limits, except probably the 
case of 100\% annihilation into the hardest channel  $\nu\bar\nu$ 
that was not considered in the AMANDA analysis~\cite{AMANDA}.

\subsection{Fully contained muons}
Fully contained muons mean $\mu^\pm $ that are created inside the detector 
and that remain inside the detector, such that it is possible to measure their initial energy.
We compute their energy spectra convoluting the neutrino fluxes plotted in fig.\fig{EarthNu}
 with the cross section in the detector volume.
They are shown in fig.\fig{EarthMuC}, where the results are normalized 
assuming the  DM annihilation rate of eq.\eq{EarthNorm},
and considering a detector with active mass times live-time equal to
a Mton$\cdot$year.  

The extent to which muons can fit into this category depends of course on the size and geometry of the detector, because more energetic muons travel a longer distance, and on the possibility to 
apply containment requirements. For instance, a km$^3$ detector in ice or water (mass 1000 Mton) contains muons up to about $100\GeV$.
However detectors of such large sensitive mass are being built with the focus on discovery:
such sizes inevitably impose to sacrifice the granularity of the detector, implying higher energy thresholds and poorer energy resolution. 
Neutrino telescope detectors may have an insufficient
granularity of photo-tubes to allow a safe containment cut.
ANTARES attempted a study~\cite{carloganu} of the energy reconstruction from the muon range
for contained events but the efficiency at sub-TeV energies is
affected by luminous backgrounds in the sea. 
In {\sc IceCube}-like detectors a good energy reconstruction
(of the order of $\pm 30\%$) is achieved above the  TeV range, 
which leaves small  room for WIMP fluxes.

A better energy resolution would help in discriminating the signal from the atmospheric background (concentrated at lower energies) and would allow to study the properties of the signal.
For example, the smaller SK detector 
achieved a 2\% in the energy resolution of charged particles, and measured quite precisely the energy 
for the single ring contained events, with $E_\mu\circa{<}10\GeV$~\cite{SKres}.
 At higher energies neutrino collisions are dominated by deep inelastic scattering
 interactions, that produce multiple final state particles
 making more difficult to tag the event and to  measure their energies.
In principle their total energy is more strongly correlated to the incoming neutrino energy;
we here compute the energy spectrum of muons only.
A water \v{C}erenkov Mton detector could isolate fully contained events
up to $(20\div30)\GeV$ (depending on its geometry)
achieving a similar energy resolution as SK.

\subsection{Showers}
So far we considered the traditional signals generated by CC $\nubarnu_\mu$ scatterings.
We now explore the shower events generated by:
\begin{itemize}
\item[(1)] CC scatterings of $\nubarnu_e$. 
We assume that the total energy of the shower is  $E_{\nu_e}$.
This is a simplistic and optimistic assumption:
the appropriate definition is detector-dependent.
One can hope that showers allow to reconstruct an energy
which is more closely related to the  incoming neutrino energy 
than what happens in the case of the  $\mu^\pm$ energy.

\item[(2)] NC scatterings of $\nubarnu_{e,\mu,\tau}$.
At given energy NC cross sections are about 3 times lower than CC cross section.
We assume that the energy of the shower is equal to the energy of the scattered hadrons.

\item[(3)] CC scatterings of $\nubarnu_\tau$, generating $\tau^\pm$ that
promptly decay into hadrons. We assume that the shower energy
is given by the sum of energies of all visible particles:
$E_{\rm shower}\equiv E_{\nu_\tau} - \sum E_\nu$.
We computed the energy spectra of final-state neutrinos 
in section~\ref{CCsec}, see fig.\fig{TauReg}.
(Sometimes $\tau^\pm$ decay into $\mu^\pm$ giving
a shower accompanied by a muon: some
detectors might be able of tagging this class of events).
\end{itemize}
Unlike the case of $\mu^\pm$, shower events can be considered as fully contained at any energy.
Water \v{C}erenkov detectors with a high photo-multiplier coverage,
such as SuperKamiokande, can identify $e$-like events most of which are due to 
$\nu_e$ interactions and separate them from $\mu$-like events that in almost all of the cases are due to $\nu_\mu$ CC interactions.
NC and other CC interactions from $\nu_e$ can be separated by the above 
topologies only on statistical basis.
The energy threshold is lower than a GeV.
Similar capabilities might be reached by a future Mton water \v{C}erenkov detector.
The largest and least granular planned detectors cannot distinguish 
$e^\pm$ from $\tau^\pm$ from hadrons:
all of them are seen as showers and
at the moment the energy threshold is around a TeV.
In conclusion, it seems possible to
measure the energy and the direction of the shower
with experimental uncertainties comparable to the ones for muons.

As in the case of fully contained $\mu^\pm$, we give
 the number of shower event for an ideal detector with Mton$\cdot$year exposure.
The search in real detector requires to include the efficiencies of detection. 
In \v{C}erenkov
detectors, it is relatively difficult to see high energy $\nu_e$, so this
issue is particularly important for showers. In the energy range relevant for the
search of neutrinos from DM annihilation, it is possible to
reach a $20 \%$ efficiency  at least~\cite{bkpc}.

We compute the spectrum of showers summing the
three sources listed above.
Indeed assuming that oscillations are fully known,
measuring the two classes of events
that we consider ($\mu^\pm$ and showers)
is enough for reconstructing the two kinds of primary neutrino fluxes produced
by DM annihilations: $\nubarnu_\tau$ and $\nubarnu_{e,\mu}$.
Fig.\fig{EarthShower} shows the energy spectrum of showers.
By comparing it with the corresponding plot for fully contained muons, fig.\fig{EarthMuC},
one notices that showers have a rate about 2 times larger, and that
retain better the features of the primary neutrino spectra
(at least in the idealized approximation we considered).

\medskip

{}From the point of view of atmospheric background, $\nubarnu_e$ and $\nubarnu_\tau$ are more favorable than $\nubarnu_\mu$ due to two factors:
I. the flux of atmospheric $\nubarnu_e$ drops more rapidly above $E_\nu \circa{>}10\GeV$
(because only at low energy atmospheric $\mu^\pm$ decay in the atmosphere
producing  $\nubarnu_e$ rather than colliding with the Earth).
II. atmospheric $\nubarnu_\tau$ are generated almost only through
atmospheric oscillations, that at baseline $2R_\oplus$ and at 
large energies $E_\nu$ give a small $P_{\mu\tau}\approx (40\GeV/E_\nu)^2$.

\begin{figure}[t]
\centerline{$\nu_{e,\mu,\tau}$ from the Sun}\vspace{-9mm}
$$\hspace{-8mm}\includegraphics[width=18cm]{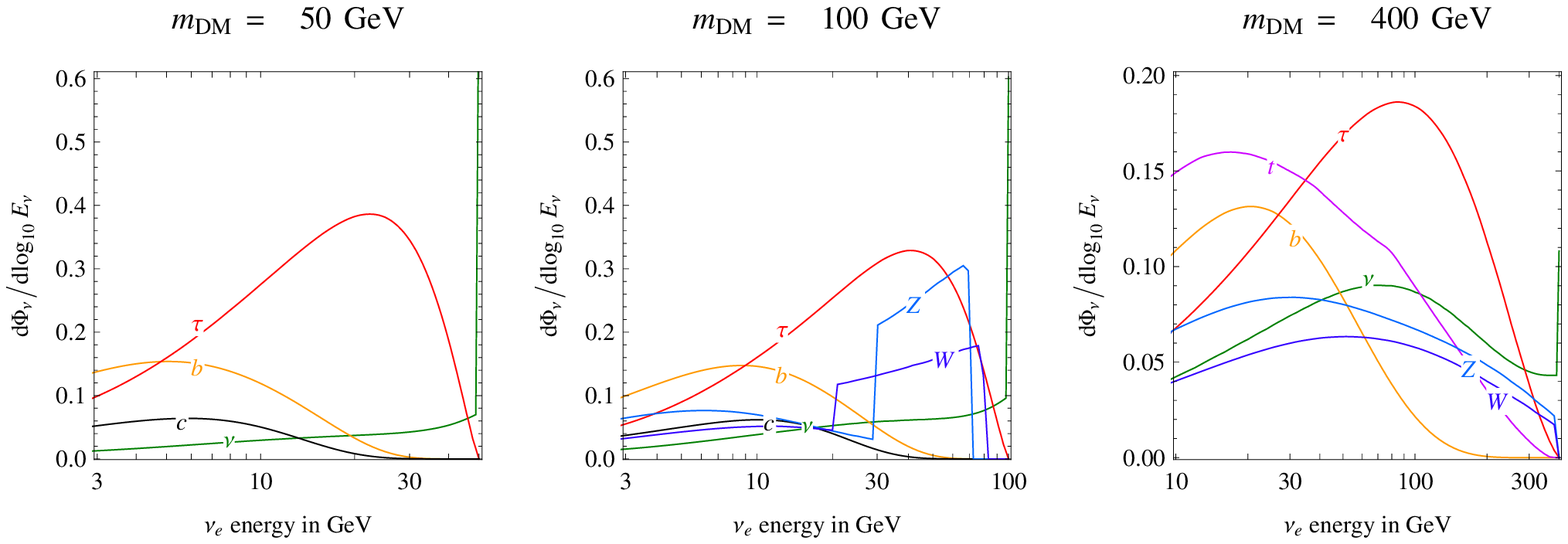}$$\vspace{-10mm}
$$\hspace{-8mm}\includegraphics[width=18cm]{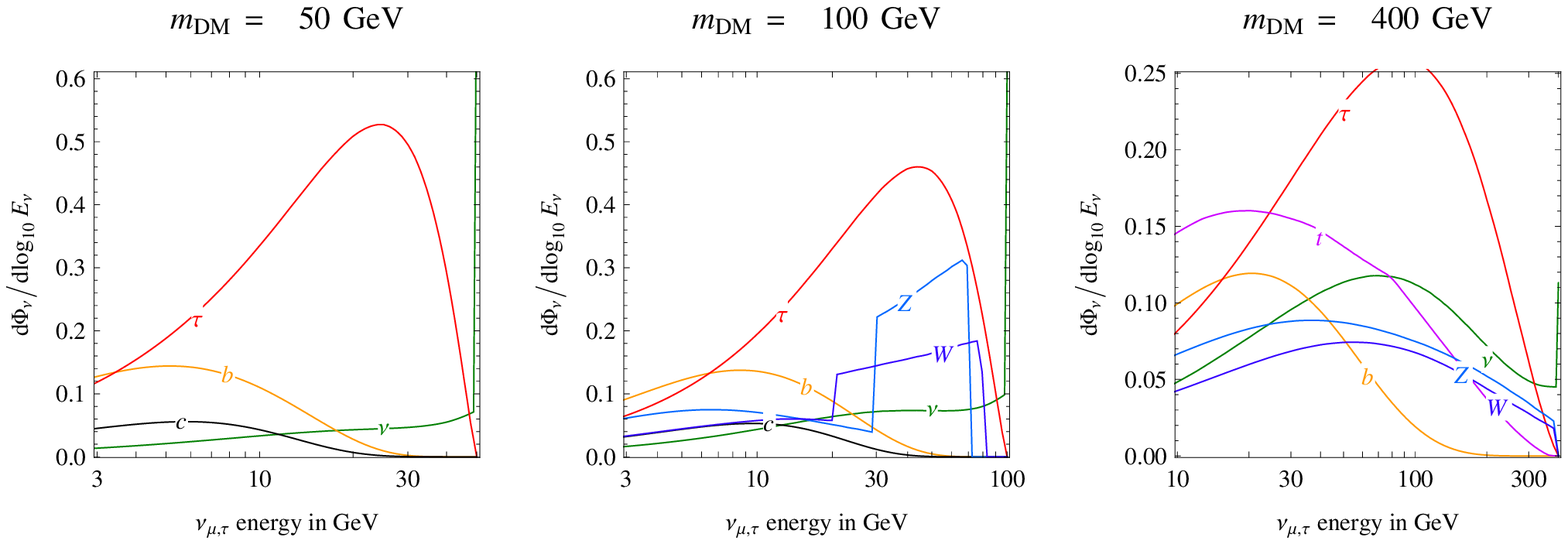}$$
\caption[X]{\em Neutrino fluxes generated by DM annihilations
around the center of Sun.
Upper row: $\nu_e$ fluxes.
Lower row: the almost equal fluxes of $\nu_{\mu}$ and $\nu_\tau$.
We plot the combination $(2\Phi_\nu + \Phi_{\bar\nu})/3$.
All spectra are significantly different from those at production point (not shown here).
For the ${\rm DM}\,{\rm DM}\to\nu\bar\nu$ channel at $E_\nu=m_{\rm DM}$ we plotted the survival probability rather than the energy spectrum, because it is not possible to plot a Dirac $\delta$ function.
\label{fig:SunNu}}
\end{figure}

\begin{figure}[p]\vspace{-1cm}
\centerline{Through-going $\mu^\pm$ from the Sun}\vspace{-10mm}
$$\includegraphics[width=17cm]{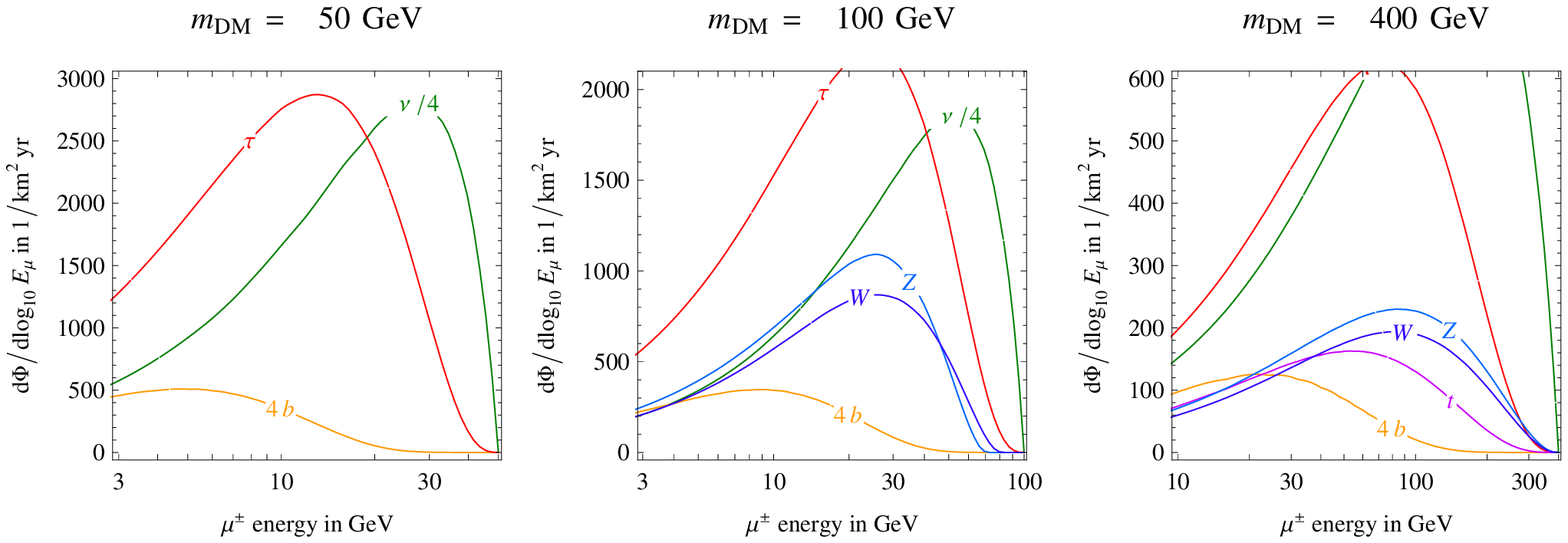}$$\vspace{-10mm}
\caption{\em Spectra of through-going $\mu^+$ and $\mu^-$ (summed) generated by DM annihilations
around the center of Sun.
The plots assume the DM annihilation rate of eq.\eq{SunNorm},
different DM masses and show the main
annihilation channels.
Some channels have been rescaled by the indicated factor for better illustration.
\label{fig:SunMuTG}}\vspace{2mm}
\centerline{Fully contained $\mu^\pm$ from the Sun}\vspace{-1cm}
$$\includegraphics[width=17cm]{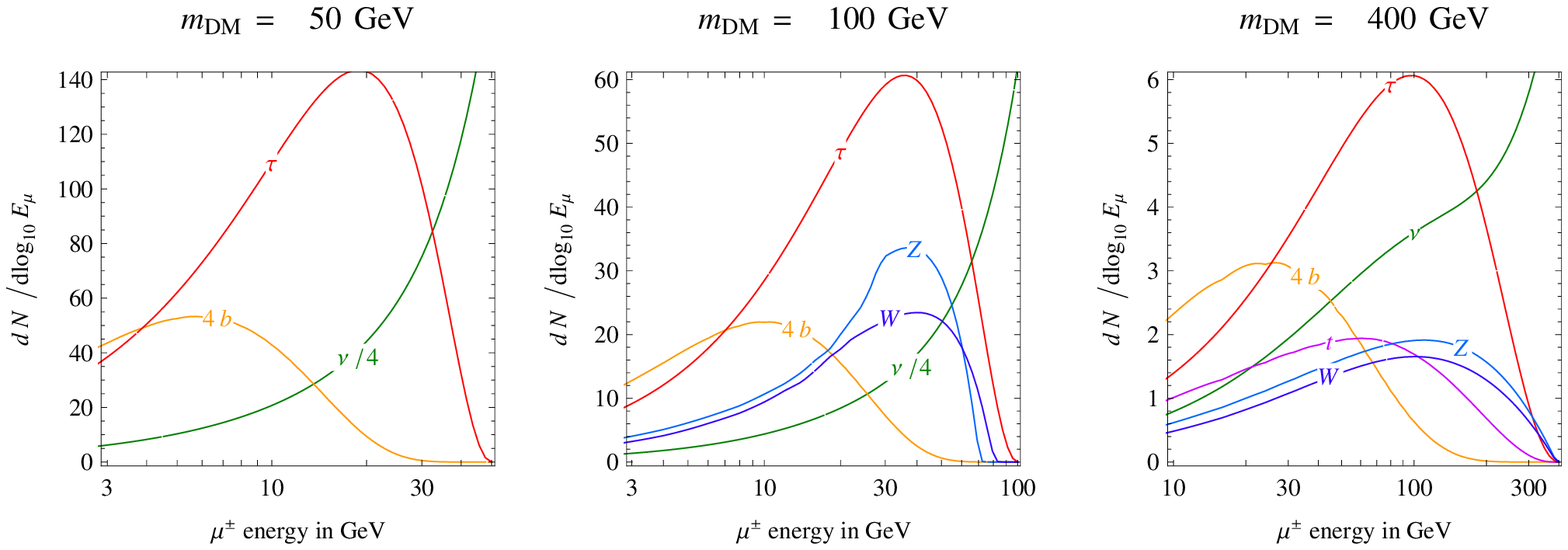}$$\vspace{-1cm}
\caption{\em Spectra of fully contained $\mu^+$ and $\mu^-$ (summed) generated by DM annihilations
around the center of Sun.
We assume the DM annihilation rate of eq.\eq{SunNorm}  and a detector
with {\rm Mton$\cdot$year} exposure.
\label{fig:SunMuC}}
\vspace{1mm}
\centerline{Showers from the Sun}\vspace{-9mm}
$$\includegraphics[width=17cm]{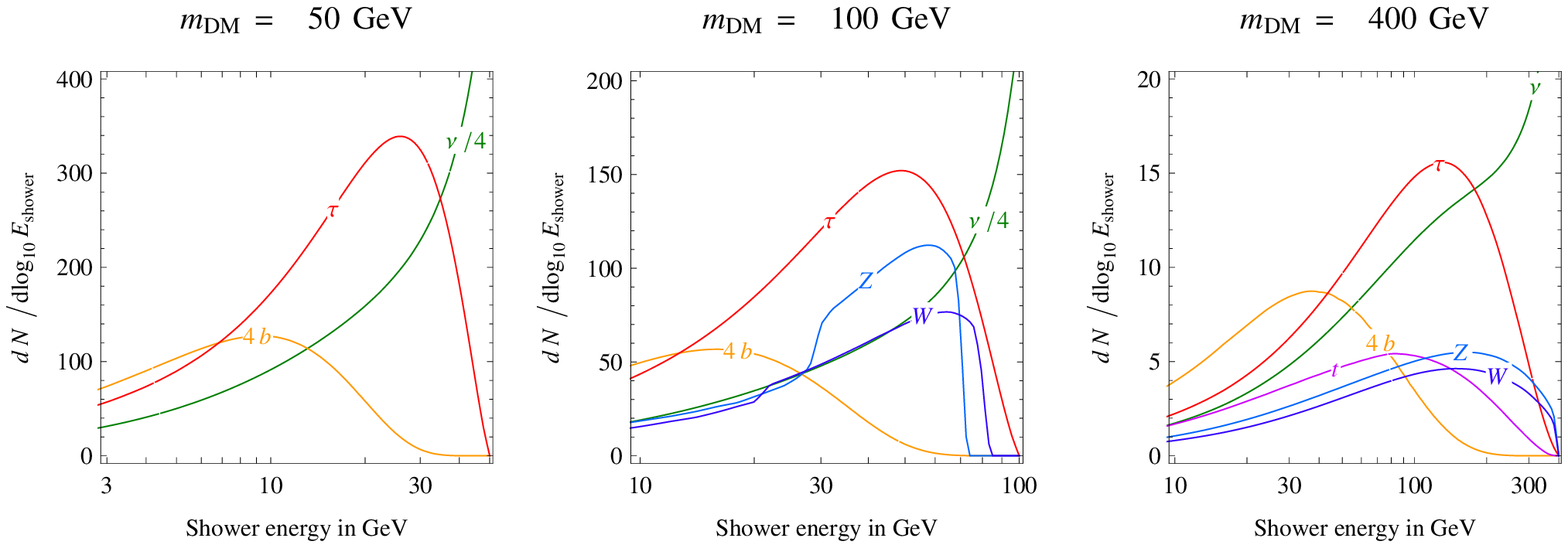}$$\vspace{-1cm}
\caption{\em (Idealized) energy spectra of showers generated by DM annihilations
around the center of Sun, in a detector with {\rm Mton$\cdot$year} exposure and unit efficiency, assuming the DM annihilation rate of eq.\eq{SunNorm}. 
\label{fig:SunShower}}
\end{figure}

\section{Neutrinos from DM annihilations in the Sun}\label{Sun}
In this section we show the results concerning the signal from Dark Matter annihilations around the center of the Sun.
The potential signal from the Sun is expected to be as promising
as the Earth signal and less subject to model dependent assumptions on  DM capture rates.

We present our results showing the same kinds of plots previously employed in the Earth case.
Fig.s\fig{SunNu} show the DM$\nu$ fluxes\footnote{These DM$\nu$ fluxes are available at~\cite{www}.}. The main topologies of events that detectors can discriminate are the ones already discussed in the Earth case: 
fig.\fig{SunMuTG} shows the spectra of through-going $\mu^\pm$,
fig.\fig{SunMuC} the spectra of fully contained $\mu^\pm$
and fig.\fig{SunShower} those of showers.
For brevity we will not here repeat the features that these two cases have in common, and we focus on their differences.

\medskip

The neutrino fluxes in fig.\fig{SunNu} are computed in three steps.
1) Evolution inside the Sun needs the formalism presented in section~\ref{formalism}.
2) Oscillations in the space between the Sun and the Earth average to zero
the coherencies among different neutrino mass eigenstates.
The density matrix becomes therefore diagonal in the mass eigenstate basis.
3) Neutrinos can be detected after having crossed the Earth,
so that we computed the functions 
$P(\nu_i\to \nu_\ell)$ and $\bar{P}(\bar\nu_i\to\bar\nu_\ell)$
(where $\nu_i=\{\nu_1,\nu_2,\nu_3\}$ are mass eigenstates and
and $\nu_\ell=\{\nu_e,\nu_\mu,\nu_\tau\}$ are flavor eigenstates)
taking into account Earth matter effects.
These functions depend on neutrino energy and neutrino path.
All plots are done assuming that neutrinos cross the center of the Earth.
We also assume $\theta_{13}=0$, such that the actual path of the neutrino
across the Earth is unimportant. Indeed for $\theta_{13}=0$
$P$ and $\bar{P}$ marginally differ from the value they achieve in the limit of
averaged vacuum oscillations:
\beq P(\nu_i\to \nu_\ell)\simeq P(\bar\nu_i\to \bar\nu_\ell) \simeq 
|V_{\ell i}|^2.\eeq
Oscillations driven by $\Delta m^2_{\rm sun}$
inside the Earth give some correction only below a few GeV.
If instead $\theta_{13}\neq 0$ a dedicated path-dependent computation is needed
in the energy range $(2\div 50)\GeV$.

\medskip

When needed, we assume the following rate of DM annihilations inside the Sun:
\beq\label{eq:SunNorm}\left.\Gamma_{\rm ann}\right|_{\rm Sun}=\left(\frac{r_{\rm SE}}{R_\oplus}\right)^2
\left.\Gamma_{\rm ann}\right|_{\rm Earth}.\eeq
where $r_{\rm SE}$ is the Sun-Earth distance, $R_\oplus$ is the radius of the Earth
and $\Gamma_{\rm ann}|_{\rm Earth}$ is the rate we assumed inside the Earth,
given in eq.\eq{EarthNorm}.
This amounts to assume that the DM annihilation rate in the Sun is bigger than 
the annihilation rate in the Earth (because the Sun is larger and more massive than the Earth)
by a factor that precisely compensates for the larger distance from the source, $r_{\rm SE}\gg R_\oplus$.
In SUSY models where the DM particle is a neutralino, samplings of the MSSM parameter space
show that this compensation is a typical outcome.
So our assumption of eq.\eq{SunNorm} is again realistically optimistic.

\paragraph{Backgrounds.}
The atmospheric neutrino background  depends on the orientation of the Sun relative to the detector, 
so that we do not show it in the figures as shaded regions, and discuss it here.
The overall magnitude of the atmospheric  flux has only a ${\cal O}(1)$ dependence on the zenith angle, 
so that in first approximation the shaded areas of fig.\fig{EarthNu} remain similar in the solar case.
However, atmospheric oscillations affect $\nubarnu_\mu$ and $\nubarnu_\tau$ coming from below (at least at energies of ${\cal O}(10)\GeV$)
and not neutrinos coming from above.
The case of $\nubarnu_\tau$ is qualitatively important: during the day 
the signal of down-going solar DM$\nubarnu_\tau$ is virtually free
from atmospheric background. 
Indeed, the direct production of atmospheric $\nubarnu_\tau$ is negligible~\cite{atm nutau} and 
so is the flux from other possible astrophysics sources~\cite{cosmic nutau}.
Unfortunately, tagging $\nubarnu_\tau$ is a difficult task, and none of the proposed experiments seems able to do it.

Furthermore there is a new background due to `corona neutrinos': high energy neutrinos produced by cosmic rays interactions in the solar corona (i.e.\ the solar analog of atmospheric neutrinos)~\cite{corona}.
Their flux is however of limited importance: terrestrial atmospheric neutrinos, restricted to the small cone of eq.\eq{cut} centered on the Sun, remain a more significant background at the neutrino energies $E_\nu\circa{<}\TeV$ where a solar DM$\nu$ signal can arise.

\paragraph{The effect of oscillations and interactions.} As discussed in section~\ref{Production}, the higher density of the Sun
mildly affects the neutrino spectra at production.
Propagation effects are instead significantly different,
as illustrated in fig.\fig{OscNoOsc}.
DM$\nu$ from the Earth are affected only by `atmospheric' oscillations:
only $\nu_\mu$ and $\nu_\tau$ are significantly affected, and only below $E_\nu\circa{<}100\GeV$.
DM$\nu$ from the Sun are instead affected by both `atmospheric' and `solar'
oscillations: in the whole plausible energy range
oscillations are averaged and all flavors are involved.
Furthermore absorption exponentially suppresses DM$\nu$ at $E_\nu $ above $100\GeV$.
This effect is partly compensated by $\nu_\tau$ and NC regeneration,
that re-inject more neutrinos below about $100\GeV$.
This explains the main features of fig.\fig{OscNoOsc}.

It is easier to see these effects at work looking at the ${\rm DM~DM}\to\nu\bar\nu$ channel.
We assume that the initial flux is equally distributed among flavors, 
so that oscillations alone would have no effect:
indeed in the Earth case the neutrino spectra remain a monochromatic line at $E_\nu = m_{\rm DM}$.
On the contrary in the Sun there is a reduced line at $E_\nu =m_{\rm DM}$,
plus a tail of regenerated neutrinos at $E_\nu < m_{\rm DM}$.
For $m_{\rm DM}\ll 100\GeV$ the line is unsuppressed 
and the tail contains a small number of neutrinos.
As $m_{\rm DM}$ increases the line becomes progressively more suppressed and
the fraction of neutrinos in the tail increases.
For $m_{\rm DM}\gg 100\GeV$ the line disappears and all neutrinos 
are in the tail, that approaches a well defined energy spectrum.

The  same phenomenon happens for the other DM annihilation channels:
the final flux is a combination of `initial' and `regenerated' neutrinos,
and the `regenerated' contribution becomes dominant for $m_{\rm DM}\gg 100\GeV$.
In section~\ref{limit} we will provide an analytical understanding
of the main features of this phenomenon.

\medskip

Table~\ref{tab:RatioFluxes} summarizes the
 effect of propagation on the total through-going $\mu^\pm$ rate:
  for $m_{\rm DM}\circa{<}100$ $\GeV$ oscillations give an ${\cal O}(1)$ correction
  (e.g.\ an enhancement by a factor 3 in the case of the $\tau\bar\tau$ channel);
  for $m_{\rm DM}\gg 100\GeV$ absorption gives a significant depletion
  (e.g.\ for $m_{\rm DM} = 1000 \GeV$ the depletion factor is 
  $0.1\div0.01$ depending on the channel).

  Table~\ref{tab:Emean} shows the average energies of DM$\nu$:
  as we now discuss solar DM$\nu$ cannot have energies much above $100\GeV$,
a scale set by the interactions in the Sun.

\begin{figure}[t]
$$\includegraphics[width=16cm]{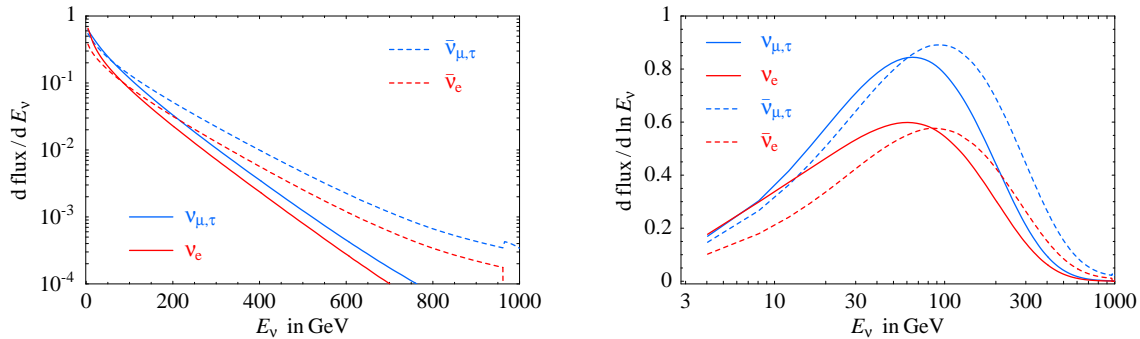}$$
\caption[X]{
\label{fig:Attractor}\em  Limit neutrino energy spectra at the exit from the Sun, occurring in the case of very energetic initial neutrinos.}
\end{figure}

\subsection{`Heavy Dark Matter' and the `limit spectrum'}\label{limit}
The numerical results presented above indicate that for $m_{\rm DM}\gg\TeV$ (`Heavy DM') the  
fluxes of DM$\nu$ from the Sun approach a well defined and simple `limit spectrum', which is essentially independent on the features of the initial fluxes.
 In this section we describe the properties of such spectrum and develop an analytic insight into them. 

There are a number of interesting cases that fall into the category of `Heavy DM'.
For instance, non-thermally produced super-massive dark matter with mass $m_{\rm DM}\sim 10^{10}\GeV$ has been considered in~\cite{superheavy, Crotty},
where the `limit spectrum' was first studied.
But also thermal relics of strongly interacting particles yield the observed DM abundance for $m_{\rm DM}\sim (10\div 100)\TeV$.\footnote{E.g.\
technicolor models can contain stable techni-baryons (analogous to the proton) that make up Dark Matter~\cite{techniDM}. TC models with a characteristic scale around a TeV
were originally proposed as a natural solution for electroweak symmetry breaking,
but now constraints from precision data push the scale at the much higher energies
also suggested by thermal DM abundance.}
Even in the case of supersymmetry, $m_{\rm DM}\circa{>}\TeV$ is possible
if the lightest sparticle is a higgsino or in coannihilation funnels,
although this requires sparticles much heavier than what suggested by naturalness considerations.

\medskip

The `limit spectrum' arises because the annihilations of very heavy DM particles produce very energetic
neutrinos which undergo many interactions inside the Sun: the interactions wash out the initial features of the neutrino fluxes (that only control the overall magnitude of the final flux) 
and determine almost universal flavor ratios and energy spectra at the exit. 
In the Sun interactions are relevant at $E_\nu\circa{>}100\GeV$ and the
limit spectrum is attained for primary neutrino energies $E_\nu\gg 1\TeV$.
In the Earth an analogous limit spectrum is attained for $E_\nu\gg100\TeV$.

Fig.\fig{Attractor} shows the outcome of a typical numerical run. 
The $\nu_e:\nu_\mu:\nu_\tau:\bar\nu_e:\bar\nu_\mu:\bar\nu_\tau$
flavor composition is determined by the effect of $\nu_\tau$-regeneration (which gives more $\nubarnu_\tau$ than $\nubarnu_{e,\mu}$) and by the effect of oscillations 
(which equate $\nubarnu_\mu$ and $\nubarnu_\tau$ and generate some $\nubarnu_e$), 
essentially independently on the original values.\footnote{A possible exception occurring if the initial flux contains only $\nubarnu_e$, that do not undergo CC regeneration nor oscillations in the Sun (due to matter suppression). We do not consider this peculiar case here.} 
The energy spectra are well approximated by exponentials $e^{-E_\nu/\mathscr{E}}$ with  slopes given by $\mathscr{E} \approx 100\GeV$ for $\nu$ and $\mathscr{E} \approx 140\GeV$ for $\bar\nu$.

In order to understand such features, it is useful to consider a simplified version of the equations described in section~\ref{formalism}, which captures the main points of the full problem.
Namely, under the assumptions that:
(a) oscillations and regeneration roughly equidistribute neutrinos among the different flavors, 
so that we can replace the flavor density matrix $\rhob(E)$ with a single density $\rho(E)$;
(b) the re-injection spectrum from NC scatterings and $\nu_\tau$-regeneration can be taken flat in $E'_\nu/E_\nu$ (although this is not an accurate description especially for $\nu_\tau$-regeneration);  
(c) the cross sections are proportional to the neutrino energy in the relevant energy range ($\sigma_{\rm CC,NC} \propto E_\nu$);
the full equations of section~\ref{formalism} reduce to a single integro-differential equation with an absorption term and a re-injection integral:
\beq
\label{eq:df} 
\mathscr{E} \frac{\partial \rho(x,E)}{\partial x} = - E\ \rho(x,E) + \int_{E}^\infty \rho(x,E')\ dE'.
\eeq
The spatial variable $r$ has been rescaled here to the quantity $x/\mathscr{E}$. 
The `neutrino optical depth'
$x$ spans (0,1), where $x=0$ corresponds to the production point in the center of the Sun and $x=1$ to the exit from it. $\mathscr{E}$ therefore incorporates the radius of the Sun and all the numerical constants that appear in the cross sections.
Using the explicit numbers for the CC and NC cross sections, we compute values for $\mathscr{E}$ that are in good agreement with the numerical results quoted above. We checked that, dropping the assumption of the flatness of the $\nu_\tau$-regeneration spectrum, the agreement actually becomes optimal. 

Eq.\eq{df} can be analytically solved:
\beq\label{eq:fsol}
\rho (x,E)= \exp(-E x/\mathscr{E}) \bigg[ \rho(0,E) + \frac{x}{\mathscr{E}} \int_{E}^\infty \rho(0,E')\ dE'\bigg]\eeq
as one can verify either directly or passing to the variable $\tilde{\rho} = e^{{E x}/{\mathscr{E}}}\rho$. 
The first term of eq.\eq{fsol} describes the initial neutrino spectrum $\rho(x=0,E)$, which suffers from 
an exponential suppression; 
the second term is the contribution of regeneration,
proportional to the traversed portion $x$.
Now, consider neutrinos with an initial energy $E_0$
(namely $\rho(0,E)=\delta(E-E_0)$): 
the first term becomes less and less relevant 
as neutrinos proceed to $x\gg \mathscr{E}/E_0$;
the second term (that reads $x/\mathscr{E}\,\theta(E_0-E)\,\exp(-Ex/\mathscr{E})$) conversely  becomes dominant over the first as $x$ increases.
More generally, after a path $x$ a `limit spectrum' of exponential shape
\beq\label{eq:exp}
\rho (x,E) \propto \exp (-Ex/\mathscr{E})
\eeq
is approached irrespectively of the initial spectrum $\rho(0,E)$
provided that the injection spectrum $\rho(0,E)$ is concentrated at $E\gg \mathscr{E}/x$.
This is a simple non-trivial result.
In all cases of `Heavy DM', such conditions are verified and indeed the exponential spectra at the exit from the Sun ($x=1$) are well visible in the outcome of the numerical computations shown in fig.\fig{Attractor}\footnote{Our `limit spectra' in fig.\fig{Attractor} 
agree with the corresponding fig.~16 of~\cite{Crotty}, within their uncertainties.
However, in that and other works, such spectra are approximated with a log-normal function,
apparently with the motivation that the central limit theorem might play some r\^ole in determining the out-coming spectrum after many random interactions. 
We find instead that a log-normal does not fit the numerical result better than an exponential, and that a log-normal does not arise from the analytical argument presented above (deviations from the exponential form of eq.\eq{exp} arise mainly 
because the CC re-injection spectra are not flat in $E'_\nu/E_\nu$).}.

\medskip

\section{Reconstructing the DM properties}\label{banana}
We now study how measurements of the spectra  of the previously discussed classes of events
can be used to reconstruct the DM properties: its mass and its branching ratios
into the various annihilation channels.
Even without a good energy resolution, some of the channels give different enough 
spectra that it seems possible to experimentally discriminate them. 
This is e.g.\ the case of $\nu\bar\nu$ versus $b\bar{b}$ and (to a lesser extent) versus $\tau\bar\tau$.
Other annihilation channels instead produce too similar energy spectra  (e.g.\ $W^+W^-$ and $ZZ$)
so that distinguishing them seems too hard.
This issue depends significantly on whether $m_{\rm DM}$ is in the energy range 1.\, 2.\ or 3., defined at page~\pageref{123}.

\begin{figure}
$$\hspace{-8mm}\includegraphics[width=7cm]{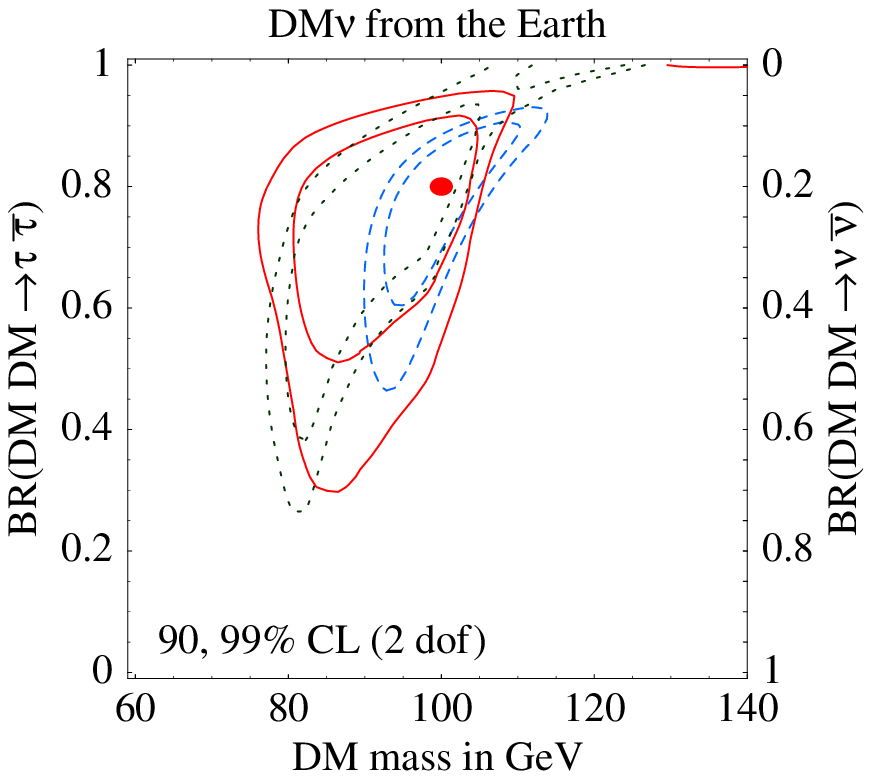}\qquad\qquad
\includegraphics[width=7cm]{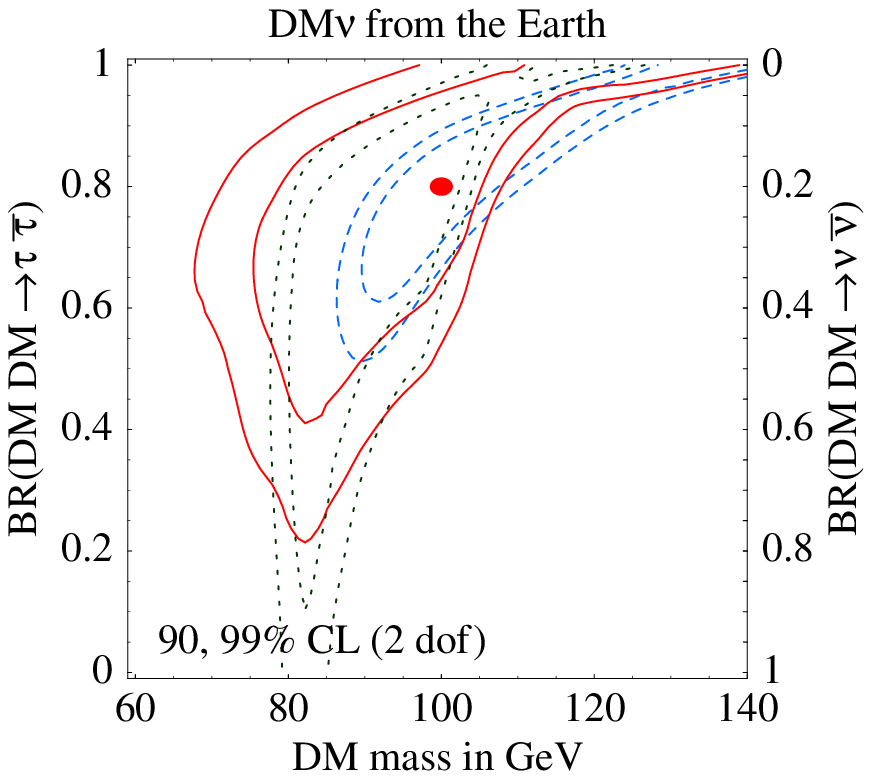}$$
\caption{\em Discrimination capabilities fitting simulated data
with a statistics of either 1000 through-going muons (dashed lines) 
or 100 fully contained muons (continuous)
or 200 showers (dotted)
measured in energy bins of width $\Delta E = 30\GeV$.
The true point is indicated by a dot.
\label{fig:banana}}
\end{figure}

Fig.s\fig{banana}a (Earth case) and\fig{banana}b (Sun case) illustrate more quantitatively the discrimination capabilities in a specific example. 
We assumed a DM  particle with mass $m_{\rm DM}=100\GeV$ annihilating into $\tau\bar\tau$ and $\nu\bar\nu$ only, with ${\rm BR}({\rm DM~DM}\to\tau\bar\tau)=0.8$ and ${\rm BR}({\rm DM~DM}\to\nu\bar\nu)=0.2$.
The contours identify the regions that would be selected at $90,99\%$ C.L.\ (2 dof) with the collection of 1000 through-going muons (dashed lines) or 100 fully contained muons (continuous)
or 200 showers (dotted).  No information on the rate at which this statistics of events is collected is assumed.
An actual experiment will have an energy-dependent energy resolution,
that we approximate in the following semi-realistic way:
we group events into  energy bins $n=1,2,3,\ldots$ with energy
$30(n\pm 1/\sqrt{12})\GeV$.
With a flat probability distribution this would correspond to
ranges $30(n-1/2)\GeV < E <30(n+1/2)\GeV$;
we assume a Gaussian probability.
No energy threshold is assumed, and it is effectively set by the atmospheric 
background, that  below $15\GeV$ dominates over the signal.
We computed the best-fit regions that would be obtained in an experiment
where the rate measured in each energy bin equals its theoretical prediction.
In general this is not true, and the  best-fit regions experience statistical fluctuations:
we computed their average position
(see~\cite{deGouveaMura} for a discussion of this point).

The banana-shaped best-fit regions in fig.\fig{banana} arise
because it is difficult to discriminate a harder channel from a heavier DM particle.
In particular, in the Sun the spectra of neutrinos with energy $E_\nu\circa{>}100\GeV$ are affected
by absorption and regeneration that tends to wash-out their initial features, see section~\ref{limit}.
{}From the point of view of the reconstruction of the DM properties, this wash-out  is clearly an unpleasant feature. 
In the limit $m_{\rm DM}\gg \TeV$ solar DM$\nu$ approach the spectrum in fig.\fig{Attractor}
irrespectively of the DM properties.

\medskip

Fig.\fig{banana} also illustrates that  different classes of events can have comparable capabilities.
Indeed
\begin{itemize}
\item  Through-going $\mu^\pm$ give the highest statistics
if DM$\nu$ have energies $E_\nu\circa{>}50\GeV$.
However, for the purpose of spectral reconstruction, 
they are less powerful than contained events:
their energy is less correlated to the energy of the scattered neutrino so that
all annihilation channels produce similar bell-shaped energy spectra
and discriminating the annihilation channel becomes harder.

\item Fully contained $\mu^\pm$ better trace the parent neutrino spectra,
but can only be observed up to a maximal energy determined by the size of the detector. 
Below this energy their rate is comparable to the rate of through-going $\mu^\pm$.
Thus, contained events are a competitive signal of
relatively light DM particles.\footnote{ In our figures,
this can be verified by comparing
fig.~\ref{fig:EarthMuTG} with fig.s~\ref{fig:EarthMuC} and
fig.~\ref{fig:EarthShower}, evaluating the number of through-going muons in
one year with an area $A=(0.1~\mbox{km})^2$, as appropriate for a Mton
detector. We see that, especially for light DM particles, the number of
contained events is comparably large.

It is useful for orientation to write the ratio
of contained-to-through-going events considering $\mu^\pm$ continuous energy
losses $d E_\mu/dx\approx -\alpha$ 
(with $\alpha\sim 2\cdot 10^{-3}$ GeV/cm) as: 
$$\frac{N_{\rm cont}}{N_{\rm through}}\approx
\frac{\alpha\cdot \langle \epsilon \rangle V/A }{m_{\rm DM}}\cdot
F\left(\frac{E_{\rm th}}{m_{\rm DM}}\right) $$ 
where $F$ is an adimensional function, and where we consider a water \v{C}erenkov detector 
with threshold $E_{\rm th}$, volume-to-area ratio $V/A$ and an average efficiency of detection
$\langle \epsilon \rangle$.
To give a concrete example of our expectations,
   for a DM candidate of $m_{\rm DM}=50$~GeV
   that annihilates preferentially into taus we expect about
   8 contained $\mu$ (fully or partially) and 16 shower events for
   each through-going muon event coming from the Earth (or the Sun),
   when we adopt as detector parameters $E_{\rm th}=15~\GeV$,
   $A=(0.1\km)^2$ and $V=(0.1\km)^3$.}

\item Showers are fully contained in all the relevant energy range
(making more difficult to tag them)
and  they can efficiently trace the parent neutrino spectra,
depending on how the detector can measure their energy.
Furthermore, the showers/muons event ratio
(not considered in our fit) allows to discriminate annihilation channels
that produce neutrinos with different flavour proportions.
\end{itemize}


\begin{figure}
$$\includegraphics[width=17cm]{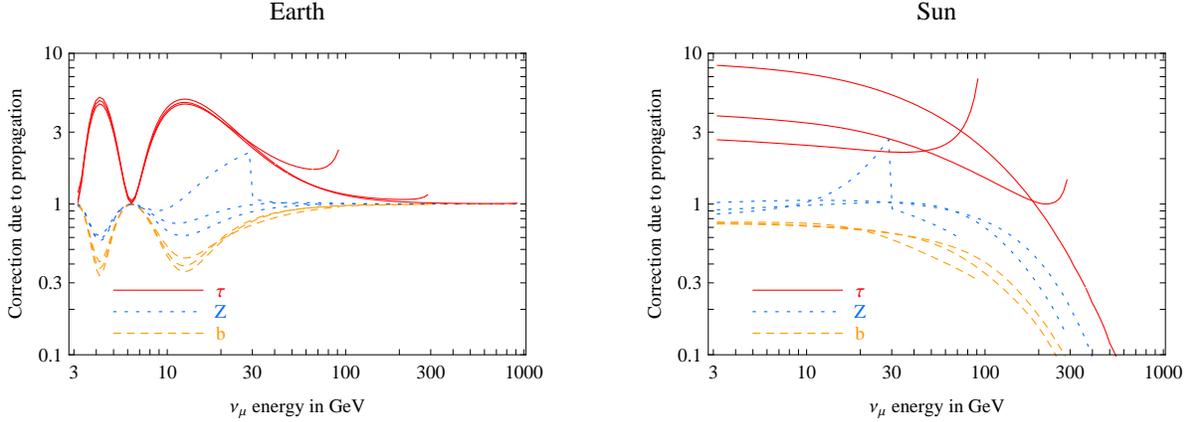}$$
\caption[X]{\label{fig:OscNoOsc}\em 
Modifications of neutrino fluxes due to propagation. The figures show the ratio of $\nu_\mu$ fluxes `with'  over `without' the effects of neutrino propagation (oscillations, absorptions, regeneration). The lines refer to neutrinos from DM annihilations into $\tau\bar\tau$ (continuous line), $ZZ$ (dotted) and $b\bar{b}$ (dashed), for $m_{\rm DM}= \{100,1000\}\GeV$ (distinguishable by the corresponding maximum neutrino energy).}
\end{figure}

\section{Conclusions}
We performed a phenomenological analysis of neutrinos of all flavors 
generated by annihilations of  DM particles (`DM$\nu$') with weak-scale mass 
accumulated inside the Earth or the Sun.
Our analysis  is valid for any  DM candidate. Indeed, 
the DM$\nu$ signal depends only on the following parameters:
the DM mass $m_{\rm DM}$, the  DM annihilation rate,  
the branching ratios ${\rm BR}({\rm DM}\,{\rm DM}\to f)$
for the various annihilation channels $f$.
A given underlying model (e.g.\ supersymmetry) predicts these quantities:
the total rates suffer a sizable astrophysical uncertainty and typically
DM$\nu$ are a promising DM signal.
The other parameters determine the expected spectra of DM$\nu$.
We therefore computed the DM$\nu$ signal as functions of these parameters
and studied how they can be reconstructed from a possible future
measurement of DM$\nu$ spectra.

\medskip

We considered annihilation channels into presently known particles:
$\nu\bar\nu$, $b\bar b$, $\tau\bar\tau$,  $ZZ$, $W^+W^-$.
We also considered annihilations into $c\bar{c}$, lighter quarks and gluons:
their contribution at $E_\nu\ll m_{\rm DM}$  is not completely negligible.
Taking into account the different energy losses of primary particles inside the Earth
and inside the Sun,
in section~\ref{Production} we computed the two independent spectra at the production point:
for $\nubarnu_\tau$ and for $\nubarnu_{e,\mu}$.
These spectra are modified by propagation:
flavor oscillations, absorption, regeneration.
The necessary formalism is presented in section~\ref{formalism}
and appendix~\ref{B} shows an example of features that
simplified approaches cannot catch.
Their combined effect is illustrated in fig.\fig{OscNoOsc} on the $\nubarnu_\mu$ flux for some selected cases, and amounts to a ${\cal O}(0.1\div 10)$ correction:
\begin{itemize}
\item DM$\nu$ from the Earth are affected only by atmospheric $\nubarnu_\mu\leftrightarrow\nubarnu_\tau$
oscillations at energies $E_\nu\circa{<}100\GeV$.
\item DM$\nu$ from the Sun of any flavor and any energy
are affected by averaged `solar' and `atmospheric' oscillations.
Furthermore, absorption suppresses neutrinos with
$E_\nu\circa{>}100\GeV$, that are partially converted
(by NC and by $\nubarnu_\tau$ regeneration) into
lower energy neutrinos.
In section~\ref{limit} we analytically studied how and when
neutrinos with energy $E_\nu\gg 100\GeV$ approach the well-defined
limit spectrum shown in fig.\fig{Attractor}.
\end{itemize}
Our result for DM$\nu$ of all flavors from the Earth (Sun) are shown in fig.\fig{EarthNu} (\ref{fig:SunNu}).
The comparison with the atmospheric background (shaded regions) is performed
assuming the realistically optimistic
annihilation rate in eq.\eq{EarthNorm}.
Table~\ref{tab:Emean} at page~\pageref{tab:Emean} summarizes the average final neutrino energies.
Table~\ref{tab:RatioFluxes} at page~\pageref{tab:RatioFluxes} summarizes how propagation modifies the total rate of through-going $\mu^\pm$ generated by $\nubarnu_\mu$.

The latter is often considered the most promising event topology for present detectors.
We also considered other topologies of events: fully contained $\mu^\pm$
and showers ($e^\pm$ generated by $\nubarnu_e$
and hadrons generated by all neutrinos).
They have a lower rate if $m_{\rm DM}\circa{>}100\GeV$ but, even in this case, these classes of events are important because
(1) their energy is more strongly
correlated to the incoming neutrino energy.
(2) there are two independent DM$\nu$ spectra `at origin' to be measured,
so that at least two classes of events are necessary.

Finally, fig.\fig{banana} illustrates quantitatively how measuring the DM$\nu$ energy spectra of these classes of events 
can allow us to reconstruct the basic properties of the DM particle: its mass and some annihilation branching ratios.

Existing detectors and those under construction will likely not have the necessary capabilities, because SK is too small, and much bigger `neutrino telescopes' are optimized for more energetic neutrinos (the large volume is obtained at the expense of granularity, resulting in high energy thresholds ($\sim 50\GeV$) and poor energy resolution ($\sim \pm 30\%$)). Increasing the instrumentation density goes in the direction of solving this issue. If a DM$\nu$ signal is discovered, it will be then interesting to tune the planned future detectors (or project a dedicated detector) to DM$\nu$, with presumable energy $E_\nu\sim 100\GeV$.

\medskip

\paragraph{Note added:}
In the present version 5 of hep-ph/0506298 a bug in the propagated fluxes of antineutrinos from the Sun has been fixed, leading to corrections of the order of 10\% in the fluxes presented in Fig.~\ref{fig:SunNu} and, as a consequence, in the spectra presented in Figures~\ref{fig:SunMuTG} $\to$  \ref{fig:SunShower}.
In previous versions, an erroneous double counting of the prompt neutrino yield in $W$-boson decays and a numerical bug in the implementation of the boost for top quark decays had been fixed. These modifications affected the $W^+W^-$ and $t\bar t$ channels in the fluxes at production of Figure~\ref{fig:Prim} as well as (as a consequence) the propagated fluxes presented in Figures~\ref{fig:EarthNu} $\to$ \ref{fig:SunShower} and Tables~\ref{tab:RatioFluxes} and \ref{tab:Emean}.
Furthermore the values of some parameters had been updated, leading to very minor changes.\\
Overall, these corrections and refinements amount to adjustments of the order of 10\% to 20\% at most in the numerical results. 
All physics discussions and conclusions are always unchanged.
Updated results are available in electronic form from~\cite{www}.


\medskip

\paragraph{Acknowledgments}
We thank Giuseppe Battistoni and Ed Kearns for useful conversations. 
We thank Joakim Edsj\"o, Tommy Ohlsson, Mattias Blennow and Chris Savage for cross-checking the results of their novel calculation with ours, from which a few errors in our first releases were found.
The work of M.C. is supported in part by the USA DOE-HEP Grant
DE-FG02-92ER-40704. The work of N.F. is supported by a
joint Research Grant of the Italian Ministero dell'Istruzione,
dell'Universit\`a e della Ricerca (MIUR) and of the Universit\`a di
Torino within the {\sl Astroparticle Physics Project} and by a Research
Grant from INFN.


\appendix

\section{Neutrino spectra per annihilation event}\label{app:boost}
The neutrino differential spectrum per annihilation event is defined
in the rest frame of the annihilating DM, since the annihilation
process occurs at rest. The spectrum can be calculated by following
analytically the decay chain of the annihilation products until a
$\tau$, a quark or a gluon is produced. The neutrino spectrum is then
obtained by using the Monte Carlo modeling of the quark and gluon
hadronization, or $\tau$ decay.  We produced the $\nu_e=\nu_\mu$
and $\nu_\tau$ differential distributions for $h=\tau,q,c,b,{\rm
gluon}$ at various injection energies for each $h$ ($q=u,d,s$ stands
for a light quark). Whenever we need the $\nu$ distribution for
an injection energy different from the produced ones, we perform an
interpolation. In order to obtain the neutrino differential
distribution in the DM rest frame we perform the necessary boosts on
the MC spectra.

For instance, let us consider neutrino production from a chain of this
type:
\begin{equation}
{\rm DM}\,{\rm DM} \rightarrow A \rightarrow a \rightarrow h \rightsquigarrow \nu \;\;.
\label{eq:decay}
\end{equation}
The neutrino differential energy
spectrum per annihilation event is obtained by the product of the
branching ratios for the production of $A$, $a$ and $h$ in the decay
chain, with the differential distribution of neutrinos produced by
the hadronization of an $h$ injected at an energy $E_{\rm prod}$
(defined in the rest frame of the $a$ decaying particle) double
boosted to the DM reference frame:
\begin{equation}
\frac{dN_\nu}{dE} =
{\rm BR}({\rm DM}\,{\rm DM} \rightarrow A)\cdot{\rm BR}(A \rightarrow a)\cdot{\rm BR}(a \rightarrow h)
\cdot  \left[\left({dN^h_{\nu}\over dE}
        \right)_{{\rm boost~}a\rightarrow A}
   \right]_{{\rm boost~}A\rightarrow{\rm DM}}\;\;.
\end{equation}
The first boost transforms the spectrum from the rest frame of $a$ (in
which $h$ is injected with energy $E_{\rm prod}$) to the rest frame of
$A$. The second brings the distribution to the rest frame of
DM. Each boost is obtained by the following expression:
\begin{equation}
\frac{dN_\nu}{dE} = \frac{1}{2} \int_{E'_-} ^{E'_+}
\left. \left(\frac{dN^h_{\nu}}{dE'}\right) \right |_{E_{\rm prod}}
\, \frac{dE'}{\gamma \beta\ E'}\qquad \hbox{with}\qquad
E'_{\pm} = \min\left[
E_{\rm prod}, \gamma E
\left( 1\pm \beta\right)
\right]\;\;.
\end{equation}
where $E$ denotes the energy of neutrinos,
$\gamma$ and $\beta$ are the Lorentz factors of the boost.

\section{Cross sections}\label{Cross}

The DIS NC differential cross section on an average nucleus $N$ is
$$\frac{d\hat{\sigma}}{dE'_\nu}(\nu N\to \nu'X) = \sum_{q=\{u,d\}} \frac{2G_{\rm F}^2m}{\pi} \bigg[p_q
\bigg(g_{Lq}^2+g_{Rq}^2\frac{E^{\prime 2}_\nu}{E_\nu^2}\bigg)+p_{\bar q}
\bigg(g_{Rq}^2+g_{Lq}^2\frac{E^{\prime 2}_\nu}{E_\nu^2}\bigg)
\bigg]$$
$$\frac{d\hat{\sigma}}{dE'_{\nu}}(\bar\nu N\to \bar\nu'X) = \sum_{q=\{u,d\}} \frac{2G_{\rm F}^2m}{\pi} \bigg[p_q
\bigg(g_{Rq}^2+g_{Lq}^2\frac{E^{\prime 2}_\nu}{E_\nu^2}\bigg)+p_{\bar q}
\bigg(g_{Lq}^2+g_{Rq}^2\frac{E^{\prime 2}_\nu}{E_\nu^2}\bigg)
\bigg]$$
where $0<E'_\nu <E_\nu$ is the energy of the scattered neutrino or anti-neutrino
and $p_u,p_d,p_{\bar{u}}, p_{\bar{d}}$ are the fractions of nucleon momentum carried by
up and down-type quarks and anti-quarks.
In a medium that contains neutrons and protons with densities $N_n$ and $N_p$
\beq\label{eq:pdf}
\begin{array}{ll}\displaystyle
p_u = \frac{0.25 N_p+0.15 N_n}{N_p+N_n}  \qquad & \displaystyle
p_{\bar u} = \frac{0.03 N_p + 0.06 N_n}{N_p+N_n},\\[3mm] \displaystyle
p_d =  \frac{0.25 N_n+0.15 N_p}{N_p+N_n} & \displaystyle
p_{\bar d} = \frac{0.03 N_n + 0.06 N_p}{N_p+N_n}
\end{array}\eeq
The $Z$-couplings of quarks are
$$g_{Lu}=\frac{1}{2}-\frac{2}{3}\sW^2 \qquad
g_{Ru}= -\frac{2}{3}\sW^2 ,\qquad
g_{Ld}= -\frac{1}{2}+\frac{1}{3}\sW^2,\qquad
g_{Rd}= \frac{1}{3}\sW^2. $$

The DIS CC differential cross sections are
\begin{eqnsystem}{sys:nuqCC}
\frac{d\hat{\sigma}}{dy}(\nu_\ell d\to \ell u) &=&
\frac{d\hat{\sigma}}{dy}(\bar\nu_\ell \bar d\to \bar \ell \bar u) =
\frac{G_{\rm F}^2\hat{s}}{\pi} ,\\
\frac{d\hat{\sigma}}{dy}(\nu_\ell \bar{u}\to \ell \bar{d}) &=&
\frac{d\hat{\sigma}}{dy}(\bar\nu_\ell u\to \bar\ell d)  = 
\frac{G_{\rm F}^2\hat{s}}{\pi} (1-y)^2 
\end{eqnsystem}
where $y\equiv -\hat{t}/\hat{s}$ ($0\le y \le 1$),
The total quark CC cross sections are
\beq\hat{\sigma}(\nu_\ell d\to \ell u) =
\hat{\sigma}(\bar\nu_\ell \bar d\to \bar \ell \bar u)=
3\hat{\sigma}(\bar\nu_\ell u\to \bar\ell d) =
3 \hat{\sigma}(\nu_\ell \bar{u}\to \ell \bar{d}) = 
\frac{G_{\rm F}^2\hat{s}}{\pi}. \eeq
$\sqrt{\hat{s}}$ is the center-of-mass energy of the quark sub-processes.
It is given by
$\hat{s}=sx$ , where
$x$ is the fraction of the total nucleon momentum $P$ carried by a quark, $\hat{p} = x P$.

\section{Oscillation and absorption in constant matter}\label{B}
Following~\cite{formalism} we described neutrino propagation
by writing a differential evolution equation for the neutrino density matrix.
Ref.~\cite{Crotty} employed an alternative approach,
merging the usual analytical treatment of oscillations with
a MC code that accounts for absorption.
We here illustrate the non trivial interplay of oscillations 
and absorption, by solving our evolution equation
 in a simple semi-realistic case
where analytic solutions can be obtained.
 Let us consider the $\nu_\mu/\nu_\tau$ system 
where oscillations are due to $\Delta m^2 = \Delta m^2_{\rm atm}$ 
and absorption is due to matter with constant density.
We assume maximal atmospheric mixing, an initial
neutrino state with energy $E_\nu$, and 
study final neutrinos at the same energy $E_\nu$.
Thus, regeneration does not contribute and 
the full system of equations\eq{drho} reduces to
a system of ordinary differential equations.
The explicit solution 
in matrix form\footnote{To derive the 
solution, we found it convenient to rewrite
the evolution equation for the density 
matrix $\rho_{ab}=\psi_a\psi_b^*$ as a Schroedinger-like
equation for the wavefunction $\psi_a$,  
where the absorption term acts as half a decay width. This gives 
a non-hermitian hamiltonian that can be easily diagonalized
and therefore exponentiated.} is:
\begin{equation}\label{eq:rhosol}
\rho(r)=e^{-r/\lambda_{\rm abs}} \;
(\; \mb{U}(r)\; \rho(0) + \rho(0)\; \mb{U}^\dagger(r)\; )\ 
\end{equation}
with
\begin{equation}
\mb{U}=
\left(
\begin{array}{cc}
c-s\; {(\Gamma_\mu-\Gamma_\tau)\lambda_{\rm osc}}/{(4\pi)} &
-i\; s\; {\Delta m^2 \lambda_{\rm osc}}/{(4\pi E_\nu)} \\
-i\; s\; {\Delta m^2 \lambda_{\rm osc}}/{(4\pi E_\nu)} &
c+s\; {(\Gamma_\mu-\Gamma_\tau)\lambda_{\rm osc}}/{(4\pi)} 
\end{array}
\right)
\end{equation}
where we used the shorthands
$c=\cos(\pi r/\lambda_{\rm osc})$ and 
$s=\sin(\pi r/\lambda_{\rm osc})$. 
The factor $\Gamma_\mu$ describes
absorption of $\nu_\mu$ and is approximatively given by
$\Gamma_\mu\approx 0.14 E_\nu\rho G_{\rm F}^2$
in an iso-scalar material with density $\rho$.
Numerically $\Gamma_\mu^{-1}\approx 10^7\,{\rm km}$ for 
the typical density of the Earth mantle $\rho\approx 2.4~{\rm ton}/{\rm m}^3$ 
and for $E_\nu=100\GeV$. 
The two relevant scales that compare with the pathlength $r$
are the absorption
and oscillation lengths:
\beq
\lambda_{\rm abs} = {2}/{(\Gamma_\mu + \Gamma_\tau)},\qquad
\lambda_{\rm osc} =4\pi/  \sqrt{({\Delta m^2}/{E_\nu})^2 - 
({\Gamma_\mu-\Gamma_\tau})^2 }.
\eeq
Notice that absorption reduces the oscillation wavelength.
When $\Gamma_\mu-\Gamma_\tau$ is large enough
$\lambda_{\rm osc}$ becomes imaginary i.e.\
the system becomes `overdamped' , and oscillations disappear. 
While absorption and
$\Gamma_\mu-\Gamma_\tau$ can be neglected in terrestrial experiments
(e.g., in future precise long-baseline experiments)
their effects are relevant for neutrinos with energy $E_\nu\circa{>}100\GeV$
produced by DM annihilation in the Sun; note that in this case
oscillations into $\nu_e$ are suppressed. 
{}From the explicit solution in eq.\eq{rhosol}
one finds the expressions of survival or appearance probabilities
that resemble known things closely,
e.g.,
$$
P_{\mu\tau}=\frac{e^{-r/\lambda_{\rm abs}}}{1-\epsilon^2}
{\sin^2\left(\frac{\Delta m^2\; r}{4E_\nu}\sqrt{1-\epsilon^2}\right)},\qquad
\epsilon\equiv 
\frac{({\Gamma_\mu-\Gamma_\tau}) 
E_\nu}{\Delta m^2}
\approx 0.2
        \left( \frac{\mbox{eV}^2}{\Delta m^2} \right)
        \left( \frac{\Gamma_\mu-\Gamma_\tau}{\mbox{km}^{-1}} \right)
        \left( \frac{E_\nu}{\mbox{GeV}} \right).
$$


\frenchspacing\footnotesize\begin{multicols}{2}
\end{multicols}


\begin{thebibliography}{22}


\bibitem{review} For recent reviews see
\art{G. Jungman, M. Kamionkowski, K. Griest}{Phys. Rep.}{267}{195}{1996},
\art[hep-ph/0404175]{G. Bertone, D. Hooper, J. Silk}{Phys.\ Rept.}{405}{279}{2005}.

\bibitem{GRS} See e.g.\ fig. 3 of
\art[hep-ph/9811386]{L. Giusti, A. Romanino, A. Strumia}{\NP}{B550}{3}{1999}.

\bibitem{Kane}
\hepart[hep-ph/0501262]{J.L. Bourjaily and G.L. Kane}.

\bibitem{idea}
\art{J. Silk, K. A. Olive and M. Srednicki}{\PRL}{55}{257}{1985};
\art{M. Srednicki, K. A. Olive and J. Silk}{\NP}{B279}{804}{1987};
\art{L.M. Krauss, K. Freese, D.N. Spergel, W.H. Press}{Astrophys. J.}{299}{1001}{1985};
\art{K. Freese}{\PL}{B167}{295}{1986};
\art{L. M. Krauss, M. Srednicki and F. Wilczek}{\PR}{D33}{2079}{1986};
\art{T. K. Gaisser, G. Steigman and S. Tilav}{\PR}{D34}{2206}{1986}.


\bibitem{previous}
Several aspects of the neutrino signal from DM annihilations have been studied, mainly with focus on the expected rates in particular models:
\art{G. B. Gelmini, P. Gondolo and E. Roulet}{\NP}{B351}{623}{1991};
  \art{M. Kamionkowski}{\PR}{D44}{3021}{1991};
  \art{F. Halzen, T. Stelzer and M. Kamionkowski}{\PR}{D45}{4439}{1992};
  \art{A. Bottino, V. de Alfaro, N. Fornengo, G. Mignola and M. Pignone}{\PL}{B265}{57}{1991};
  \art[hep-ph/9603342]{V. Berezinsky, A. Bottino, J. R. Ellis, N. Fornengo, G. Mignola and S. Scopel}{Astropart. Phys.}{5}{333}{1996};
\art[hep-ph/9607237]{L. Bergstrom, J. Edsjo and P. Gondolo}{\PR}{D55}{1765}{1997};
  \art[astro-ph/9702037]{L. Bergstrom, J. Edsjo and M. Kamionkowski}{Astropart. Phys.}{7}{147}{1997};
 \art[hep-ph/9806293]{L. Bergstrom, J. Edsjo and P. Gondolo}{\PR}{D58}{103519}{1998};
\art[hep-ph/9809239]{A. Bottino, F. Donato, N. Fornengo and S. Scopel}{Astropart. Phys.}{10}{203}{1999};
 \art[astro-ph/0008115]{J. L. Feng, K. T. Matchev and F. Wilczek}{\PR}{D63}{045024}{2001};
\art[hep-ph/0105182]{V. D. Barger, F. Halzen, D. Hooper and C. Kao}{\PR}{D65}{075022}{2002};
\art[hep-ph/0204135]{V. Bertin, E. Nezri and J. Orloff}{Eur. Phys. J.}{C26}{111}{2002};
 \art[hep-ph/0210034]{V. Bertin, E. Nezri and J. Orloff}{JHEP}{0302}{046}{2003};
\art[hep-ph/0405210]{H. Baer, A. Belyaev, T. Krupovnickas and J. O'Farrill}{JCAP}{0408}{005}{2004}.
\art{K. M. Belotsky, M. Y. Khlopov, K. I. Shibaev}{Phys. Atom. Nucl.}{65}{382}{2002} [{\it Yad. Fiz.} 65 (2002) 407];
\art[astro-ph/0201314]{K. M. Belotsky, T. Damour, M. Y. Khlopov}{\PL}{B529}{10}{2002};
\art{D. Fargion et al.}{Mod. Phys. Lett.}{A11}{1363}{1996};
\art[hep-ph/9906345]{A. E. Faraggi, K. A. Olive, M. Pospelov}{Astropart.\ Phys.}{13}{31}{2000}.

\bibitem{IMB} 
\art{J. M. LoSecco et al.}{\PL}{B188}{388}{1987}.


\bibitem{Kamiokande}
\art{M. Mori et al.}{\PR}{D48}{5505}{1993}.


\bibitem{Baksan}
M. M. Boliev et al., Proc. of Int. Workshop on Aspects of Dark Matter in 
Astrophysics and Particle Physics, Heidelberg, Germany, 16-20 Sep 1996.
Published in {\it Heidelberg 1996, Dark matter in astro- and particle physics},
711-717.


\bibitem{MACRO}
\art[hep-ex/9812020]{M. Ambrosio et al.}{\PR}{D60}{082002}{1999}.



\bibitem{SK}
\art[hep-ex/0404025]{S. Desai et al.}{\PR}{D70}{083523}{2004}, Erratum: {\it ibid.}D70, 109901 (2004).


\bibitem{AMANDA}
\art[astro-ph/0202370]{J. Ahrens et al.}{\PR}{D66}{032006}{2002};
Also M. Ackermann et al., ``Limits to the muon flux from neutralino annihilations in the Sun with the AMANDA detector'', submitted to {\it Phys. Rev. Lett.} on march 2005.


\bibitem{BAIKAL}
\art{V. Aynutdinov et al.}
{Nucl. Phys. Proc. Suppl.}{143}{335-342}{2005}.


\bibitem{ANTARES}
The ANTARES collaboration, 
``A Deep Sea Telescope for High Energy Neutrinos - Proposal for a 0.1km$^2$ Neutrino Telescope'', 31 May 1999,
antares.in2p3.fr/Publications/proposal/proposal99/\-proposal.pdf


\bibitem{ICECUBE}
See \art[astro-ph/0305196]{J. Ahrens {\it et al.} [IceCube collaboration]}{Astropart. Phys.}{20}{507}{2004} 
    and references therein.


\bibitem{NEMO}
P. Piattelli for the NEMO collaboration 
{\it Proceedings of 8th International Workshop on Topics in Astroparticle and 
Underground Physics (TAUP 2003)}, Seattle, Washington, 5-9 Sep 2003.
Published in {\it Nucl. Phys. Proc. Suppl.} 138 (2005) 191.
See also the web page NEMOweb.lns.infn.it.


\bibitem{NESTOR}
Web page www.nestor.org.gr.


\bibitem{Mton}
See the UNO Whitepaper (pre-print: SBHEP01-3) at  ale.physics.Sunysb.edu/uno and  
\hepart[hep-ex/0005046]{C. K. Jung};
  a summary of the Hyper-Kamiokande proposal in 
\art{K. Nakamura}{Int. J. Mod. Phys.}{A18}{4053}{2003}; 
and a presentation of the MEMPHYS project at the NNN05 meeting in 
nnn05.in2p3.fr/schedule.html, where the other set-ups are also reviewed.


\bibitem{angdistribution}
\art[hep-ph/9504283]{J. Edsjo, P. Gondolo}{\PL}{B357}{595}{1995}.
Notice that, due to the radial distribution of DM inside the Earth, the precise effect of oscillations is in principle different at different zenith angles. However these are a small modifications. In particular, oscillations negligibly alter the zenith-angle DM$\nu$ distributions.


\bibitem{nuDMosc}
The impact of oscillations has also been partially addressed in:
\art[hep-ph/9506221]{E. Roulet}{\PL}{B356}{264}{1995};
 \hepart[hep-ph/9904351]{N. Fornengo};
\art[hep-ph/0009183]{M. Kovalski}{\PL}{B511}{119}{2001};
\art[hep-ph/0006157]{A. de Gouvea}{\PR}{D63}{093003}{2001}.


\bibitem{nuDMinterac}
The effect of neutrino interactions in the dense matter of the Sun was included in an effective way in 
\art[hep-ph/9407351]{G. Jungman and M. Kamionkowski}{\PR}{D51}{328}{1985} 
and later in 
\art[hep-ph/9504205]{J. Edsjo}{Nucl. Phys. Proc. Suppl.}{43}{265}{1995};
\hepart[TSL-ISV-93-0091]{J. Edsjo}




\bibitem{Gould:1987}
\art{A. Gould}{Astrophys. J.}{321}{571}{1987}.


\bibitem{capture1}
\art{A. Gould}{Astrophys. J.}{321}{561}{1987}.
\art{A. Gould}{Astrophys. J.}{328}{919}{1988}.


\bibitem{edsjo}
\art{J. Lundberg, J. Edsj\"o}{\PR}{D69}{123505}{2004}.


\bibitem{helio} 
\art{A. Bottino, G. Fiorentini,N. Fornengo, B. Ricci, S. Scopel,
F.L. Villante}{\PR}{D66}{053005}{2002}.


\bibitem{DMastro} 
\art{P. Belli, R. Cerulli, N. Fornengo and S. Scopel}{\PR}{D66}{043503}{2002}.


\bibitem{n(r)}  
\art{K. Griest and D. Seckel}{\NP}{B283}{681}{1987} [Erratum: {\it ibid.}\ B296 (1988) 1034].
\art{W.H. Press and D.N. Spergel}{Astrophys. J.}{296}{679}{1985}.


\bibitem{ritz}
\art{S. Ritz, D. Seckel}{\NP}{B304}{877}{1988}.


\bibitem{pythia}
\art{T. Sj\"ostrand}{Comput Phys. Commun.}{135}{238}{2001}.


\bibitem{cross}
\art{B. Povh, J. H\"ufner}{\PL}{B245}{653}{1990}.




\bibitem{www} 
Web pages: \\ www.to.infn.it/$\sim$fornengo/DMnu.html, www.cern.ch/astrumia/DMnu.html,
www.marcocirelli.net/DMnu.html.



\bibitem{formalism}
The formalism for the propagation of interacting and oscillating neutrino has been fully presented in 
\art[hep-ph/9209276]{G. Raffelt, G. Sigl, L. Stodolsky}{\PRL}{70}{2363}{1993}, 
 \art{G. Sigl, G. Raffelt}{\NP}{B406}{423}{1993}
  although it was pioneeringly introduced in 
\art{A. D. Dolgov}{Sov.\ J.\ Nucl.\ Phys.}{33}{700}{1981} (\art{A. D. Dolgov}{Yad.\ Fiz.}{33}{1309}{1981})   
and 
  \art{R. Barbieri, A. Dolgov}{\NP}{B349}{743}{1991}.
 It has been described also in M. Kowalski in~\cite{nuDMosc} and 
\hepart[hep-ph/0502223]{M. C. Gonzalez-Garcia, F. Halzen and M. Maltoni}.


\bibitem{MSW}
\art{L. Wolfenstein}{\PR}{D17}{2369}{1978};
\art{S.P. Mikheyev, A. Yu Smirnov}{Sovietic Journal
Nucl. Phys.}{42}{913}{1986}.

For a review see
\art{T.K. Kuo, J. Pantaleone}{Rev. Mod. Phys.}{61}{937}{1989}.

\bibitem{BP00}
\art[astro-ph/0010346]{J.N. Bahcall, S. Basu, M.H. Pinsonneault}{Astrophys. J.}{555}{990}{2001}. 
See www.sns.ias.edu/$\sim$jnb/SNdata/sndata.html


\bibitem{PREM}
\art{A.M. Dziewonski, D.L. Anderson}{Phys. Earth Planet. Interior}{25}{297}{1981}.


\bibitem{Vmutau}
\art{F.J. Botella, C.S. Lim, W.J. Marciano}{\PR}{D35}{896}{1987}.


\bibitem{NuFit}
\hepart[hep-ph/0503246]{A. Strumia and F. Vissani}.


\bibitem{taureg idea}
\art[hep-ph/9804354]{F. Halzen and D. Saltzberg}{\PRL}{81}{4305}{1998}
 \art[astro-ph/0111482]{J. F. Beacom, P. Crotty and E. W. Kolb}{\PR}{D66}{021302}{2002}.



\bibitem{taureg comput}
\art[hep-ph/0312295]{E. Bugaev, T. Montaruli, Y. Shlepin and I. Sokalski}{Astropart.\ Phys.}{21}{491}{2004} and references therein.
  For background information, see also \book{A. Stahl}{Physics with $\tau$ leptons}{Springer Verlag}{Heidelberg}{2000}, where spectra of $\tau$ decay products are reviewed. Some analytical approximations can also be found e.g. in \art[hep-ph/0005310]{S. I. Dutta, M. H. Reno and I. Sarcevic}{\PR}{D62}{123001}{2000}.


\bibitem{sigmaonelectrons}
See e.g. R. Gandhi {\it et al.} in~\cite{sigmatau}.


\bibitem{sigmatau}
Cross sections for $\nu_\tau N\to \tau X$ have been computed in
\art[hep-ph/0208187]{S. Kretzer, M.H. Reno}{\PR}{D66}{113007}{2002}.
\art[hep-ph/0305324]{K. Hagiwara, K. Mawatari, H. Yokoya}{\NP}{B668}{364}{2003}.
\art[hep-ph/0407275]{K. M. Graczyk}{\NP}{A748}{313}{2005} and
  \hepart[hep-ph/0407283]{K.M. Graczyk}.
For a simple summary of results see
\hepart[hep-ph/0407371]{J.M. L\'evy}.
See also \art[hep-ph/9512364]{R. Gandhi, C. Quigg, M. H. Reno and I. Sarcevic}{Astropart.\ Phys.}{5}{81}{1996}.



\bibitem{Crotty} 
\art[hep-ph/0205116]{P. R. Crotty}{\PR}{D66}{063504}{2002};
  See also P. R. Crotty, 
FERMILAB-THESIS-2002-54.




\bibitem{FLUKA}
\art[hep-ph/9907408]{G. Battistoni, A. Ferrari, P. Lipari, T. Montaruli, P. R. Sala and T. Rancati}{Astropart. Phys.}{12}{315}{2000}
  and 
\art[hep-ph/0207035]{G. Battistoni, A. Ferrari, T. Montaruli and P. R. Sala}{Astropart. Phys.}{19}{269}{2003},
\art{Erratum}{ibidem}{19}{291}{2003}.
   The data tables can be downloaded from www.mi.infn.it/$\sim$battist/neutrino.html.


\bibitem{muonEloss}
For the computation of the energy loss process for muons we use the precise results of 
\art[hep-ph/0010322]{I. A. Sokalski, E. V. Bugaev and S. I. Klimushin}{\PR}{D64}{074015}{2001}

\bibitem{carloganu}
C. Carloganu, 
 ``Caract\'erisation des performances \`a basse \'energie du futur t\'elescope 
sous-marin \`a neutrinos ANTARES et leur application \`a l'\'etude des 
oscillations des neutrinos atmosph\'eriques  
(1999)''
in antares.in2p3.fr/Publications/index.html\#thesis

\bibitem{SKres}
\hepart[hep-ex/0501064]{SK collaboration}, page 13.

\bibitem{bkpc} 
Giuseppe Battistoni, Ed Kearns, private communications.

\bibitem{atm nutau}
\art[hep-ph/9811268]{L. Pasquali, M. Reno}{Phys. Rev.}{D59}{093003}{1999}.


\bibitem{cosmic nutau}
\art[hep-ph/0407182]{H. Athar, C. Kim}{Phys. Lett.}{B598}{1}{2004}.


\bibitem{corona}
\art[hep-ph/9604288]{G. Ingelman, M. Thunman}{\PR}{D54}{4385}{1996}. 



\bibitem{superheavy}
Neutrinos from superheavy DM: \art[hep-ph/0009017]{I. F. M. Albuquerque, L. Hui and E. W. Kolb}{\PR} {D64}{083504} {2001}.


\bibitem{techniDM}
\art{S. Nussinov}{Phys. Lett.}{B279}{1992}{111}, 
    \art[hep-ph/9310290]{J. Bagnasco, M. Dine and S. Thomas}{Phys. Lett.}{B320}{99}{1994}. 
  
  \bibitem{deGouveaMura}
\art[hep-ph/9904399]{A. de Gouvea, A. Friedland, H. Murayama}{\PR}{D60}{093011}{1999}.


\end{thebibliography}
\end{document}